\documentclass[12pt]{article}
\usepackage{stmaryrd}
\usepackage{bbm}
\usepackage{mathrsfs}
\usepackage{amsfonts}
\usepackage{amssymb}
\usepackage{amsmath}
\usepackage{amsthm}

\newtheorem{theorem}{Theorem}
\newtheorem{proposition}{Proposition}
\newtheorem{lemma}{Lemma}
\newtheorem{corollary}{Corollary}

\allowdisplaybreaks[4]
\begin{document}

\title{Classification of classical and non-local symmetries of fourth-order nonlinear evolution
equations \thanks{Supported by NSF-China grants 10671156 and
10771170.} }
\author{{Qing Huang$^{1,2}$, C. Z. Qu$^{1,2}$ and R. Zhdanov$^3$
\thanks{Corresponding author. Email: renat.zhdanov@bio-key.com}}\\
{\small 1. Department of Mathematics, Northwest University, Xi'an
710069},\\ {\small People's Republic of China}\\ {\small 2. Center
for Nonlinear Studies, Northwest University, Xi'an 710069},\\
{\small People's Republic of China}\\ {\small 3. BIO-key
International, 55121 Eagan, MN, USA}}
\date{}
\maketitle

\author{}
\date{}
\maketitle

\begin{abstract}
In this paper, we consider group classification of local and
quasi-local symmetries for a general fourth-order evolution
equations in one spatial variable. Following the approach developed
in \cite{ZhdL99}, we construct all inequivalent evolution equations
belonging to the class under study which admit either semi-simple
Lie groups or solvable Lie groups. The obtained lists of invariant
equations (up to a local change of variables) contain both the
well-known equations and a variety of new ones possessing rich
symmetry. Based on the results on the group classification for local
symmetries, the group classification for quasi-local symmetries of
the equations is also given.
\end{abstract}

\vskip 0.5in

\section*{\small\bf 1. Introduction}

The purpose of this article is to classify local and nonlocal
symmetries of fourth-order evolution equations of the form
\begin{equation}\label{eq}
    u_t=F(t,x,u,u_x,u_{xx},u_{xxx})u_{xxxx}+G(t,x,u,u_x,u_{xx},u_{xxx}).
\end{equation}
Here $F$ and $G$ are arbitrary smooth functions and $F\neq0$.
Hereafter we adopt notations $u=u(t,x),\ u_t=\partial u/\partial t,\
u_x=\partial u/\partial x,\ u_{xx}=\partial^2 u/\partial x^2$, etc.

This paper is a farther development of the the methods
classification approach suggested in \cite{ZhdL99} and successfully
applied to various equations of mathematical and theoretical physics
in \cite{BasGL08, BasLZ01, GunLZ04, Huang09, LahZ05, ZhdL99,
ZhdL07}. It also generalizes the results of \cite{Huang09}.

The class partial differential equations of Eq. \eqref{eq} contains
a number of important mathematical physics equations. Among the
particular cases of Eq.\eqref{eq} there are, the
Kuramoto-Sivashinsky (KS) equation
\begin{equation}\label{ks}
    u_t=-u_{xxxx}-u_{xx}-\frac12u_x^2,
\end{equation}
the extended Fisher-Kolmogorov (eFK) equation
\begin{equation}\label{efk}
    u_t=-u_{xxxx}+u_{xx}-u^3+u,
\end{equation}
the Swift-Hohenberg (SH) equation
\begin{equation}\label{sh}
    u_t=-u_{xxxx}-2u_{xx}-u^3+(\kappa-1)u,\ \ \kappa\in\mathbf{R}.
\end{equation}
It also contains the equation describing surface tension driven thin
film flows
\begin{equation*}
    u_t=-\Big( u^3u_{xxx}+f(u,u_x,u_{xx})\Big)_x,
\end{equation*}
whose appropriate forms have been used to model fluid flows in
physical situations such as coating, draining of foams, and the
movement of contact lenses \cite{Mye98}. Another interesting
particular case is the equation
\begin{equation*}
    u_t=-(f(u)u_{xxx})_x+(g(u)u_x)_x,
\end{equation*}
which appears in the context of surface dominated motion of thin
viscous films and spreading droplets or plasticity (for further
details, see, \cite{Ber98} and the references therein). If
$f(u)=u^n,\ g(u)\equiv0$, it is the generalized lubrication
equation. When $n=3$, it is the capillary-driven flow. Equations
with $n<3$ can be used to describe slip models in the vicinity of
$u\rightarrow0$. Another case of practical importance is $n=1$,
which describes the thickness of a thin bridge between two masses of
fluid in a Hele-Shaw cell \cite{Con93}.

In \cite{Huang09}, we perform preliminary group classification of
the equations
\begin{equation}\label{eq1}
    u_t=-u_{xxxx}+G(t,x,u,u_x,u_{xx},u_{xxx}),
\end{equation}
obtained from Eq.\eqref{eq} by putting $F=-1$. Note that the so
obtained lists of invariant equations contain equations \eqref{eq1}
as particulars cases.

Lie symmetries play the central role in the modern theory of
differential equations. One of the primary questions of theory of
Lie symmetries is whether a given equation admits nontrivial
symmetry. If the answer is positive one can apply the whole wealth
of approaches utilizing symmetries to analyze and solve the equation
in question. So that group classification of equations \eqref{eq} is
necessary first step in utilizing any symmetry methods and
techniques. In the present paper, we perform preliminary group
classification of Eq.\eqref{eq} and describe all possible forms of
the functions $F$ and $G$ such that Eq.\eqref{eq} admit symmetry
groups of the dimension $n\leq4$.

The history of group classification methods goes back to Lie
himself. In \cite{Lie81} he proved that a linear two-dimensional
second-order partial differential equation (PDE) can admit at most
three-parameter invariance group (apart from the trivial
infinite-parameter symmetry group, which is due to linearity). The
modern formulation of the problem of group classification of PDEs
has been suggested by Ovsyannikov \cite{Ovs59}. He introduced a
regular method based on the concept of equivalence group (we will
refer to it as the Lie-Ovsyannikov method). Ovsyannikov's approach
works at its best when the equivalence group of an equation under
study is finite-dimensional and is not very efficient otherwise. So
that the equations with arbitrary elements depending on two and more
arguments cannot be efficiently handled with the Lie-Ovsyannikov
method. To overcome this difficulty, Zhdanov and Lahno developed a
different purely algebraic approach enabling to classify classes of
PDEs having infinite-dimensional equivalence groups \cite{ZhdL99}.

The method suggested in \cite{ZhdL99} has been applied to classify
broad classes of heat conductivity \cite{BasLZ01}, Schr\"odinger
\cite{ZhdR00}, KdV-type evolution \cite{GunLZ04}, nonlinear wave
\cite{LahZ05}, general second-order quasi-linear evolution
\cite{ZhdL07}, third-order nonlinear evolution \cite{BasGL08} and
fourth-order evolution \cite{Huang09} equations. Here we follow the
approach of \cite{ZhdL99} in order to describe fourth-order
nonlinear evolution equations \eqref{eq} having non-trivial symmetry
properties.

Our classification algorithm for local symmetries is a combination
of the standard Lie infinitesimal algorithm, equivalence group and
the structures of abstract Lie algebra. It consists of three major
steps. We give a brief description of the steps involved. Further
details can be found in \cite{ZhdL99}.

The first step is to compute the most general symmetry group of
\eqref{eq} together with the classifying equations for $F$ and $G$.
In addition, we calculate the maximal local equivalence group
admitted by Eq.\eqref{eq} under consideration. The second step is
essentially based on the explicit forms of commutation relations of
low dimensional abstract Lie algebras \cite{BarR86,PatSWZ76,Tur88}.
Using these we construct all inequivalent realizations of symmetry
algebras by basis infinitesimal operators admitted by Eq.\eqref{eq}.
The third step is inserting the canonical forms of symmetry
generators into the classifying equations, solving them, and
deriving the invariant equations. Also we need to make sure that the
corresponding symmetry algebras are maximal in Lie's sense.

Since Lie symmetry is not always the best answer to all challenges
of the modern theory of nonlinear differential equations, one has
always been looking for ways to generalize it. A generalization is
to allow for infinitesimals depending on the global behavior of the
dependent variable, which is just the way the non-local symmetries
arose. The most well studied non-local symmetries are those of
linear PDEs \cite{FuN1,FuN2}, while for those of nonlinear PDEs,
much less is known. One possible approach has been suggested by
Bluman \cite{Blu93,BluRK88}. He derived non-local symmetries
(potential symmetries) admitted by a given differential equation by
realizing such symmetries as local symmetries of an associated
auxiliary system. There is another way to constructing non-local
symmetries which is to apply a non-local transformation to an
equation admitting nontrivial Lie symmetries. Then some of Lie
symmetries of the initial equation will remain Lie symmetries of the
transformed equation, while the others become non-local ones, which
are called quasi-local symmetries. The term quasi-local has been
introduced independently in \cite{AkhGI91} and \cite{MeiPS97} to
distinguish non-local symmetries that are equivalent to local ones
through non-local transformation. It was noted in \cite{BasLZ01}
that results of the group classification for local symmetries can be
utilized to derive quasi-local symmetries of PDEs under study.
Zhdanov and Lahno obtained some nontrivial examples of second-order
evolution equations with quasi-local symmetries in \cite{ZhdL05}.
recently, Zhdanov suggested a regular group-theoretical approach to
the problem of classification of evolution equations that admit
quasi-local symmetries \cite{Zhd09}.

The principal motivation for the present paper is a need for a
unified group classification approach enabling to obtain both
local/Lie and quasi-local/non-local symmetries of equations of the
form \eqref{eq}. This approach is essentially based on the methods
developed in the above mentioned papers \cite{ZhdL99, Zhd09}.

The outline of this paper is as follows. In Section 2, we obtain the
classifying equations for the functions $F$, $G$ and give the
equivalence group of Eq.\eqref{eq}. In Section 3, we provide a
detailed description of equations of the form \eqref{eq} invariant
with respect to semi-simple algebras. Section 4 is devoted to the
classification of those evolution equations \eqref{eq} which admit
solvable Lie algebras. In Section 5, utilizing the results obtained
while classifying for local symmetries, we derive the realizations
of Lie algebras leading to equations admitting quasi-local
symmetries.

\section*{\small\bf 2. Preliminary group analysis of Eq.\eqref{eq} }

In this section we derive the classifying equations for coefficients
of ininitesimal operators of symmetry group of Eq.\eqref{eq}. Next,
we obtain its most general equivalence group. It is a common
knowledge that the Lie transformation group admitted by \eqref{eq}
is generated by the infinitesimal operators of the form
\begin{equation}\label{V}
    V=\tau(t,x,u)\partial_t+\xi(t,x,u)\partial_x+\eta(t,x,u)\partial_u,
\end{equation}
where $\tau, \xi, \eta$ are arbitrary, real-valued smooth functions
defined in some subspace of the space $V={\mathbb{R}^2 \times
\mathbb{R}^1}$ of the independent $\langle t, x \rangle$ and the
dependent $\langle u \rangle$ variables.

The vector field \eqref{V} generates one-parameter invariance group
of \eqref{eq} iff its coefficients $\tau, \xi, \eta$ satisfy the
determining equation (Lie's invariance criterion)
\begin{equation*}
\begin{split}
    &\eta^t-[\tau F_t+\xi F_x+\eta F_u+\eta^x F_{u_x}+\eta^{xx}
    F_{u_{xx}}+\eta^{xxx} F_{u_{xxx}}]u_{xxxx}-\eta^{xxxx}F\\[2mm]
    &-[\tau G_t+\xi F_x+\eta G_u+\eta^x G_{u_x}+\eta^{xx}
    G_{u_{xx}}+\eta^{xxx} G_{u_{xxx}}]|_{u_t=Fu_{xxxx}+G}=0,
\end{split}
\end{equation*}
where
\begin{align*}
        \eta^t      &= D_t(\eta)-u_tD_t(\tau)-u_xD_t(\xi), \\[2mm]
        \eta^x      &= D_x(\eta)-u_tD_x(\tau)-u_xD_x(\xi), \\[2mm]
        \eta^{xx}   &= D_x(\eta^x)-u_{xt}D_x(\tau)-u_{xx}D_x(\xi), \\[2mm]
        \eta^{xxx}  &= D_x(\eta^{xx})-u_{xxt}D_x(\tau)-u_{xxx}D_x(\xi), \\[2mm]
        \eta^{xxxx} &= D_x(\eta^{xxx})-u_{xxxt}D_x(\tau)-u_{xxxx}D_x(\xi).
\end{align*}
and the operators $D_t$ and $D_x$ denote the total derivatives with
respect to the variables $t$ and $x$ respectively,
\begin{align*}
  D_t &=
  \partial_t+u_t\partial_u+u_{tt}\partial_{u_t}+u_{xt}\partial_{u_x}+\dots,
  \\[2mm] D_x &=
  \partial_x+u_x\partial_u+u_{tx}\partial_{u_t}+u_{xx}\partial_{u_x}+\dots.
\end{align*}
Equating coefficients of linearly independent terms of invariance
condition above to zero yields an over-determined system of linear
PDEs. Solving it, we arrive at the following assertion.
\begin{proposition}
The most general symmetry group of \eqref{eq} is generated by the
infinitesimal operators
\begin{equation}\label{sym}
    V=\tau(t)\partial_t+\xi(t,x,u)\partial_x+\eta(t,x,u)\partial_u,
\end{equation}
where $\tau, \xi$ and $\eta$ are real-valued functions satisfying
the classifying equations
\begin{equation}\label{class}
\begin{split}
    &( 4\xi_u u_x+4\xi_x-\dot{\tau} )F-\tau F_t-\xi F_x-\eta
     F_u+( u_x\xi_x-u_x\eta_u+u_x^2\xi_u-\eta_x )F_{u_x}\\[2mm]
    &\qquad ( -u_{xx}\eta_u-\eta_{xx}+u_x^3\xi_{uu}-u_x^2\eta_{uu}-2u_x\eta_{xu}+2u_{xx}\xi_x+u_x\xi_{xx}+2u_x^2\xi_{xu}\\[2mm]
    &\qquad +3u_xu_{xx}\xi_u )F_{u_{xx}}+( -3u_x^2\eta_{x,u,u}-\eta_{xxx}+u_x^4\xi_{uuu}-3u_x\eta_{xxu}-3u_{xx}\eta_{xu}\\[2mm]
    &\qquad +3u_{xxx}\xi_x+3u_{xx}^2\xi_u+3u_{xx}\xi_{xx}+3u_x^3\xi_{xuu}+u_x\xi_{xxx}+6u_x^2u_{xx}\xi_{uu}-u_x^3\eta_{uuu}\\[2mm]
    &\qquad +9u_xu_{xx}\xi_{xu}-u_{xxx}\eta_u+4u_xu_{xxx}\xi_u+3u_x^2\xi_{xxu}-3u_xu_{xx}\eta_{uu} )F_{u_{xxx}}=0,\\[2mm]
    &( -12u_xu_{xx}\eta_{xuu}-u_x^4\eta_{uuuu}+u_x\xi_{xxxx}+u_x^5\xi_{uuuu}+4u_x^4\xi_{xuuu}-6u_x^2\eta_{xxuu}\\[2mm]
    &\qquad +6u_x^3\xi_{xxuu}-4u_{xxx}\eta_{xu}-6u_{xx}\eta_{xxu}-4u_x\eta_{xxxu}+12u_{xx}^2\xi_{xu}-4u_x^3\eta_{xuuu}\\[2mm]
    &\qquad -3u_{xx}^2\eta_{uu}+4u_x^2\xi_{xxxu}+4u_{xx}\xi_{xxx}+6u_{xxx}\xi_{xx}-\eta_{xxxx}-6u_x^2u_{xx}\eta_{uuu}\\[2mm]
    &\qquad 16u_xu_{xxx}\xi_{xu}-4u_xu_{xxx}\eta_{uu}+10u_x^2u_{xxx}\xi_{uu}+15u_xu_{xx}^2\xi_{uu}+10u_x^3u_{xx}\xi_{uuu}\\[2mm]
    &\qquad +10u_{xx}u_{xxx}\xi_u+24u_x^2u_{xx}\xi_{xuu}+18u_xu_{xx}\xi_{xxu} )F+( \eta_u-\dot{\tau}-u_x\xi_u )G\\[2mm]
    &\qquad -\tau G_t-\xi G_x-\eta G_u+( u_x\xi_x-u_x\eta_u+u_x^2\xi_u-\eta_x )G_{u_x}+( -u_{xx}\eta_u-\eta_{xx}\\[2mm]
    &\qquad +u_x^3\xi_{uu}-u_x^2\eta_{uu}-2u_x\eta_{xu}+2u_{xx}\xi_x+u_x\xi_{xx}+2u_x^2\xi_{xu}+3u_xu_{xx}\xi_u )G_{u_{xx}}\\[2mm]
    &\qquad ( -3u_x^2\eta_{xuu}-\eta_{xxx}+u_x^4\xi_{uuu}-3u_x\eta_{xxu}-3u_{xx}\eta_{xu}+3u_{xxx}\xi_x+3u_{xx}^2\xi_u\\[2mm]
    &\qquad +3u_{xx}\xi_{xx}+3u_x^3\xi_{xuu}+u_x\xi_{xxx}+6u_x^2u_{xx}\xi_{uu}-u_x^3\eta_{uuu}+9u_xu_{xx}\xi_{xu}\\[2mm]
    &\qquad -u_{xxx}\eta_u+4u_xu_{xxx}\xi_u+3u_x^2\xi_{xxu}-3u_xu_{xx}\eta_{uu}
    )G_{u_{xxx}}-u_x\xi_t+\eta_t=0,
\end{split}
\end{equation}
Here and after the dot over a symbol stands for differentiation with
respect to its argument.
\end{proposition}

If there are no restrictions on $F$ and $G$, then \eqref{class}
should be satisfied identically, which is only possible when the
symmetry group is the trivial group of identity transformations. Our
goal is finding all specific forms of $F$ and $G$ for which
Eq.\eqref{eq} admits non-trivial symmetry groups. The basic idea, as
suggested in \cite{ZhdL99}, is to utilize the fact that for
arbitrarily fixed $F$ and $G$, all admissible vector fields form a
Lie algebra. So that we can use the results of classification of
abstract low-dimensional Lie algebras to construct all Lie
symmetries of the form \eqref{sym} admitted by \eqref{eq}. The next
step is integrating the classifying equations, which yields the
explicit form of the functions $F,\ G$.

As a first step we need to compute the equivalence group of
Eq.\eqref{eq}. To this end we have to construct all possible
invertible changes of variables
\begin{equation*}
    \bar{t}=T(t,x,u),\quad \bar{x}=X(t,x,u),\quad
    \bar{u}=U(t,x,u),\qquad \frac{D(T,X,U)}{D(t,x,u)}\neq 0,
\end{equation*}
which do not alter the form of Eq.\eqref{eq}. We present the final
result of our calculations without detailed proof.
\begin{proposition}\rm\cite{BasGL08}
The maximal equivalence group of Eq.\eqref{eq} reads as
\begin{equation}\label{tranf}
    \bar{t}=T(t),\quad \bar{x}=X(t,x,u),\quad
    \bar{u}=U(t,x,u),
\end{equation}
where $\dot{T}\neq0$ and $\frac{D(X,U)}{D(x,u)}\neq 0$.
\end{proposition}

We use equivalence transformation \eqref{tranf} to transform vector
field \eqref{sym} to a simplest possible (canonical) form.
\begin{lemma}
Within the point transformation \eqref{tranf}, vector field
\eqref{sym} is equivalent to one of the following canonical
operators
\begin{equation}\label{1-sym}
    \partial_t,\qquad \partial_x.
\end{equation}
\end{lemma}
\proof Transformation \eqref{tranf} converts operator \eqref{sym}
into
\begin{equation*}
    \bar{V}=\tau\dot{T}\partial_{\bar{t}}
    +(\tau X_t+\xi X_x+\eta X_u)\partial_{\bar{x}}
    +(\tau U_t+\xi U_x+\eta U_u)\partial_{\bar{u}}.
\end{equation*}

The are two cases $\tau\neq 0$ and $\tau=0$ to consider.

If $\tau\neq 0$, choosing in \eqref{tranf} the function $T$ to
satisfy $\tau\dot{T}=1$ and the functions $X$ and $U$ to be any two
independent solutions of the equation
\begin{equation*}
    \tau Y_t+\xi Y_x+\eta Y_u=0,\qquad Y=Y(t,x,u),
\end{equation*}
leads to the canonical form $\partial_t$.

Suppose now that $\tau=0$ and $\xi^2+\eta^2\neq 0$ (otherwise
operator \eqref{sym} vanishes identically), then \eqref{sym} is
transformed to
\begin{equation*}
    \bar{V}=(\xi X_x+\eta X_u)\partial_{\bar{x}}+(\xi U_x+\eta U_u)\partial_{\bar{u}}.
\end{equation*}

If $\xi\neq 0$, then choosing in \eqref{tranf} a particular solution
of $\xi X_x+\eta X_u=1$ as the function $X$ and a fundamental
solution of $\xi U_x+\eta U_u=0$ as $U$, transforms \eqref{sym} into
$\partial_x$.

If $\xi=0,\ \eta\neq 0$, by making the transformation \eqref{tranf}
with $\bar{t}=t,\ \bar{x}=u,\ \bar{u}=x$, we get the case $\xi\neq
0$, which has been already considered.

It is straightforward to verify that there is no transformation
\eqref{tranf} which can transform $\partial_t$ and $\partial_x$ into
another. \qed

Consequently, we have two inequivalent realizations of
one-dimensional symmetry algebras $\partial_t$ and $\partial_x$.
Integrating the classifying equations for each symmetry operator
yields the corresponding inequivalent equations of class \eqref{eq}.

In what follows, we use the notation $A_{k.i}=\langle
V_1,V_2,\cdots,V_k\rangle$ to denote a $k$-dimensional Lie algebra
with basis elements $V_i$ $(i=1,2,\cdots,k)$, the index $i$ standing
for the number of the class to which the given algebra belongs.
\begin{theorem}
There are two inequivalent equations \eqref{eq} invariant under
one-para\-me\-ter symmetry groups:
\begin{equation*}
\begin{array}{ll}
A^1_1=\langle \partial_t\rangle:
&u_t=F(x,u,u_x,u_{xx},u_{xxx})u_{xxxx}+G(x,u,u_x,u_{xx},u_{xxx}),\\[2mm]
A^2_1=\langle \partial_x\rangle:
&u_t=F(t,u,u_x,u_{xx},u_{xxx})u_{xxxx}+G(t,u,u_x,u_{xx},u_{xxx}).
\end{array}
\end{equation*}
Here $F$ and $G$ are arbitrary smooth functions. Furthermore, the
associated symmetry algebra is maximal in Lie's sense.
\end{theorem}

\section*{\small\bf 3. Classification of equations invariant under
semi-simple Lie algebras}

In this section, we classify all the equations of the form
\eqref{eq} admitting semi-simple Lie algebras. The lowest order real
semi-simple Lie algebras are isomorphic to one of the following two
three-dimensional algebras:
\begin{equation*}
\begin{array}{cl}
    {\it{so}}(3): &[V_1,V_2]=V_3,\ [V_1,V_3]=-V_2,\ [V_2,V_3]=V_1;\\[2mm]
    {\it{sl}}(2,\mathbb{R}): &[V_1,V_2]=2V_2,\ [V_1,V_3]=-2V_3,\
    [V_2,V_3]=V_1.
\end{array}
\end{equation*}

We begin by studying realizations of the algebras {\it{so}}(3)
within the class of operators \eqref{sym}.

Taking into account of our preliminary classification, we can assume
without any loss of generality that one of the basis operators, say
$V_1$, is reduced to one of the canonical forms $\partial_t$ and
$\partial_x$.

Let $V_1=\partial_t$ and $V_2$, $V_3$ be of the form \eqref{sym},
i.e., $V_i=\tau_i(t)\partial_t + \xi_i(t,x,u)\partial_x +
\eta_i(t,x,u)\partial_u,\ i=2,3$. Using $[V_1,V_2]=V_3$ we find that
$\dot{\tau_2}=\tau_3$. Nest, it follows from $[V_1,V_3]=-V_2$ that
$\dot{\tau_3}=-\tau_2$. With these facts in hand and using the
relation $[V_2,V_3]=V_1$ we arrive at $\tau_2\dot{\tau_3} -
\tau_3\dot{\tau_2}=1$, which in turn yields
equation$\tau_2^2+\dot{\tau_2}^2=-1$. This equation has no real
solutions, so there are no realizations of {\it{so}}(3) with
$V_1=\partial_t$.

Turn now to the case $V_1=\partial_x$. A similar analysis yileds a
unique realization of {\it{so}}(3)
\begin{equation*}
    \langle \partial_x,\tan u\sin x\partial_x+\cos x\partial_u,\tan u\cos x\partial_x-\sin x\partial_u \rangle
\end{equation*}
However, this algebra cannot be symmetry algebra of an equation of
the form \eqref{eq}.

\begin{theorem}
There are no {\it{so}}(3)-invariant equations of the form
\eqref{eq}.
\end{theorem}

In a similar way we analyze possible realizations of the algebra
${\it{sl}}(2,\mathbb{R})$ and get the following assertion.
\begin{theorem}\label{semi}
There are six inequivalent realizations of ${\it{sl}}(2,\mathbb{R})$
by operators \eqref{sym}, which are admitted by Eq.\eqref{eq}. These
realizations and the corresponding forms of the functions $F,\ G$
are presented below
\begin{align*}
& {\it{sl}}^1(2,\mathbb{R})=\langle 2t\partial_t+x\partial_x,-t^2\partial_t-tx\partial_x+x^2\partial_u,\partial_t \rangle:\\[2mm]
& \qquad\qquad              F=x^2F(\omega_1,\omega_2,\omega_3),\ G=\frac{G(\omega_1,\omega_2,\omega_3)}{x^2}+\frac{xuu_x-u^2}{x^2},\\[2mm]
& \qquad\qquad\qquad        \omega_1=xu_x-2u,\omega_2=x^2u_{xx}-2u,\ \omega_3=x^3u_{xxx},\\[2mm]
& {\it{sl}}^2(2,\mathbb{R})=\langle 2t\partial_t+x\partial_x,-t^2\partial_t+(x^3-tx)\partial_x,\partial_t \rangle:\\[2mm]
& \qquad\qquad              F=\frac{F(u,\omega_1,\omega_2)}{x^3u_x^5},\\[2mm]
& \qquad\qquad              G=\frac{18x^2u_{xxx}+87xu_{xx}+105u_x}{x^6u_x^5}F(u,\omega_1,\omega_2)+\frac{G(u,\omega_1,\omega_2)}{x^3u_x}-\frac{u_x}{4x},\\[2mm]
& \qquad\qquad\qquad        \omega_1=\frac{xu_{xx}+3u_x}{xu_x^2},\omega_2=\frac{x^2u_{xxx}+9xu_{xx}+15u_x}{x^2u_x^3},\\[2mm]
& {\it{sl}}^3(2,\mathbb{R})=\langle 2x\partial_x-u\partial_u,-x^2\partial_x+xu\partial_u,\partial_x \rangle:\\[2mm]
& \qquad\qquad              F=\frac{F(t,\omega_1,\omega_2)}{u^8},\ G=-8\frac{2u_{xxx}u^2-9u_xu_{xx}u+9u_x^3}{u^{11}}F(t,\omega_1,\omega_2)+u G(t,\omega_1,\omega_2),\\[2mm]
& \qquad\qquad\qquad        \omega_1=\frac{uu_{xx}-2u_x^2}{u^6},\omega_2=\frac{u_{xxx}u^2-9u_xu_{xx}u+12u_x^3}{u^9},\\[2mm]
& {\it{sl}}^4(2,\mathbb{R})=\langle 2x\partial_x,-x^2\partial_x,\partial_x \rangle:\\[2mm]
& \qquad\qquad              F=\frac{F(t,u,\omega)}{u_x^4},\ G=6\frac{u_{xx}}{u_x^6}(u_{xx}^2-u_xu_{xxx})F(t,u,\omega)+u G(t,u,\omega),\\[2mm]
& \qquad\qquad\qquad        \omega=\frac{3u_{xx}^2-2u_xu_{xxx}}{u_x^4},\\[2mm]
& {\it{sl}}^5(2,\mathbb{R})=\langle 2x\partial_x-u\partial_u,(\frac1{u^4}-x^2)\partial_x+xu\partial_u,\partial_x \rangle:\\[2mm]
& \qquad\qquad              F=\frac{u^4}{(u^6+4u_x^2)^2}F(t,\omega_1,\omega_2),\\[2mm]
& \qquad\qquad              G=-\frac{u}{4(u^6+4u_x^2)^4}[146880u_x^8+192u(641u^5-167u_{xx})u_x^6-896u^2u_{xxx}u_x^5\\[2mm]
& \qquad\qquad\               +96u^2(417u^{10}-208u_{xx}u^5+28u_{xx}^2)u_x^4+32u^3u_{xxx}(u^5+20u_{xx})u_x^3\\[2mm]
& \qquad\qquad\               +6u^3(855u^{15}-1092u_{xx}u^{10}-176u_{xx}^2u^5-160u_{xx}^3)u_x^2+32u^9u_{xxx}(2u^5+5u_{xx})u_x\\[2mm]
& \qquad\qquad\               +3u^{14}(77u^{10}-162u_{xx}u^5+36u_{xx}^2)]F(t,\omega_1,\omega_2)+\frac{(u^6+4u_x^2)^\frac12}{u^2}G(t,\omega_1,\omega_2),\\[2mm]
& \qquad\qquad\qquad        \omega_1=\frac{u^3}{(u^6+4u_x^2)^\frac32}(u^6-2uu_{xx}+10u_x^2),\\[2mm]
& \qquad\qquad\qquad        \omega_2=\frac{u^3}{(u^6+4u_x^2)^3}[u_{xxx}u^8-9u_xu_{xx}u^7+12u_x^3u^6+4u_x(u_xu_{xxx}-3u_{xx}^2)u^2\\[2mm]
& \qquad\qquad\qquad\         +36u_x^3u_{xx}u-60u_x^5],\\[2mm]
& {\it{sl}}^6(2,\mathbb{R})=\langle 2x\partial_x-u\partial_u,-(x^2+\frac1{u^4})\partial_x+xu\partial_u,\partial_x \rangle:\\[2mm]
& \qquad\qquad              F=\frac{u^4}{(u^6-4u_x^2)^2}F(t,\omega_1,\omega_2),\\[2mm]
& \qquad\qquad              G=-\frac{u}{4(u^6-4u_x^2)^4}[146880u_x^8-192u(641u^5+167u_{xx})u_x^6-896u^2u_{xxx}u_x^5\\[2mm]
& \qquad\qquad\               +96u^2(417u^{10}+208u_{xx}u^5+28u_{xx}^2)u_x^4-32u^3u_{xxx}(u^5-20u_{xx})u_x^3\\[2mm]
& \qquad\qquad\               -6u^3(855u^{15}+1092u_{xx}u^{10}-176u_{xx}^2u^5+160u_{xx}^3)u_x^2+32u^9u_{xxx}(2u^5-5u_{xx})u_x\\[2mm]
& \qquad\qquad\               +3u^{14}(77u^{10}+162u_{xx}u^5+36u_{xx}^2)]F(t,\omega_1,\omega_2)+\frac{(4u_x^2-u^6)^\frac12}{u^2}G(t,\omega_1,\omega_2),\\[2mm]
& \qquad\qquad\qquad        \omega_1=\frac{u^3}{(4u_x^2-u^6)^\frac32}(u^6+2uu_{xx}-10u_x^2),\\[2mm]
& \qquad\qquad\qquad        \omega_2=\frac{u^3}{(u^6-4u_x^2)^3}[u_{xxx}u^8-9u_xu_{xx}u^7+12u_x^3u^6+4u_x(3u_{xx}^2-u_xu_{xxx})u^2\\[2mm]
& \qquad\qquad\qquad\         -36u_x^3u_{xx}u+60u_x^5].
\end{align*}
The algebras ${\it{sl}}^i(2,\mathbb{R}),\ (i=1,2,\dots,6)$ are the
maximal invariance algebras of the corresponding PDEs, provided the
functions $F$ and $G$ are arbitrary.
\end{theorem}

\begin{theorem}
The invariant equations listed in Theorem \ref{semi} exhaust the
list of all possible inequivalent PDEs \eqref{eq}, whose invariance
algebras are semi-simple.
\end{theorem}
\proof It is a common knowledge that the lowest dimensional
semi-simple Lie algebras admit the following isomorphisms
\begin{equation*}
    {\it{so}}(3)\sim{\it{su}}(2)\sim{\it{sp}}(1),\qquad
    {\it{sl}}(2,\mathbb{R})\sim{\it{su}}(1,1)\sim{\it{so}}(2,1)
    \sim{\it{sp}}(1,\mathbb{R})
\end{equation*}
(see, e.g., \cite{BarR86}). Hence it immediately follows that the
realizations of the algebra ${\it{sl}}(2,\mathbb{R})$ exhaust the
set of all possible inequivalent realizations of three-dimensional
semi-simple Lie algebras admitted by \eqref{eq}.

The next admissible dimension for classical semi-simple Lie algebras
is six. There are four non-isomorphic semi-simple Lie algebras over
the real number field, namely, ${\it{so}}(4)$,
${\it{so}}^{\ast}(4)$, ${\it{so}}(3,1)$, and ${\it{so}}(2,2)$. As
${\it{so}}(4)\sim{\it{so}}(3)\oplus{\it{so}}(3)$,
${\it{so}}^{\ast}(4)\sim{\it{so}}(3)\oplus{\it{sl}}(2,\mathbb{R})$,
and the algebra ${\it{so}}(3,1)$ contains ${\it{so}}(3)$ as a
subalgebra, there are no realization of these algebras by operator
\eqref{sym}. Therefore ${\it{so}}(2,2)$ is the only candidate for
six-dimensional semi-simple symmetry algebra of \eqref{eq}.

In view of ${\it{so}}(2,2) \sim{\it{sl}}(2,\mathbb{R})
\oplus{\it{sl}}(2,\mathbb{R})$, we can choose
${\it{so}}(2,2)=\langle Q_i,K_i|i=1,2,3\rangle$, where $\langle
Q_1,Q_2,Q_3\rangle$ and $\langle K_1,K_2,K_3\rangle$ are all
${\it{sl}}(2,\mathbb{R})$ algebras, and $[Q_i,K_j]=0,\ (i,j=1,2,3)$.
Taking $Q_1,\ Q_2,\ Q_3$ to be the basis operators of the
realizations of ${\it{sl}}(2,\mathbb{R})$ given in the Theorem
\ref{semi}, and $K_1,\ K_2,\ K_3$ be of the general form
\eqref{sym}, we establish that realizations of
${\it{sl}}(2,\mathbb{R})$ cannot be extended to a realization of
${\it{so}}(2,2)$. Hence no equation of form \eqref{eq} is invariant
under six-dimensional semi-simple Lie algebras.

The similar reasoning yields that there are no realizations of
eight-dimensional semi-simple Lie algebras
${\it{sl}}(3,\mathbb{R})$, ${\it{su}}(3)$ and ${\it{su}}(2,1)$.

The same assertion holds true for the algebras $A_{n-1}\ (n>1)$,
$D_n\ (n>1)$, $B_n\ (n\geq1)$, $C_n\ (n\geq1)$ and the exceptional
semi-simple Lie algebras $G_2$, $F_4$, $E_6$, $E_7$, $E_8$.

The theorem is proved.\qed

\section*{\small\bf 4. Classification of equations invariant under
solvable Lie algebras}

Using the concept of compositional series for a solvable algebra we
can construct all possible realizations of solvable Lie algebras
admitted by Eq.\eqref{eq} starting from the one-dimensional ones and
proceeding to the solvable Lie algebras of the dimension
$2,3,4,\ldots$ (for more details, see \cite{BasLZ01}).

In this section, we describe inequivalent equations of the form
Eq.\eqref{eq} which are invariant under solvable Lie algebras of the
dimension up to four. Since equations invariant with respect to
one-dimensional algebras have already been constructed, we start by
analyzing two-dimensional solvable algebras.

\subsection*{\small\bf 4.1. Equations with two-dimensional Lie algebras}

There are two non-isomorphic two-dimensional Lie algebras
\begin{equation*}
    A_{2.1}:\ \ [V_1,V_2]=0,\qquad A_{2.2}:\ \ [V_1,V_2]=V_2.
\end{equation*}

As both $A_{2.1}$ and $A_{2.2}$ contain the algebra $A_1$, we can
assume that the basis operator of the latter is reduced to one of
the canonical forms.

We consider in detail the case of $A_{2.1}$, while for the case of
algebra $A_{2.2}$ we present the final results only.

Let $V_1=\partial_t$ and $V_2$ be an operator of the most general
form \eqref{sym}
\begin{equation*}
V_2=\tau(t)\partial_t+ \xi(t,x,u)\partial_x+ \eta(t,x,u)\partial_u.
\end{equation*}
Then the commutation relation implies that $\dot{\tau}=
\xi_t=\eta_t=0$. Therefore $\tau$ is a constant and $\xi=\xi(x,u),\
\eta=\eta(x,u)$. So without any loss of generality we can put
$V_2=\xi(x,u)\partial_x+\eta(x,u)\partial_u$. Before simplifying
$V_2$ with equivalence transformations \eqref{tranf}, we seek those
equivalent transformations which preserve the basis operator $V_1$
\begin{equation*}
V_1 \to \bar V_1=\dot T \partial_{\bar t} + X_t\partial_{\bar x}
+U_t\partial_{\bar u} =\partial_{\bar t}.
\end{equation*}
Hence, $\dot T=1$ and $X_t=U_t=0$. Consequently, we can take $T=t, \
X=X(x,u)$ and $U=U(x,u)$. Performing this transformation yields
\begin{equation*}
  V_2\rightarrow\bar{V_2}=(\xi X_x+\eta X_u)\partial_{\bar{x}}
  +(\xi U_x+\eta U_u)\partial_{\bar{u}}.
\end{equation*}

If $\eta=0$, we choose $U=U(u)$ and $\xi$ satisfying $\xi X_x=1$
thus getting $\bar{V_2}= \partial_{\bar{x}}$. Provided $\eta\neq 0$,
we can take $X$ as a solution of $\xi X_x+\eta X_u=1$ and select $U$
to satisfy $\xi U_x+\eta U_u=0$, get $\partial_{\bar{x}}$ again.
Thus within the action of equivalence group of Eq.\eqref{eq}, we get
the following two-dimensional invariance algebra $\langle
\partial_t, \partial_x \rangle$.

Consider now the case when $V_1=\partial_x$ and $V_2$ is an operator
of the form \eqref{sym}. Inserting $V_1$ and $V_2$ into the
commutation relation yields $V_2=\tau(t)\partial_t
+\xi(t,u)\partial_x+ \eta(t,u)\partial_u$. The equivalence
transformation, which leaves $V_1$ invariant, reads as
\begin{equation*}
    \bar x = x+X(t,u),  \ \bar u = U(t,u)
\end{equation*}
with $U_u\neq 0$. This transformation reduces $V_2$ to the form
\begin{equation*}
   \bar V_2=\tau\dot{T}\partial_{\bar t}+(\tau X_t+\xi+\eta X_u)
   \partial_{\bar x}+(\tau U_t+\eta U_u)\partial_{\bar u}.
\end{equation*}

We consider the cases $\tau=0$ and $\tau\neq 0$ separately.

If $\tau=0$ and $\eta=0$, then $\bar{V_2}= \xi \partial_{\bar{x}}$.
If $\xi_u\neq 0$, choosing $U=\xi$ yields
$\bar{u}\partial_{\bar{x}}$. If $\xi_u=0$ and $\xi_t=0$, then
$\bar{V_2}=\partial_{\bar{x}}$. And if $\xi_u=0$ and $\xi_t\neq0$,
selecting $T=\xi$ yields $\bar{V_2}=\bar{t}\partial_{\bar{x}}$. It
is straightforward to verify that the so obtained two-dimensional
Lie algebra, $\langle \partial_x,t\partial_x\rangle$ cannot be
invariance algebras of an equation of the form \eqref{eq}. Now if
$\tau=0$ and $\eta\neq0$, we choose $U$ and $X$ satisfying $\eta
U_u=1$ and $\xi+\eta X_u=0$ and get $\partial_{\bar{u}}$.

Turn next to the remaining case $\tau\neq0$. Choosing of $T$, $X$
and $U$ as the solutions of the equations $\tau\dot{T}=1$, $\tau
X_t+\xi+\eta X_u=0$ and $\tau U_t+\eta U_u=0$, respectively, we
arrive at the operator $\partial_{\bar{t}}$. This yields the already
known realization $\langle \partial_x,\partial_t\rangle$.

This completes analysis of the realizations of the algebra
$A_{2.1}$. The case of $A_{2.2}$ is treated similarly.

Substituting the so obtained basis operators into the classifying
equations and solving the latter yields the corresponding invariant
equations.

\begin{theorem}\label{2D}
There exist three Abelian and four non-Abelian two-dimensional
symmetry algebras admitted by \eqref{eq}. These algebras and the
corresponding invariant equations are given below
\begin{center}
\begin{tabular*}{148mm}{@{\extracolsep\fill}lll}\hline
\qquad\ \  Algebra                                         & \qquad\qquad$F$                   & \qquad\qquad$G$          \\[0.10mm]\hline
$A^1_{2.1}\langle \partial_t,\partial_x\rangle$        & $F(u,u_x,u_{xx},u_{xxx})$        & $G(u,u_x,u_{xx},u_{xxx})$\\[3mm]
$A^2_{2.1}\langle \partial_x,\partial_u\rangle$        & $F(t,u_x,u_{xx},u_{xxx})$        & $G(t,u_x,u_{xx},u_{xxx})$\\[3mm]
$A^3_{2.1}\langle \partial_u,x\partial_u\rangle$       & $F(t,x,u_{xx},u_{xxx})$          & $G(t,x,u_{xx},u_{xxx})$\\[3mm]
$A^1_{2.2}\langle -t\partial_t-x\partial_x,\partial_t\rangle$
                                                       & $x^3F(u,xu_x,x^2u_{xx},x^3u_{xxx})$ &$x^{-1}G(u,xu_x,x^2u_{xx},x^3u_{xxx})$\\[3mm]
$A^2_{2.2}\langle -t\partial_t-x\partial_x,\partial_x\rangle$
                                                       & $t^3F(u,tu_x,t^2u_{xx},t^3u_{xxx})$ &$t^{-1}G(u,tu_x,t^2u_{xx},t^3u_{xxx})$\\[3mm]
$A^3_{2.2}\langle -x\partial_x-u\partial_u,\partial_x\rangle$
                                                       & $u^4F(t,u_x,uu_{xx},u^2u_{xxx})$    &$uG(t,u_x,uu_{xx},u^2u_{xxx})$\\[3mm]
$A^4_{2.2}\langle -x\partial_x,\partial_x\rangle$
                                                       & $u_x^{-4}F(t,u,u_x^{-2}u_{xx},u_x^{-3}u_{xxx})$   & $G(t,u,u_x^{-2}u_{xx},u_x^{-3}u_{xxx})$\\[.10cm]\hline
\end{tabular*}
\end{center}
\end{theorem}

\subsection*{\small\bf 4.2. Equations admitting three-dimensional solvable Lie
algebras}

We split three-dimensional solvable Lie algebras into two classes.
One contains the algebras which are direct sums of lower dimensional
algebras, the other includes the remaining algebras. We consider the
decomposable and non-decomposable Lie algebras separately.

\subsubsection*{\small\bf 4.2.1 Three-dimensional decomposable
algebras} There exist two non-isomorphic three-dimensional
decomposable Lie algebras, $A_{3.1}$ and $A_{3.2}$,
\begin{equation*}
    A_{3.1}=A_1\oplus A_1\oplus A_1,
\end{equation*}
with commutation relations
\begin{equation*}
    [V_i,V_j]=0\ (i,j=1,2,3),
\end{equation*}
and
\begin{equation*}
    A_{3.2}=A_{2.2}\oplus A_1,
\end{equation*}
with commutation relations
\begin{equation*}
    [V_1,V_2]=V_2,\ [V_1,V_3]=0,\ [V_2,V_3]=0.
\end{equation*}

It is a common knowledge that any three-dimensional solvable Lie
algebra contains two-dimensional solvable algebra as a subalgebra.
So that to describe all possible realizations of three-dimensional
solvable algebras admitted by Eq.\eqref{eq} it suffices to consider
all possible extensions of two dimensional algebras listed in
Theorem \ref{2D} by vector fields $V_3$, of the form \eqref{sym}.
Then for each of the so obtained realizations we simplify $V_3$
using equivalence transformations which preserve the operators $V_1$
and $V_2$. Having performed these two steps we obtain the following
list of invariant equations \eqref{eq}:

$A_{3.1}-$ invariant equations
\begin{align*}
& A^1_{3.1}=\langle \partial_t,\partial_x,\partial_u \rangle:\\[2mm]
& \qquad\qquad             F=F(u_x,u_{xx},u_{xxx}),\quad G=G(u_x,u_{xx},u_{xxx}),\\[2mm]
& A^2_{3.1}=\langle\partial_t,\partial_u,x\partial_u\rangle:\\[2mm]
& \qquad\qquad             F=F(x,u_{xx},u_{xxx}),\quad G=G(x,u_{xx},u_{xxx}),\\[2mm]
& A^3_{3.1}=\langle \partial_u,x\partial_u,f(t,x)\partial_u \rangle,\ f_{xx}\neq0:\\[2mm]
& \qquad\qquad             F=F(t,x,\omega),\quad G=-\frac{f_{xxxx}}{f_{xx}}u_{xx}F(t,x,\omega)+G(t,x,\omega)+\frac{f_t}{f_{xx}}u_{xx},\\[2mm]
& \qquad\qquad\qquad         \omega=u_{xxx}-\frac{f_{xxx}}{f_{xx}}u_{xx}.\\[2mm]
\end{align*}

$A_{3.2}-$ invariant equations
\begin{align*}
&A^1_{3.2}=\langle -t\partial_t-x\partial_x,\partial_t,\partial_u\rangle:\\[2mm]
&\qquad\qquad                        F=x^3F(\omega_1,\omega_2,\omega_3),\quad G=x^{-1}G(\omega_1,\omega_2,\omega_3)\\[2mm]
&\qquad\qquad\qquad                        \omega_1=xu_x,\ \omega_2=x^2u_{xx},\ \omega_3=x^3u_{xxx},\\[2mm]
&A^2_{3.2}=\langle -t\partial_t-u\partial_u,\partial_t,xu\partial_u\rangle:\\[2mm]
&\qquad\qquad                        F=u^{-1}\mathrm{e}^{x\sigma_1}F(x,\omega_1,\omega_2),\\[2mm]
&\qquad\qquad                        G=\mathrm{e}^{x\sigma_1}[(6\sigma_1^2\sigma_2-4\sigma_1\sigma_3-3\sigma_1^4)F(x,\omega_1,\omega_2)+G(x,\omega_1,\omega_2)],\\[2mm]
&\qquad\qquad\qquad                        \sigma_1=u^{-1}u_x,\ \sigma_2=u^{-1}u_{xx},\ \sigma_3=u^{-1}u_{xxx},\\[2mm]
&\qquad\qquad\qquad                        \omega_1=\sigma_2-\sigma_1^2,\ \omega_2=\sigma_3+2\sigma_1^3-3\sigma_1\sigma_2,\\[2mm]
&A^3_{3.2}=\langle -t\partial_t-x\partial_x,\partial_x,tu\partial_x\rangle:\\[2mm]
&\qquad\qquad                        F=tu_x^{-4}F(u,\omega_1,\omega_2),\\[2mm]
&\qquad\qquad                        G=-5t^{-1}u_x^{-6}u_{xx}(2u_xu_{xxx}-3u_{xx}^2)F(u,\omega_1,\omega_2)+u_xG(u,\omega_1,\omega_2)-t^{-1}u,\\[2mm]
&\qquad\qquad\qquad                        \omega_1=t^{-1}u_x^{-3}u_{xx},\ \omega_2=t^{-1}u_x^{-5}(u_xu_{xxx}-3u_{xx}^2),\\[2mm]
&A^4_{3.2}=\langle -t\partial_t-x\partial_x,\partial_x,\partial_u\rangle:\\[2mm]
&\qquad\qquad                        F=t^3F(\omega_1,\omega_2,\omega_3),\quad G=t^{-1}G(\omega_1,\omega_2,\omega_3),\\[2mm]
&\qquad\qquad\qquad                        \omega_1=tu_x,\ \omega_2=t^2u_{xx},\ \omega_3=t^3u_{xxx},\\[2mm]
&A^5_{3.2}=\langle -t\partial_t-x\partial_x,\partial_x,t\partial_t\rangle:\\[2mm]
&\qquad\qquad                        F=t^{-1}u_x^{-4}F(u,\omega_1,\omega_2),\quad G=t^{-1}G(u,\omega_1,\omega_2),\\[2mm]
&\qquad\qquad\qquad                        \omega_1=u_x^{-2}u_{xx},\ \omega_2=u_x^{-3}u_{xxx},\\[2mm]
&A^6_{3.2}=\langle -t\partial_t-x\partial_x,\partial_x,t\partial_t+\partial_u\rangle:\\[2mm]
&\qquad\qquad                        F=t^3\mathrm{e}^{-4u}F(\omega_1,\omega_2,\omega_3),\quad G=t^{-1}G(\omega_1,\omega_2,\omega_3), \\[2mm]
&\qquad\qquad\qquad                        \omega_1=t\mathrm{e}^{-u}u_x,\ \omega_2=t^2\mathrm{e}^{-2u}u_{xx},\ \omega_3=t^3\mathrm{e}^{-3u}u_{xxx},\\[2mm]
&A^7_{3.2}=\langle -x\partial_x-u\partial_u,\partial_x,u\partial_x \rangle:\\[2mm]
&\qquad\qquad                        F=u^4{u_x}^{-4}F(t,\omega_1,\omega_2),\\[2mm]
&\qquad\qquad                        G=-5u^4u_x^{-6}u_{xx}(2u_xu_{xxx}-3u_{xx}^2)F(t,\omega_1,\omega_2)+uu_xG(t,\omega_1,\omega_2),\\[2mm]
&\qquad\qquad\qquad                        \omega_1=uu_x^{-3}u_{xx},\ \omega_2=u^2u_x^{-5}(u_xu_{xxx}-3u_{xx}^2),\\[2mm]
&A^8_{3.2}=\langle -x\partial_x-u\partial_u,\partial_x,\partial_t \rangle:\\[2mm]
&\qquad\qquad                        F=u^4F(\omega_1,\omega_2,\omega_3),\quad G=uG(\omega_1,\omega_2,\omega_3), \\[2mm]
&\qquad\qquad\qquad                        \omega_1=u_x,\ \omega_2=uu_{xx},\ \omega_3=u^2u_{xxx}, \\[2mm]
&A^9_{3.2}=\langle -x\partial_x-u\partial_u,\partial_x,tu\partial_u  \rangle:\\[2mm]
&\qquad\qquad                        F=u^4u_x^{-4}F(t,\omega_1,\omega_2),\quad G=uG(t,\omega_1,\omega_2)+t^{-1}u\ln{|u_x|},\\[2mm]
&\qquad\qquad\qquad                        \omega_1=uu_x^{-2}u_{xx},\ \omega_2=u^2u_x^{-3}u_{xxx},\\[2mm]
&A^{10}_{3.2}=\langle -x\partial_x,\partial_x,\partial_t \rangle:\\[2mm]
&\qquad\qquad                        F=u_x^{-4}F(t,\omega_1,\omega_2),\quad G=G(t,\omega_1,\omega_2),\\[2mm]
&\qquad\qquad\qquad                        \omega_1=u_x^{-2}u_{xx},\ \omega_2=u_x^{-3}u_{xxx},\\[2mm]
&A^{11}_{3.2}=\langle -x\partial_x,\partial_x,\partial_u \rangle:\\[2mm]
&\qquad\qquad                        F=u_x^{-4}F(u,\omega_1,\omega_2),\quad G=G(u,\omega_1,\omega_2),\\[2mm]
&\qquad\qquad\qquad                        \omega_1=u_x^{-2}u_{xx},\ \omega_2=u_x^{-3}u_{xxx}.\\[2mm]
\end{align*}

\subsubsection*{\small\bf 4.2.2 Three-dimensional non-decomposable
algebras} The list of inequivalent three-dimensional
non-decomposable Lie algebras consists of the following seven Lie
algebras (only nonzero commutation relations are given),
\begin{align*}
&A_{3.3}:\ \ [V_2,V_3]=V_1;\\[2mm]
&A_{3.4}:\ \ [V_1,V_3]=V_1,\quad [V_2,V_3]=V_1+V_2;\\[2mm]
&A_{3.5}:\ \ [V_1,V_3]=V_1,\quad [V_2,V_3]=V_2;\\[2mm]
&A_{3.6}:\ \ [V_1,V_3]=V_1,\quad [V_2,V_3]=-V_2;\\[2mm]
&A_{3.7}:\ \ [V_1,V_3]=V_1,\quad [V_2,V_3]=qV_2,\ (0<|q|<1);\\[2mm]
&A_{3.8}:\ \ [V_1,V_3]=-V_2,\quad [V_2,V_3]=V_1;\\[2mm]
&A_{3.9}:\ \ [V_1,V_3]=qV_1-V_2,\quad [V_2,V_3]=V_1+qV_2,\ (q>0).
\end{align*}
All these algebras contain a two-dimensional Abelian ideal as a
subalgebra. Thus, we can utilize results of  classification of
$A_{2.1}-$ invariant equations to construct Eq.\eqref{eq} which
admit non-decomposable three-dimensional solvable Lie algebras. As a
result, we arrive at the following invariance equations:

$A_{3.3}-$ invariant equations,
\begin{align*}
&A^1_{3.3}=\langle \partial_u,\partial_t,\partial_x+t\partial_u \rangle:\\[2mm]
&\qquad\qquad                        F=F(u_x,u_{xx},u_{xxx}),\quad G=x+G(u_x,u_{xx},u_{xxx}),\\[2mm]
&A^2_{3.3}=\langle \partial_u,\partial_t,(t+x)\partial_u \rangle:\\[2mm]
&\qquad\qquad                        F=F(x,u_{xx},u_{xxx}),\quad G=u_x+G(x,u_{xx},u_{xxx}),\\[2mm]
&A^3_{3.3}=\langle \partial_u,\partial_x,t\partial_x+x\partial_u \rangle:\\[2mm]
&\qquad\qquad                        F=F(t,u_{xx},u_{xxx}),\quad G=-\frac{u_x^2}2+G(t,u_{xx},u_{xxx}),\\[2mm]
&A^4_{3.3}=\langle \partial_u,\partial_x,\partial_t+x\partial_u \rangle:\\[2mm]
&\qquad\qquad                        F=F(u_x-t,u_{xx},u_{xxx}),\quad G=G(u_x-t,u_{xx},u_{xxx}),\\[2mm]
&A^5_{3.3}=\langle \partial_u,x\partial_u,-\partial_x \rangle:\\[2mm]
&\qquad\qquad                        F=F(t,u_{xx},u_{xxx}),\quad G=G(t,u_{xx},u_{xxx}),\\[2mm]
&A^6_{3.3}=\langle \partial_u,x\partial_u,\partial_t-\partial_x \rangle:\\[2mm]
&\qquad\qquad                        F=F(x+t,u_{xx},u_{xxx}),\quad G=G(x+t,u_{xx},u_{xxx}),\\[2mm]
&A^7_{3.3}=\langle x\partial_u,\partial_u,x^2\partial_x+xu\partial_u \rangle:\\[2mm]
&\qquad\qquad                        F=x^8F(t,\omega_1,\omega_2),\quad G=4x^6(3u_{xx}+2xu_{xxx})F(t,\omega_1,\omega_2)+xG(t,\omega_1,\omega_2),\ \\[2mm]
&\qquad\qquad\qquad                     \omega_1=x^3u_{xx},\ \omega_2=x^4(3u_{xx}+xu_{xxx}),\\[2mm]
&A^8_{3.3}=\langle x\partial_u,\partial_u,\partial_t+x^2\partial_x+xu\partial_u \rangle:\\[2mm]
&\qquad\qquad                        F=x^8F(\omega_1,\omega_2,\omega_3),\quad G=4x^6(3u_{xx}+2xu_{xxx})F(\omega_1,\omega_2,\omega_3)-xG(\omega_1,\omega_2,\omega_3),\\[2mm]
&\qquad\qquad\qquad                     \omega_1=t+x^{-1},\ \omega_2=x^3u_{xx},\ \omega_3=x^4(3u_{xx}+xu_{xxx}).\\[2mm]
\end{align*}

$A_{3.4}-$ invariant equations,
\begin{align*}
& A^1_{3.4}=\langle \partial_u,\partial_t,t\partial_t+\partial_x+(t+u)\partial_u \rangle:\\[2mm]
& \qquad\qquad                       F=\mathrm{e}^{-x}F(\omega_1,\omega_2,\omega_3),\quad G=x+G(\omega_1,\omega_2,\omega_3),\\[2mm]
& \qquad\qquad\qquad                     \omega_1=\mathrm{e}^{-x}u_x,\ \omega_2=\mathrm{e}^{-x}u_{xx},\ \omega_3=\mathrm{e}^{-x}u_{xxx},\\[2mm]
& A^2_{3.4}=\langle \partial_u,\partial_t,t\partial_t+(t+u)\partial_u \rangle:\\[2mm]
& \qquad\qquad                       F=u_x^{-1}F(x,\omega_1,\omega_2),\quad G=\ln{|u_x|}+G(x,\omega_1,\omega_2), \\[2mm]
& \qquad\qquad\qquad                     \omega_1=u_x^{-1}u_{xx},\ \omega_2=u_x^{-1}u_{xxx},\\[2mm]
& A^3_{3.4}=\langle \partial_u,\partial_x,\partial_t+x\partial_x+(x+u)\partial_u\rangle:\\[2mm]
& \qquad\qquad                       F=\mathrm{e}^{4t}F(\omega_1,\omega_2,\omega_3),\quad G=\mathrm{e}^tG(\omega_1,\omega_2,\omega_3),\\[2mm]
& \qquad\qquad\qquad                     \omega_1=u_x-t,\ \omega_2=\mathrm{e}^tu_{xx},\ \omega_3=\mathrm{e}^{2t}u_{xxx},\\[2mm]
& A^4_{3.4}=\langle \partial_u,\partial_x,x\partial_x+(x+u)\partial_u\rangle:\\[2mm]
& \qquad\qquad                       F=\mathrm{e}^{4u_x}F(t,\omega_1,\omega_2),\quad G=\mathrm{e}^{u_x}G(t,\omega_1,\omega_2),\\[2mm]
& \qquad\qquad\qquad                     \omega_1=\mathrm{e}^{u_x}u_{xx},\ \omega_2=\mathrm{e}^{2u_x}u_{xxx},\\[2mm]
& A^5_{3.4}=\langle \partial_u,x\partial_u,-\partial_x+u\partial_u \rangle:\\[2mm]
& \qquad\qquad                       F=F(t,\omega_1,\omega_2),\quad G=\mathrm{e}^{-x}G(t,\omega_1,\omega_2),\\[2mm]
& \qquad\qquad\qquad                     \omega_1=\mathrm{e}^{x}u_{xx},\ \omega_2=\mathrm{e}^{x}u_{xxx},\\[2mm]
& A^6_{3.4}=\langle \partial_u,x\partial_u,\partial_t-\partial_x+u\partial_u \rangle:\\[2mm]
& \qquad\qquad                       F=F(\omega_1,\omega_2,\omega_3),\quad G=\mathrm{e}^tG(\omega_1,\omega_2,\omega_3),\\[2mm]
& \qquad\qquad\qquad                     \omega_1=x+t,\ \omega_2=\mathrm{e}^{-t}u_{xx},\ \omega_3=\mathrm{e}^{-t}u_{xxx},\\[2mm]
& A^7_{3.4}=\langle x\partial_u,\partial_u,x^2\partial_x+(1+x)u\partial_u \rangle:\\[2mm]
& \qquad\qquad                       F=x^8F(t,\omega_1,\omega_2),\quad G=4x^6(3u_{xx}+2xu_{xxx})F(t,\omega_1,\omega_2)+x\mathrm{e}^{-\frac1x}G(t,\omega_1,\omega_2),\\[2mm]
& \qquad\qquad\qquad                     \omega_1=x^3\mathrm{e}^{\frac1x}u_{xx},\ \omega_2=x^4\mathrm{e}^{\frac1x}(3u_{xx}+xu_{xxx}),\\[2mm]
& A^8_{3.4}=\langle x\partial_u,\partial_u,\partial_t+x^2\partial_x+(1+x)u\partial_u \rangle:\\[2mm]
& \qquad\qquad                       F=x^8F(\omega_1,\omega_2,\omega_3),\quad G=4x^6(3u_{xx}+2xu_{xxx})F(\omega_1,\omega_2,\omega_3)-x\mathrm{e}^tG(\omega_1,\omega_2,\omega_3),\\[2mm]
& \qquad\qquad\qquad                     \omega_1=t+\frac1x,\ \omega_2=x^3\mathrm{e}^{-t}u_{xx},\ \omega_3=x^4\mathrm{e}^{-t}(3u_{xx}+xu_{xxx}).\\[2mm]
\end{align*}

$A_{3.5}-$ invariant equations,
\begin{align*}
& A^1_{3.5}=\langle \partial_t,\partial_u,t\partial_t+u\partial_u \rangle:\\[2mm]
& \qquad\qquad                       F=u_x^{-1}F(x,u_x^{-1}u_{xx},u_x^{-1}u_{xxx}),\quad G=G(x,u_x^{-1}u_{xx},u_x^{-1}u_{xxx})\\[2mm]
& A^2_{3.5}=\langle \partial_t,\partial_u,t\partial_t+\partial_x+u\partial_u \rangle:\\[2mm]
& \qquad\qquad                       F=\mathrm{e}^{-x}F(\omega_1,\omega_2,\omega_3),\quad G=G(\omega_1,\omega_2,\omega_3),\\[2mm]
& \qquad\qquad\qquad                      \omega_1=\mathrm{e}^{-x}u_x,\ \omega_2=\mathrm{e}^{-x}u_{xx},\ \omega_3=\mathrm{e}^{-x}u_{xxx},\\[2mm]
& A^3_{3.5}=\langle \partial_x,\partial_u,x\partial_x+u\partial_u \rangle:\\[2mm]
& \qquad\qquad                       F=u_{xx}^{-4}F(t,u_x,u_{xx}^{-2}u_{xxx}),\quad G=u_{xx}^{-1}F(t,u_x,u_{xx}^{-2}u_{xxx}),\\[2mm]
& A^4_{3.5}=\langle \partial_x,\partial_u,\partial_t+x\partial_x+u\partial_u \rangle:\\[2mm]
& \qquad\qquad                       F=\mathrm{e}^{4t}F(\omega_1,\omega_2,\omega_3),\quad G=\mathrm{e}^tG(\omega_1,\omega_2,\omega_3),\\[2mm]
& \qquad\qquad\qquad                      \omega_1=u_x,\ \omega_2=\mathrm{e}^tu_{xx},\ \omega_3=\mathrm{e}^{2t}u_{xxx},\\[2mm]
& A^5_{3.5}=\langle \partial_u,x\partial_u,u\partial_u \rangle:\\[2mm]
& \qquad\qquad                       F=F(t,x,u_{xx}^{-1}u_{xxx}),\quad G=u_{xx}G(t,x,u_{xx}^{-1}u_{xxx}),\\[2mm]
& A^6_{3.5}=\langle \partial_x,u\partial_x,\partial_t+x\partial_x \rangle:\\[2mm]
& \qquad\qquad                       F=F(x,\omega_1,\omega_2),\quad G=\mathrm{e}^tG(x,\omega_1,\omega_2),\\[2mm]
& \qquad\qquad\qquad                     \omega_1=\mathrm{e}^{-t}u_{xx},\ \omega_2=\mathrm{e}^{-t}u_{xxx}.\\[2mm]
\end{align*}

$A_{3.6}-$ invariant equations,
\begin{align*}
& A^1_{3.6}=\langle \partial_t,\partial_u,t\partial_t-u\partial_u \rangle:\\[2mm]
& \qquad\qquad                        F=u_xF(x,u_x^{-1}u_{xx},u_x^{-1}u_{xxx}),\quad G=u_x^2G(x,u_x^{-1}u_{xx},u_x^{-1}u_{xxx}),\\[2mm]
& A^2_{3.6}=\langle \partial_t,\partial_u,t\partial_t+\partial_x-u\partial_u \rangle:\\[2mm]
& \qquad\qquad                        F=\mathrm{e}^{-x}F(\omega_1,\omega_2,\omega_3),\quad G=\mathrm{e}^{-2x}G(\omega_1,\omega_2,\omega_3),\\[2mm]
& \qquad\qquad\qquad                      \omega_1=\mathrm{e}^xu_x,\ \omega_2=\mathrm{e}^xu_{xx},\ \omega_3=\mathrm{e}^xu_{xxx},\\[2mm]
& A^3_{3.6}=\langle \partial_x,\partial_u,x\partial_x-u\partial_u \rangle:\\[2mm]
& \qquad\qquad                        F=u_x^{-2}F(t,\omega_1,\omega_2),\quad G=u_x^{\frac12}G(t,\omega_1,\omega_2),\\[2mm]
& \qquad\qquad\qquad                      \omega_1=u_x^{-\frac32}u_{xx},\ \omega_2=u_x^{-2}u_{xxx},\\[2mm]
& A^4_{3.6}=\langle \partial_x,\partial_u,\partial_t+\partial_x-u\partial_u \rangle:\\[2mm]
& \qquad\qquad                        F=\mathrm{e}^{4t}F(\omega_1,\omega_2,\omega_3),\quad G=\mathrm{e}^{-t}G(\omega_1,\omega_2,\omega_3),\\[2mm]
& \qquad\qquad\qquad                      \omega_1=\mathrm{e}^{2t}u_x,\ \omega_2=\mathrm{e}^{3t}u_{xx},\ \omega_3=\mathrm{e}^{4t}u_{xxx},\\[2mm]
& A^5_{3.6}=\langle \partial_x,\partial_u,x\partial_x-u\partial_u \rangle:\\[2mm]
& \qquad\qquad                        F=x^4F(t,\omega_1,\omega_2),\quad G=x^{\frac12}G(t,\omega_1,\omega_2),\\[2mm]
& \qquad\qquad\qquad                      \omega_1=x^{\frac32}u_{xx},\ \omega_2=x^{\frac52}u_{xxx},\\[2mm]
& A^6_{3.6}=\langle \partial_u,x\partial_u,\partial_t+2x\partial_x+u\partial_u \rangle:\\[2mm]
& \qquad\qquad                        F=\mathrm{e}^{8t}F(\omega_1,\omega_2,\omega_3),\quad G=\mathrm{e}^tG(\omega_1,\omega_2,\omega_3),\\[2mm]
& \qquad\qquad\qquad                      \omega_1=\mathrm{e}^{-2t}x,\ \omega_2=\mathrm{e}^{3t}u_{xx},\ \omega_3=\mathrm{e}^{5t}u_{xxx}.\\[2mm]
\end{align*}

$A_{3.7}-$ invariant equations,
\begin{align*}
& A^1_{3.7}=\langle \partial_t,\partial_u,t\partial_t+qu\partial_u\rangle:\\[2mm]
& \qquad\qquad                        F=u_x^{-\frac1q}F(x,u_x^{-1}u_{xx},u_x^{-1}u_{xxx}),\quad F=u_x^{1-\frac1q}G(x,u_x^{-1}u_{xx},u_x^{-1}u_{xxx}),\\[2mm]
& A^2_{3.7}=\langle \partial_t,\partial_u,t\partial_t+\partial_x+qu\partial_u \rangle:\\[2mm]
& \qquad\qquad                        F=\mathrm{e}^{-x}F(\omega_1,\omega_2,\omega_3),\quad G=\mathrm{e}^{(q-1)x}G(\omega_1,\omega_2,\omega_3),\\[2mm]
& \qquad\qquad\qquad                      \omega_1=\mathrm{e}^{-qx}u_x,\ \omega_2=\mathrm{e}^{-qx}u_{xx},\ \omega_3=\mathrm{e}^{-qx}u_{xxx},\\[2mm]
& A^3_{3.7}=\langle \partial_x,\partial_u,x\partial_x+qu\partial_u \rangle:\\[2mm]
& \qquad\qquad                        F=u_x^{\frac4{q-1}}F(t,\omega_1,\omega_2),\quad G=u_x^{\frac q{q-1}}G(t,\omega_1,\omega_2),\\[2mm]
& \qquad\qquad\qquad                      \omega_1=u_x^{\frac{2-q}{q-1}}u_{xx},\ \omega_2=u_x^{\frac{3-q}{q-1}}u_{xxx},\\[2mm]
& A^4_{3.7}=\langle \partial_x,\partial_u,\partial_t+x\partial_x+qu\partial_u \rangle:\\[2mm]
& \qquad\qquad                        F=\mathrm{e}^{4t}F(\omega_1,\omega_2,\omega_3),\quad G=\mathrm{e}^{qt}G(\omega_1,\omega_2,\omega_3),\\[2mm]
& \qquad\qquad\qquad                      \omega_1=\mathrm{e}^{(1-q)t}u_x,\ \omega_2=\mathrm{e}^{(2-q)t}u_{xx},\ \omega_3=\mathrm{e}^{(3-q)t}u_{xxx},\\[2mm]
& A^5_{3.7}=\langle \partial_x,u\partial_x,x\partial_x+(1-q)u\partial_u \rangle:\\[2mm]
& \qquad\qquad                        F=x^4F(t,\omega_1,\omega_2),\quad G=x^{\frac 1{1-q}}G(t,\omega_1,\omega_2),\\[2mm]
& \qquad\qquad\qquad                      \omega_1=x^{\frac{2q-1}{q-1}}u_{xx},\ \omega_2=x^{\frac{3q-2}{q-1}}u_{xxx},\\[2mm]
& A^6_{3.7}=\langle \partial_x,u\partial_x,\partial_t+x\partial_x+(1-q)u\partial_u \rangle:\\[2mm]
& \qquad\qquad                        F=\mathrm{e}^{4(1-q)t}F(\omega_1,\omega_2,\omega_3),\quad G=\mathrm{e}^tG(\omega_1,\omega_2,\omega_3),\\[2mm]
& \qquad\qquad\qquad                      \omega_1=\mathrm{e}^{(q-1)t}x,\ \omega_2=\mathrm{e}^{(1-2q)t}u_{xx},\ \omega_3=\mathrm{e}^{(2-3q)t}u_{xxx}.\\[2mm]
\end{align*}

$A_{3.8}-$ invariant equations,
\begin{align*}
& A^1_{3.8}=\langle \partial_x,\partial_u,u\partial_x-x\partial_u \rangle:\\[2mm]
& \qquad\qquad                        F=(1+u_x^2)^{-2}F(t,\omega_1,\omega_2),\\[2mm]
& \qquad\qquad                        G=-5u_xu_{xx}(1+u_x^2)^{-4}(2u_x^2u_{xxx}-3u_xu_{xx}^2+2u_{xxx})F(t,\omega_1,\omega_2)\\[2mm]
& \qquad\qquad\qquad                      +(1+u_x^2)^{\frac12}G(t,\omega_1,\omega_2),\\[2mm]
& \qquad\qquad\qquad                      \omega_1=(1+u_x^2)^{-\frac32}u_{xx},\ \omega_2=(1+u_x^2)^{-3}(u_x^2u_{xxx}-3u_xu_{xx}^2+u_{xxx}),\\[2mm]
& A^2_{3.8}=\langle \partial_x,\partial_u,\partial_t+u\partial_x-x\partial_u \rangle:\\[2mm]
& \qquad\qquad                        F=(1+u_x^2)^{-2}F(\omega_1,\omega_2,\omega_3),\\[2mm]
& \qquad\qquad                        G=5\frac{u_{xx}}{(1+u_x^2)^4}[ 6u_{xx}^2\arctan^2 u_x+(2u_x^2u_{xxx}+12tu_{xx}^2-6u_xu_{xx}^2+2u_{xxx})\arctan u_x\\[2mm]
& \qquad\qquad\qquad                    +6t^2u_{xx}^2-6tu_xu_{xx}^2+2tu_x^2u_{xxx}+3u_x^2u_{xx}^2+2tu_{xxx}-2u_x^3u_{xxx}-2u_xu_{xxx}  ]\\[2mm]
& \qquad\qquad\qquad                     F(\omega_1,\omega_2,\omega_3)+(1+u_x^2)^{\frac12}G(\omega_1,\omega_2,\omega_3),\\[2mm]
& \qquad\qquad\qquad                      \omega_1=\arctan u_x +t,\ \omega_2=(1+u_x^2)^{-\frac32}u_{xx},\\[2mm]
& \qquad\qquad\qquad                      \omega_3=(1+u_x^2)^{-3}(u_x^2u_{xxx}-3u_xu_{xx}^2+u_{xxx}+3u_{xx}^2\omega_1),\\[2mm]
& A^3_{3.8}=\langle \partial_u,x\partial_u,-(x^2+1)\partial_x-xu\partial_u \rangle:\\[2mm]
& \qquad\qquad                        F=(1+x^2)^4F(t,\omega_1,\omega_2),\\[2mm]
& \qquad\qquad                        G=4x(1+x^2)^2(2x^2u_{xxx}+3xu_{xx}+2u_{xxx})F(t,\omega_1,\omega_2)+(1+x^2)^{\frac12}G(t,\omega_1,\omega_2),\\[2mm]
& \qquad\qquad\qquad                      \omega_1=(1+x^2)^{\frac32}u_{xx},\ \omega_2=(1+x^2)^{\frac32}(x^2u_{xxx}+3xu_{xx}+u_{xxx}),\\[2mm]
& A^4_{3.8}=\langle \partial_u,x\partial_u,\partial_t-(x^2+1)\partial_x-xu\partial_u \rangle:\\[2mm]
& \qquad\qquad                        F=(1+x^2)^4F(\omega_1,\omega_2,\omega_3),\\[2mm]
& \qquad\qquad                        G=4x(1+x^2)^2u_{xx}(2x^2u_{xxx}+3xu_{xx}+2u_{xxx})]F(\omega_1,\omega_2,\omega_3)+(1+x^2)^\frac12G(\omega_1,\omega_2,\omega_3),\\[2mm]
& \qquad\qquad\qquad                      \omega_1=t+\arctan x,\ \omega_2=(1+x^2)^{\frac32}u_{xx},\ \omega_3=(1+x^2)^{\frac52}u_{xxx}-3\omega_2(\omega_1-x).\\[2mm]
\end{align*}

$A_{3.9}-$ invariant equations,
\begin{align*}
& A^1_{3.9}=\langle \partial_x,\partial_u,(qx+u)\partial_x+(qu-x)\partial_u \rangle:\\[2mm]
& \qquad\qquad                        F=(1+u_x^2)^{-2}\mathrm{e}^{-4q\arctan u_x}F(t,\omega_1,\omega_2),\\[2mm]
& \qquad\qquad                        G=-5u_xu_{xx}(1+u_x^2)^{-4}\mathrm{e}^{-4q\arctan u_x}(2u_x^2u_{xxx}-3u_xu_{xx}^2+2u_{xxx})F(t,\omega_1,\omega_2)\\[2mm]
& \qquad\qquad\qquad                      +(1+u_x^2)^{\frac12}\mathrm{e}^{-q\arctan u_x}G(t,\omega_1,\omega_2)\\[2mm]
& \qquad\qquad\qquad                      \omega_1=(1+u_x^2)^{-\frac32}u_{xx},\ \omega_2=(1+u_x^2)^{-3}(u_x^2u_{xxx}-3u_xu_{xx}^2+u_{xxx}),\\[2mm]
& A^2_{3.9}=\langle \partial_x,\partial_u,\partial_t+(qx+u)\partial_x+(qu-x)\partial_u \rangle:\\[2mm]
& \qquad\qquad                        F=(1+u_x^2)^{-2}\mathrm{e}^{4qt}F(\omega_1,\omega_2,\omega_3),\\[2mm]
& \qquad\qquad                        G=-5u_{xx}(1+u_x^2)^{-4}\mathrm{e}^{4qt}(2u_x^3u_{xxx}-3u_x^2u_{xx}^2+2u_xu_{xxx}+3u_{xx}^2)F(\omega_1,\omega_2,\omega_3)\\[2mm]
& \qquad\qquad\qquad                      +(1+u_x^2)^{\frac12}\mathrm{e}^{qt}G(\omega_1,\omega_2,\omega_3),\\[2mm]
& \qquad\qquad\qquad                      \omega_1=t+\arctan u_x,\ \omega_2=(1+u_x^2)^{-\frac32}\mathrm{e}^{qt}u_{xx},\\[2mm]
& \qquad\qquad\qquad                      \omega_3=(1+u_x^2)^{-3}\mathrm{e}^{2qt}(u_x^2u_{xxx}-3u_xu_{xx}^2+u_{xxx}),\\[2mm]
& A^3_{3.9}=\langle \partial_u,x\partial_u,-(x^2+1)\partial_x+(q-x)u\partial_u \rangle:\\[2mm]
& \qquad\qquad                        F=(x^2+1)^4F(t,\omega_1,\omega_2),\\[2mm]
& \qquad\qquad                        G=4x(x^2+1)^2[2(1+x^2)u_{xxx}+3xu_{xx}]F(t,\omega_1,\omega_2)+(x^2+1)^{\frac12}\mathrm{e}^{-q\arctan x}G(t,\omega_1,\omega_2),\\[2mm]
& \qquad\qquad\qquad                      \omega_1=(x^2+1)^{\frac32}\mathrm{e}^{q\arctan x}u_{xx},\\[2mm]
& \qquad\qquad\qquad                      \omega_2=(1+u_x^2)^{\frac32}\mathrm{e}^{q\arctan x}(x^2u_{xxx}+3xu_{xx}+u_{xxx}),\\[2mm]
& A^4_{3.9}=\langle \partial_x,u\partial_x,\partial_t+(q-u)x\partial_x-(u^2+1)\partial_u \rangle:\\[2mm]
& \qquad\qquad                        F=(u^2+1)^4u_x^{-4}F(\omega_1,\omega_2,\omega_3),\\[2mm]
& \qquad\qquad                        G=[-5(u^2+1)^4u_x^{-6}u_{xx}(2u_xu_{xxx}-3u_{xx}^2)+8u(u^2+1)^3u_x^{-4}u_{xx}(u_xu_{xxx}-3u_{xx}^2)\\[2mm]
& \qquad\qquad\qquad                    +12(u^2-1)(u^2+1)^2u_x^{-2}u_{xx}]F(\omega_1,\omega_2,\omega_3)+(u^2+1)^{\frac12}\mathrm{e}^{qt}u_xG(\omega_1,\omega_2,\omega_3),\\[2mm]
& \qquad\qquad\qquad                      \omega_1=t+\arctan u,\ \omega_2=(u^2+1)^{\frac32}\mathrm{e}^{-qt}u_x^{-3}u_{xx},\\[2mm]
& \qquad\qquad\qquad                      \omega_3=(u^2+1)^{\frac52}\mathrm{e}^{-qt}u_x^{-5}(u_xu_{xxx}-3u_{xx}^2)+3u(u^2+1)^\frac32\mathrm{e}^{-qt}u_x^{-3}u_{xx}.\\[2mm]
\end{align*}

\subsection*{\small\bf 4.3 Equations invariant under
four-dimensional solvable Lie algebras}

Now we perform group classification of Eq.\eqref{eq} admitting
four-dimensional solvable Lie algebras. To this end, we utilize the
well known classification of abstract four-dimensional Lie algebras
as well as the above obtained results on three-dimensional solvable
algebras. For the sake of convenience we treat the cases of
decomposable and non-decomposable algebras separately. Note that we
skip quite cumbersome calculation details presenting the final
results, symmetry algebras and invariant equations, only.

\subsubsection*{\small\bf 4.3.1
Equations with four-dimensional decomposable algebras}

The list of non-isomorphic four-dimensional decomposable Lie
algebras contains the following ten algebras:
\begin{align*}
& A_{2.2}\oplus A_{2.2}=2A_{2.2},\\[2mm]
& A_{3.1}\oplus A_1=4A_1,\\[2mm]
& A_{3.2}\oplus A_1=A_{2.2}\oplus 2A_1,\\[2mm]
& A_{3.i}\oplus A_1\ (i=3,4,\cdots,9).
\end{align*}

The exhaustive list of equations \eqref{eq} admitting one of the
above algebras is given in Appendix A.

\subsubsection*{\small\bf 4.3.2.
Equations with four-dimensional non-decomposable algebras}

There exist ten non-isomorphic four-dimensional non-decomposable Lie
algebras, $A_{4.i}$ $(i=1,2,\cdots,10)$:
\begin{align*}
& A_{4.1}:\qquad [X_2,X_4]=X_1,\quad [X_3,X_4]=X_2;\\[2mm]
& A_{4.2}:\qquad [X_1,X_4]=qX_1,\quad [X_2,X_4]=X_2,\quad
[X_3,X_4]=X_2+X_3,\ q\neq0;\\[2mm]
& A_{4.3}:\qquad [X_1,X_4]=X_1,\quad [X_3,X_4]=X_2;\\[2mm]
& A_{4.4}:\qquad [X_1,X_4]=X_1,\quad [X_2,X_4]=X_1+X_2,\quad
[X_3,X_4]=X_2+X_3;\\[2mm]
& A_{4.5}:\qquad [X_1,X_4]=X_1,\quad [X_2,X_4]=qX_2,\quad
[X_3,X_4]=pX_3,\\[2mm]
& \qquad\qquad\ \ -1\leqslant p\leqslant q\leqslant1,\ pq\neq0;\\[2mm]
& A_{4.6}:\qquad [X_1,X_4]=qX_1,\quad [X_2,X_4]=pX_2-X_3,\quad
[X_3,X_4]=X_2+pX_3,\\[2mm]
& \qquad\qquad\ \ q\neq0,\ p\geqslant0;\\[2mm]
& A_{4.7}:\qquad [X_2,X_3]=X_1,\quad [X_1,X_4]=2X_1, \quad
[X_2,X_4]=X_2,\\[2mm]
&\qquad\qquad\ \ [X_3,X_4]=X_2+X_3;\\[2mm]
& A_{4.8}:\qquad [X_2,X_3]=X_1,\quad [X_1,X_4]=(1+q)X_1,\quad
[X_2,X_4]=X_2,\\[2mm]
& \qquad\qquad\ \ [X_3,X_4]=qX_3,\ |q|\leqslant1;\\[2mm]
& A_{4.9}:\qquad [X_2,X_3]=X_1,\quad [X_1,X_4]=2qX_1,\quad
[X_2,X_4]=qX_2-X_3,\\[3mm]
& \qquad\qquad\ \ [X_3,X_4]=X_2+qX_3,\ q\geqslant0;\\[2mm]
& A_{4.10}:\qquad [X_1,X_3]=X_1,\quad [X_2,X_3]=X_2,\quad
[X_1,X_4]=-X_2,\quad [X_2,X_4]=X_1.
\end{align*}
Each of the above algebras can be decomposed into a semi-direct sum
of a three-dimensional ideal $N$ and a one-dimensional Lie algebra.
Analysis of the commutation relations above shows that $N$ is of the
type $A_{3.1}$ for the algebras $A_{4.i}\ (i=1,2,\cdots,6)$, of the
type $A_{3.3}$ for the algebras $A_{4.7},\ A_{4.8},\ A_{4.9}$, and
of the type $A_{3.5}$ for the algebra $A_{4.10}$. Thus we can
utilize the already known realizations of three-dimensional solvable
Lie algebras to obtain exhaustive descriptions of the
four-dimensional non-decomposable solvable Lie algebras admitted by
Eq.\eqref{eq}. The full list of inequivalent symmetry algebras and
the corresponding invariant equations is given in Appendix B.

\section*{\small\bf 5. Classification of equations admitting
quasi-local symmetries}

Following the scheme developed in \cite{Zhd09} we make use of local
(Lie) symmetries of the fourth-order evolution equations \eqref{eq}
to obtain their non-local symmetries.

 As established earlier, the most general infinitesimal operator,
$V$, admitted by evolution equation \eqref{eq} reads as
\begin{equation*}
    V=\tau(t)\partial_t+\xi(t,x,u)\partial_x+\eta(t,x,u)\partial_u
\end{equation*}
and the maximal equivalence group of Eq.\eqref{eq} takes the form
\begin{equation*}
    \bar{t}=T(t),\quad \bar{x}=X(t,x,u),\quad
    \bar{u}=U(t,x,u),
\end{equation*}
where $\dot{T}\neq0$ and $\frac{D(X,U)}{D(x,u)}\neq 0$.

If $\tau=0$, we have $V = \xi(t,x,u)\partial_x +
\eta(t,x,u)\partial_u$ and there is an equivalence transformation
$(t,x,u)\rightarrow(\bar{t},\bar{x},\bar{u})$ that reduces $V$ to
the canonical form $\partial_u$ (we drop the bars). Eq.\eqref{eq}
now becomes
\begin{equation}\label{1eq}
    u_t=F(t,x,u_x,u_{xx},u_{xxx})u_{xxxx}+G(t,x,u_x,u_{xx},u_{xxx})
\end{equation}
Note that the right-hand side of Eq. \eqref{1eq} does not depend
explicitly on $u$.

Differentiating \eqref{1eq} with respect to $x$ yields
\begin{equation*}
\begin{split}
    &u_{tx}=Fu_{xxxxx}+[(F_x+F_{u_x}u_{xx}+F_{u_{xx}}u_{xxx}+F_{u_{xxx}}u_{xxxx})u_{xxxx}\\[2mm]
    &\quad\quad\   +G_x+G_{u_x}u_{xx}+G_{u_{xx}}u_{xxx}+G_{u_{xxx}}u_{xxxx}].
\end{split}
\end{equation*}
Making the non-local change of variables
\begin{equation}\label{nonlocal-tranf}
    \bar{t}=t,\quad\bar{x}=x,\quad\bar{u}=u_x
\end{equation}
and dropping the bars, we finally get
\begin{equation}\label{2eq}
\begin{split}
    &u_t=Fu_{xxxx}+[(F_x+F_uu_x+F_{u_x}u_{xx}+F_{u_{xx}}u_{xxx})u_{xxx}\\[2mm]
    &\quad\quad\   +G_x+G_uu_x+G_{u_x}u_{xx}+G_{u_{xx}}u_{xxx}].
\end{split}
\end{equation}
where $F=F(t,x,u,u_x,u_{xx})$ and $G=G(t,x,u,u_x,u_{xx})$.

Thus the non-local transformation \eqref{nonlocal-tranf} preserves
the differential structure of the class of evolution equation
\eqref{1eq}.

Assume that Eq.\eqref{1eq} admits $r$-parameter Lie transformation
group
\begin{equation}\label{1group}
    t'=T(t,\vec{\theta}),\quad x'=X(t,x,u,\vec{\theta}),\quad u'=U(t,x,u,\vec{\theta})
\end{equation}
with the vector of group parameters $\vec{\theta} =
(\theta_1,\dots,\theta_r)$ and $r\geq2$. To obtain the symmetry
group of Eq.\eqref{2eq}, we need to transform \eqref{1group}
according to \eqref{nonlocal-tranf}. Computing the first
prolongation of formulas \eqref{1group}, we derive the
transformation rule for the first derivative of $u$
\begin{equation*}
    u'_{x'}=\frac{U_uu_x+U_x}{X_uu_x+X_x}.
\end{equation*}
The transformation group \eqref{1group} now reads as
\begin{equation}\label{2group}
    t'=T(t,\vec{\theta}),\quad x'=X(t,x,v,\vec{\theta}),\quad u'=\frac{U_vu+U_x}{X_vu+X_x}.
\end{equation}
with $v=\partial^{-1}u$ and $U=U(t,x,v,\vec{\theta})$. Consequently,
Eq.\eqref{2eq} admits symmetry group \eqref{2group}. Provided the
right-hand sides of the transformation group \eqref{2group} depend
explicitly on the non-local variable $v$, then \eqref{2group} is a
quasi-local symmetry of Eq.\eqref{2eq}.

Evidently, Eq.\eqref{2eq} admits quasi-local symmetry iff
transformation \eqref{2group} satisfies
\begin{equation*}
    X_v\neq0\quad \text{or} \quad
    \frac{\partial}{\partial v}\Big(\frac{U_vu+U_x}{X_vu+X_x}\Big)\neq0,
\end{equation*}
or, equivalently
\begin{equation*}
\begin{split}
    X_v\neq0\quad \text{or}\quad X_v=0,\quad\ U^2_{xv}+U^2_{vv}\neq0.
\end{split}
\end{equation*}
It is straightforward to express the above constraints in terms of
the coefficients of the corresponding infinitesimal operator of
group \eqref{1group} and obtain the following assertion.
\begin{theorem}\label{criterion}\rm\cite{Zhd09}
Evolution equation \eqref{1eq} can be transformed to Eq.\eqref{2eq}
admitting a quasi-local symmetry if \eqref{1eq} admits a Lie
symmetry whose infinitesimal generator satisfies one of the
following inequalities:
\begin{align}
& \xi_u\neq0, \label{1condi}\\[2mm]
& \xi_u=0,\quad \eta^2_{xu}+\eta^2_{uu}\neq0. \label{2condi}
\end{align}
\end{theorem}

By making a hodograph transformation
\begin{equation*}
    \bar{t}=t,\quad \bar{x}=u,\quad \bar{u}=x,
\end{equation*}
and dropping the bars, the evolution equation
\begin{equation}\label{3eq}
    u_t=F(t,u,u_x,u_{xx},u_{xxx})u_{xxxx}+G(t,u,u_x,u_{xx},u_{xxx})
\end{equation}
can be transformed to an equation of the form \eqref{1eq}. Hence we
can get the following assertion.
\begin{corollary}
Equation \eqref{3eq} can be reduced to an equation having a
quasi-local symmetry if Eq.\eqref{3eq} admits a Lie symmetry
satisfying one of the inequalities
\begin{align*}
          & \eta_x\neq0,\\[2mm]
or \qquad & \eta_x=0,\quad \xi^2_{xu}+\xi^2_{xx}\neq0.
\end{align*}
\end{corollary}

Based on the above reasonings, now we can formulate the algorithm
for constructing evolution equations of the form \eqref{eq}
admitting a quasi-local symmetry.

\begin{enumerate}

\item Select all invariant equations, whose invariance algebras
contain at least one operator of the form $V=\xi(t,x,u)\partial_x +
\eta(t,x,u)\partial_u$.

\item For each of these equations, make a suitable local equivalence
transformation reducing $V$ to the canonical form $\partial_u$, the
original equations being transformed to evolution equations of the
form \eqref{1eq}.

item For each Lie symmetry of invariance algebra admitted by
\eqref{1eq} check whether its infinitesimal generator satisfies one
of conditions \eqref{1condi}, \eqref{2condi} of Theorem
\ref{criterion}. This analysis yields the list of evolution
equations \eqref{eq} that can be reduced to those admitting
quasi-local symmetries.

item Performing the non-local change of variables
\eqref{nonlocal-tranf} transforms Eq.\eqref{1eq} to \eqref{2eq}
which has quasi-local symmetries \eqref{2group}.

\end{enumerate}

We processes in this way all the invariant equations obtained in
Sections 3, 4 and thus obtain the list of fourth-order evolution
equations \eqref{eq} admitting quasi-local symmetries. Here we give
the symmetry algebras of the corresponding invariant equations
omitting the expressions for the functions $F$ and $G$ equations for
brevity. Since we keep the same notations for Lie algebra
realizations used in the previous sections, it is straightforward to
deduce the explicit forms of the invariant equations given the form
of its Lie symmetry algebra.

Semi-simple Lie algebras:
\begin{align*}
& {\it{sl}}^3(2,\mathbb{R})=\langle 2x\partial_x-u\partial_u,-x^2\partial_x+xu\partial_u,\partial_x \rangle,\\[2mm]
& {\it{sl}}^4(2,\mathbb{R})=\langle 2x\partial_x,-x^2\partial_x,\partial_x \rangle,\\[2mm]
& {\it{sl}}^5(2,\mathbb{R})=\langle 2x\partial_x-u\partial_u,(\frac1{u^4}-x^2)\partial_x+xu\partial_u,\partial_x \rangle,\\[2mm]
& {\it{sl}}^6(2,\mathbb{R})=\langle
2x\partial_x-u\partial_u,-(x^2+\frac1{u^4})\partial_x+xu\partial_u,\partial_x\rangle.
\end{align*}

Three-dimensional solvable Lie algebras:
\begin{align*}
& A^7_{3.3}=\langle x\partial_u,\partial_u,x^2\partial_x+xu\partial_u \rangle,\\[2mm]
& A^8_{3.3}=\langle x\partial_u,\partial_u,\partial_t+x^2\partial_x+xu\partial_u \rangle,\\[2mm]
& A^7_{3.4}=\langle x\partial_u,\partial_u,x^2\partial_x+(1+x)u\partial_u \rangle,\\[2mm]
& A^8_{3.4}=\langle x\partial_u,\partial_u,\partial_t+x^2\partial_x+(1+x)u\partial_u \rangle,\\[2mm]
& A^1_{3.8}=\langle \partial_x,\partial_u,u\partial_x-x\partial_u \rangle,\\[2mm]
& A^2_{3.8}=\langle \partial_x,\partial_u,\partial_t+u\partial_x-x\partial_u \rangle,\\[2mm]
& A^3_{3.8}=\langle \partial_u,x\partial_u,-(x^2+1)\partial_x-xu\partial_u \rangle,\\[2mm]
& A^4_{3.8}=\langle \partial_u,x\partial_u,\partial_t-(x^2+1)\partial_x-xu\partial_u \rangle,\\[2mm]
& A^1_{3.9}=\langle \partial_x,\partial_u,(qx+u)\partial_x+(qu-x)\partial_u \rangle,\\[2mm]
& A^2_{3.9}=\langle \partial_x,\partial_u,\partial_t+(qx+u)\partial_x+(qu-x)\partial_u \rangle,\\[2mm]
& A^3_{3.9}=\langle \partial_u,x\partial_u,-(x^2+1)\partial_x+(q-x)u\partial_u \rangle,\\[2mm]
& A^4_{3.9}=\langle
\partial_x,u\partial_x,\partial_t+(q-u)x\partial_x-(u^2+1)\partial_u\rangle.
\end{align*}

Four-dimensional solvable Lie algebras:
\begin{align*}
& 2A^1_{2.2}=\langle -t\partial_t-u\partial_u,\partial_t,\partial_x,u\mathrm{e}^x\partial_u \rangle,\\[2mm]
& 2A^2_{2.2}=\langle -t\partial_t-x\partial_x,\partial_t,\partial_u,\mathrm{e}^u\partial_u \rangle,\\[2mm]
& 2A^3_{2.2}=\langle -t\partial_t-u\partial_u,\partial_t,\partial_x,\mathrm{e}^x\partial_x+\alpha u\mathrm{e}^x\partial_u \rangle,\ \alpha\neq 0,\\[2mm]
& 2A^{11}_{2.2}=\langle -t\partial_t-x\partial_x,\partial_x,\partial_u,\mathrm{e}^u\partial_u \rangle,\\[2mm]
& 2A^{12}_{2.2}=\langle -t\partial_t-u\partial_u,\partial_u,\partial_x,\alpha \mathrm{e}^x\partial_x+t\mathrm{e}^x\partial_u \rangle,\ \alpha\neq 0,\\[2mm]
& 2A^{13}_{2.2}=\langle -t\partial_t-x\partial_x,\partial_x,\partial_u,t\mathrm{e}^u\partial_x \rangle,\\[2mm]
& 2A^{29}_{2.2}=\langle -x\partial_x,\partial_x,\partial_u,t\mathrm{e}^u\partial_u \rangle,\\[2mm]
& A^4_{3.3}\oplus A_1=\langle \partial_x,\partial_u,\partial_t+u\partial_x,\partial_t+\alpha t\partial_x+\alpha \partial_u \rangle,\ \alpha\in\mathbb{R},\\[2mm]
& A^5_{3.3}\oplus A_1=\langle \partial_x,u\partial_x,-\partial_u,\partial_t \rangle,\\[2mm]
& A^7_{3.3}\oplus A_1=\langle x\partial_u,\partial_u,x^2\partial_x+xu\partial_u,\partial_t \rangle,\\[2mm]
& A^4_{3.4}\oplus A_1=\langle \partial_x,\partial_u,(x+u)\partial_x+u\partial_u,\partial_t \rangle,\\[2mm]
& A^7_{3.4}\oplus A_1=\langle u\partial_x,\partial_x,x(1+u)\partial_x+u^2\partial_u,u\mathrm{e}^{-\frac1u}\partial_x \rangle,\\[2mm]
& A^7_{3.4}\oplus A_1=\langle u\partial_x,\partial_x,x(1+u)\partial_x+u^2\partial_u,tu\mathrm{e}^{-\frac1u}\partial_x \rangle,\\[2mm]
& A^7_{3.4}\oplus A_1=\langle u\partial_x,\partial_x,x(1+u)\partial_x+u^2\partial_u,\partial_t \rangle,\\[2mm]
& A^8_{3.4}\oplus A_1=\langle u\partial_x,\partial_x,\partial_t+x(1+u)\partial_x+u^2\partial_u,u\mathrm{e}^tf(t+\frac1u)\partial_x \rangle,\ f''\neq 0,\\[2mm]
& A^1_{3.8}\oplus A_1=\langle \partial_x,\partial_u,u\partial_x-x\partial_u,\partial_t \rangle,\\[2mm]
& A^3_{3.8}\oplus A_1=\langle \partial_u,x\partial_u,-(x^2+1)\partial_x-xu\partial_u,\partial_t \rangle,\\[2mm]
& A^3_{3.8}\oplus A_1=\langle \partial_u,x\partial_u,-(x^2+1)\partial_x-xu\partial_u,(x^2+1)^\frac12\partial_u \rangle,\\[2mm]
& A^3_{3.8}\oplus A_1=\langle \partial_u,x\partial_u,-(x^2+1)\partial_x-xu\partial_u,t(x^2+1)^\frac12\partial_u \rangle,\\[2mm]
& A^4_{3.8}\oplus A_1=\langle \partial_x,u\partial_x,\partial_t-xu\partial_x-(u^2+1)\partial_u,(1+u^2)^\frac12f(t+\arctan{u})\partial_x \rangle,\ f+f''\neq0,\\[2mm]
& A^1_{3.9}\oplus A_1=\langle \partial_x,\partial_u,(qx+u)\partial_x+(qu-x)\partial_u,\partial_t \rangle,\\[2mm]
& A^3_{3.9}\oplus A_1=\langle \partial_u,x\partial_u,-(x^2+1)\partial_x+(q-x)u\partial_u,\partial_t \rangle,\\[2mm]
& A^3_{3.9}\oplus A_1=\langle \partial_u,x\partial_u,-(x^2+1)\partial_x+(q-x)u\partial_u,(x^2+1)^\frac12\mathrm{e}^{-q\arctan{x}}\partial_u \rangle,\\[2mm]
& A^3_{3.9}\oplus A_1=\langle \partial_u,x\partial_u,-(x^2+1)\partial_x+(q-x)u\partial_u,t(x^2+1)^\frac12\mathrm{e}^{-q\arctan{x}}\partial_u \rangle,\\[2mm]
& A^4_{3.9}\oplus A_1=\langle \partial_x,u\partial_x,\partial_t+(q-u)x\partial_x-(u^2+1)\partial_u,(u^2+1)^\frac12\mathrm{e}^{qt}\partial_x \rangle,\\[2mm]
& A^4_{3.9}\oplus A_1=\langle \partial_x,u\partial_x,\partial_t+(q-u)x\partial_x-(u^2+1)\partial_u,f(t+\arctan u)(u^2+1)^\frac12\mathrm{e}^{qt}\partial_x  \rangle,\\[2mm]
& \qquad\qquad\quad                      \ f+f''\neq 0,\\[2mm]
& A^3_{4.1}=\langle x\partial_u,\partial_u,\partial_t,x^2\partial_x+(t+xu)\partial_u \rangle,\\[2mm]
& A^1_{4.2}=\langle \partial_t,\partial_u,\partial_x,qt\partial_t+x\partial_x+(x+u)\partial_u \rangle,\\[2mm]
& A^5_{4.2}=\langle \partial_t,x\partial_u,\partial_u,qt\partial_t+x^2\partial_x+(x+1)u\partial_u \rangle,\\[2mm]
& A^1_{4.3}=\langle \partial_t,\partial_u,\partial_x,t\partial_t+x\partial_u \rangle,\\[2mm]
& A^4_{4.3}=\langle \partial_t,x\partial_u,\partial_u,t\partial_t+x^2\partial_x+xu\partial_u \rangle,\\[2mm]
& A^1_{4.4}=\langle \partial_u,\partial_x,\partial_t,t\partial_t+(t+x)\partial_x+(x+u)\partial_u \rangle,\\[2mm]
& A^3_{4.4}=\langle x\partial_u,\partial_u,\partial_t,t\partial_t+x^2\partial_x+[t+(x+1)u]\partial_u \rangle,\\[2mm]
& A^1_{4.6}=\langle \partial_t,\partial_x,\partial_u,qt\partial_t+(px+u)\partial_x+(pu-x)\partial_u \rangle,\\[2mm]
& A^2_{4.6}=\langle \partial_t,\partial_u,x\partial_u,qt\partial_t-(x^2+1)\partial_x+(p-x)u\partial_u \rangle,\\[2mm]
& A^1_{4.7}=\langle \partial_u,\partial_x,t\partial_x+x\partial_u,-\partial_t+x\partial_x+2u\partial_u \rangle,\\[2mm]
& A^2_{4.7}=\langle \partial_u,\partial_x,\partial_t+x\partial_u,t\partial_t+(t+x)\partial_x+(\frac {t^2}2+2u)\partial_u \rangle,\\[2mm]
& A^3_{4.7}=\langle \partial_u,x\partial_u,-\partial_x,x\partial_x+(u-\frac{x^2}2)\partial_u \rangle,\\[2mm]
& A^4_{4.7}=\langle \partial_u,x\partial_u,-\partial_x,\partial_t+x\partial_x+(u-\frac{x^2}2)\partial_u \rangle,\\[2mm]
& A^6_{4.7}=\langle x\partial_u,\partial_u,x^2\partial_x+xu\partial_u,-x\partial_x+(u-\frac1{2x})\partial_u \rangle,\\[2mm]
& A^7_{4.7}=\langle x\partial_u,\partial_u,x^2\partial_x+xu\partial_u,\partial_t-x\partial_x+(u-\frac1{2x})\partial_u \rangle,\\[2mm]
& A^8_{4.7}=\langle x\partial_u,\partial_u,\partial_t+x^2\partial_x+xu\partial_u,t\partial_t-x\partial_x+(t+u+\frac{t^2x}2)\partial_u \rangle,\\[2mm]
& A^3_{4.8}=\langle \partial_u,\partial_x,t\partial_x+x\partial_u,(1-q)t\partial_t+x\partial_x+(1+q)u\partial_u \rangle,\ q\neq1,\\[2mm]
& A^4_{4.8}=\langle \partial_u,\partial_x,t\partial_x+x\partial_u,x\partial_x+2u\partial_u \rangle,\ q\neq1,\\[2mm]
& A^5_{4.8}=\langle \partial_u,\partial_x,\partial_t+x\partial_u,qt\partial_t+x\partial_x+(1+q)u\partial_u \rangle,\ q\neq0,\\[2mm]
& A^6_{4.8}=\langle \partial_u,\partial_x,\partial_t+x\partial_u,x\partial_x+u\partial_u \rangle,\\[2mm]
& A^7_{4.8}=\langle \partial_u,x\partial_u,-\partial_x,-x\partial_x+t\partial_u \rangle,\\[2mm]
& A^8_{4.8}=\langle \partial_u,x\partial_u,-\partial_x,qx\partial_x+(1+q)u\partial_u \rangle,\ q\neq 1,\\[2mm]
& A^9_{4.8}=\langle \partial_u,x\partial_u,-\partial_x,x\partial_x+2u\partial_u \rangle,\\[2mm]
& A^{10}_{4.8}=\langle \partial_u,x\partial_u,-\partial_x,\partial_t+qx\partial_x+(1+q)u\partial_u \rangle,\\[2mm]
& A^{14}_{4.8}=\langle x\partial_u,\partial_u,x^2\partial_x+xu\partial_u,-qx\partial_x+u\partial_u \rangle,\ q\neq1,\\[2mm]
& A^{15}_{4.8}=\langle x\partial_u,\partial_u,x^2\partial_x+xu\partial_u,-x\partial_x+u\partial_u \rangle,\\[2mm]
& A^{16}_{4.8}=\langle x\partial_u,\partial_u,x^2\partial_x+xu\partial_u,x\partial_x+(tx+u)\partial_u \rangle,\\[2mm]
& A^{17}_{4.8}=\langle x\partial_u,\partial_u,x^2\partial_x+xu\partial_u,\partial_t-qx\partial_x+u\partial_u \rangle,\\[2mm]
& A^{18}_{4.8}=\langle u\partial_x,\partial_x,\partial_t+xu\partial_x+u^2\partial_u,qt\partial_t+x\partial_x-qu\partial_u \rangle,\ q\neq0,\\[2mm]
& A^{19}_{4.8}=\langle u\partial_x,\partial_x,\partial_t+xu\partial_x+u^2\partial_u,x\partial_x \rangle,\\[2mm]
& A^1_{4.9}=\langle \partial_u,\partial_x,t\partial_x+x\partial_u,-(1+t^2)\partial_t+(q-t)x\partial_x+(-\frac{x^2}2+2qu)\partial_u \rangle,\\[2mm]
& A^1_{4.10}=\langle \partial_x,\partial_u,x\partial_x+u\partial_u,u\partial_x-x\partial_u \rangle,\\[2mm]
& A^2_{4.10}=\langle \partial_x,\partial_u,x\partial_x+u\partial_u,\partial_t+u\partial_x-x\partial_u \rangle,\\[2mm]
& A^3_{4.10}=\langle \partial_x,\partial_u,\partial_t+x\partial_x+u\partial_u,u\partial_x-x\partial_u \rangle,\\[2mm]
& A^4_{4.10}=\langle \partial_x,\partial_u,\partial_t+x\partial_x+u\partial_u,\partial_t+u\partial_x-x\partial_u \rangle,\\[2mm]
& A^5_{4.10}=\langle \partial_u,x\partial_u,u\partial_u,-(1+x^2)\partial_x-xu\partial_u \rangle,\\[2mm]
& A^6_{4.10}=\langle \partial_u,x\partial_u,u\partial_u,\partial_t-(1+x^2)\partial_x-xu\partial_u \rangle,\\[2mm]
& A^7_{4.10}=\langle \partial_u,x\partial_u,\partial_t+u\partial_u,-(1+x^2)\partial_x-xu\partial_u \rangle,\\[2mm]
& A^8_{4.10}=\langle
\partial_u,x\partial_u,\partial_t+u\partial_u,\partial_t-(1+x^2)\partial_x-xu\partial_u
\rangle.
\end{align*}

Making a suitable local equivalence transformation to reduce basis
elements $V=\xi(t,x,u)\partial_x+\eta(t,x,u)\partial_u$ of each
algebra listed above to the canonical form $\partial_u$, yields
evolution equations of the form \eqref{1eq}. Next, differentiating
the so obtained equations with respect to $x$ and replacing $u_x$
with $u$ yields an equation of the form \eqref{2eq} that admits a
quasi-local symmetry. Here we consider two illustrative examples.

{\bf Example 1.} Consider the algebra
\begin{equation*}
{\it{sl}}^4(2,\mathbb{R})=\langle
2x\partial_x,-x^2\partial_x,\partial_x \rangle
\end{equation*}
and make a hodograph transformation
\begin{equation*}
    \bar{t}=t,\quad \bar{x}=u,\quad \bar{u}=x,
\end{equation*}
which transforms the original algebra to
\begin{equation*}
\langle 2u\partial_u,-u^2\partial_u,\partial_u \rangle.
\end{equation*}
Here we drop the bars. The invariant equation corresponding to the
above algebra reads as
\begin{equation*}
    u_t=F(t,x,\omega)u_{xxxx}+\frac{3u_{xx}^3-4u_xu_{xx}u_{xxx}}{u_x^2}F(t,x,\omega)+u_xG(t,x,\omega),
\end{equation*}
where $F,\ G$ are arbitrary smooth functions and
$\omega=(2u_xu_{xxx}-3u_{xx}^2)u_x^{-2}$. Differentiating the above
equation with respect to $x$ and replacing $u_x$ with $u$ according
to \eqref{nonlocal-tranf} we arrive at the evolution equation
\begin{equation*}
\begin{split}
    u_t=& Fu_{xxxx}+(u_{xxx}+\frac{3u_{xx}^3-4u_xu_{xx}u_{xxx}}{u_x^2})(F_x+\sigma F_\omega)\\[2mm]
        &  -\frac{4u^2(u_{xx}^2+u_xu_{xxx})-13uu_x^2u_{xx}+6u_x^4}{u^3}F+u_xG+uG_x+u\sigma G_\omega,
\end{split}
\end{equation*}
with $\omega=(2uu_{xx}-3u_x^2)u^{-2}$ and
$\sigma=2(u^2u_{xxx}-4uu_xu_{xx}+3u_x^3)u^{-3}$. This equation
admits the non-local symmetry transformation group
\begin{equation*}
    t'=t,\quad x'=x,\quad u'=\frac{u}{(\theta v+1)^2},
\end{equation*}
with the generator $-2uv\partial_u$, where $\theta$ is the group
parameter and $v=\partial^{-1}u$.

{\bf Example 2.} Consider the Lie algebra
\begin{equation*}
2A^2_{2.2}=\langle
-t\partial_t-x\partial_x,\partial_t,\partial_u,\mathrm{e}^u\partial_u
\rangle
\end{equation*}
and its corresponding invariant equation
\begin{equation*}
u_t=x^3F(\omega_1,\omega_2)u_{xxxx}-x^3(u_x^4-6u_x^2u_{xx}+4u_xu_{xxx}+3u_{xx}^2)F(\omega_1,\omega_2)+u_xG(\omega_1,\omega_2),\\[2mm]
\end{equation*}
where $\omega_1=x(u_x^2-u_{xx})u_x^{-1},\
\omega_2=x^2(u_x^3-3u_xu_{xx}+u_{xxx})u_x^{-1}$ and $F,\ G$ are
arbitrary smooth functions. Differentiating the above equation with
respect to $x$ and replacing $u_x$ with $u$ according to
\eqref{nonlocal-tranf} yield the evolution equation
\begin{equation*}
\begin{split}
    u_t=& x^3Fu_{xxxx}+3x^2Fu_{xxx}+[x^3u_{xxx}-x^3(u_x^4-6u_x^2u_{xx}+4u_xu_{xxx}+3u_{xx}^2)][F_{\omega_1}\sigma_1\\[2mm]
        & +F_{\omega_2}\sigma_2]-x^2[3u^4+4xu^3u_x-6(xu_{xx}+3u_x)u^2+(-12xu_x^2+12u_{xx}+4xu_{xxx})u\\[2mm]
        & +9u_x^2+10xu_xu_{xx}]F+u_xG+u\sigma_1 G_{\omega_1}+u\sigma_2 G_{\omega_2},
\end{split}
\end{equation*}
with
\begin{equation*}
    \omega_1=\frac{x(u^2-u_x)}u,\ \omega_2=\frac{x^2(u^3-3uu_x+u_{xx})}u,
\end{equation*}
and
\begin{align*}
&  \sigma_1=\frac{u^3+xu_xu^2-(xu_{xx}+u_x)u+xu_x^2}{u^2},\\[2mm]
&
\sigma_2=\frac{x[2u^4+2xu_xu^3-3(xu_{xx}+2u_x)u^2+(xu_{xxx}+2u_{xx})u-xu_xu_{xx}]}{u^2}.
\end{align*}

The new equation admits the non-local symmetry
\begin{equation*}
    t'=t,\quad x'=x,\quad u'=\frac{u}{1-\theta\mathrm{e}^v},
\end{equation*}
with symmetry generator
\begin{equation*}
    \mathrm{e}^vu\partial_u.
\end{equation*}

\section*{\small\bf 6. Concluding remarks}

\section*{Appendix A. Equations admitting decomposable four-dimensional solvable Lie algebras}

$2A_{2.2}-$ invariant equations,
\begin{align*}
& 2A^1_{2.2}=\langle -t\partial_t-u\partial_u,\partial_t,\partial_x,u\mathrm{e}^x\partial_u \rangle:\\[2mm]
& \qquad\qquad          F=\frac{\mathrm{e}^{\frac{u_x}u}}uF(\omega_1,\omega_2),\\[2mm]
& \qquad\qquad          G=-\frac{u_x\mathrm{e}^{\frac{u_x}u}}{u^4}[u^3+(-3u_x+6u_{xx}+4u_{xxx})u^2-6u_x(u_{xx}+u_x)u+3u_x^3]F(\omega_1,\omega_2)\\[2mm]
& \qquad\qquad\qquad      +\mathrm{e}^{\frac{u_x}u}G(\omega_1,\omega_2),\\[2mm]
& \qquad\qquad\qquad      \omega_1=\frac{(u_{xx}-u_x)u-u_x^2}{u^2},\ \omega_2=\frac{(u_{xxx}-u_x)u^2-3u_xu_{xx}u+2u_x^3}{u^3},\\[2mm]
& 2A^2_{2.2}=\langle -t\partial_t-x\partial_x,\partial_t,\partial_u,\mathrm{e}^u\partial_u \rangle:\\[2mm]
& \qquad\qquad          F=x^3F(\omega_1,\omega_2),\\[2mm]
& \qquad\qquad          G=-x^3(u_x^4-6u_x^2u_{xx}+4u_xu_{xxx}+3u_{xx}^2)F(\omega_1,\omega_2)+u_xG(\omega_1,\omega_2),\\[2mm]
& \qquad\qquad\qquad      \omega_1=xu_x^{-1}(u_x^2-u_{xx}),\ \omega_2=x^2u_x^{-1}(u_x^3-3u_xu_{xx}+u_{xxx}),\\[2mm]
& 2A^3_{2.2}=\langle -t\partial_t-u\partial_u,\partial_t,\partial_x,\mathrm{e}^x\partial_x+\alpha u\mathrm{e}^x\partial_u \rangle, \alpha\neq 0:\\[2mm]
& \qquad\qquad          F=\frac{u^{\alpha+3}}{(u_x-\alpha u)^{\alpha+4}}F(\omega_1,\omega_2),\\[2mm]
& \qquad\qquad          G=\frac{u^{\alpha+3}}{(u_x-\alpha u)^{\alpha+4}}[(\alpha^4-6\alpha^3+11\alpha^2-6\alpha)u-2(2\alpha^3-9\alpha^2+11\alpha-3)u_x\\[2mm]
& \qquad\qquad\qquad      +(6\alpha^2-18\alpha+11)u_{xx}-(4\alpha-6)u_{xxx}]F(\omega_1,\omega_2)+\frac{u^\alpha}{(u_x-\alpha u)^\alpha}G(\omega_1,\omega_2),\\[2mm]
& \qquad\qquad\qquad      \omega_1=\frac{(u_{xx}-u_x)u-u_x^2}{u^2},\ \omega_2=\frac{(u_{xxx}-u_x)u^2-3u_xu_{xx}u+2u_x^3}{u^3},\\[2mm]
& 2A^4_{2.2}=\langle -t\partial_t-x\partial_x,\partial_t,\alpha x\partial_x-u\partial_u,\partial_u \rangle:\\[2mm]
& \qquad\qquad          F=\frac{x^{3-\alpha}}{u_x^\alpha}F(\omega_1,\omega_2),\quad G=\frac{u_x^{1-\alpha}}{x^\alpha}G(\omega_1,\omega_2),\\[2mm]
& \qquad\qquad\qquad      \omega_1=\frac{xu_{xx}}{u_x},\ \omega_2=\frac{x^2u_{xxx}}{u_x},\\[2mm]
& 2A^5_{2.2}=\langle -t\partial_t-u\partial_u,\partial_t,x\partial_x,xu\partial_u \rangle:\\[2mm]
& \qquad\qquad          F=\frac{x^4}u\mathrm{e}^\frac{xu_x}{u}F(\omega_1,\omega_2),\\[2mm]
& \qquad\qquad          G=-\frac{x^4}{u^4}\mathrm{e}^\frac{xu_x}{u}u_x(4u_{xxx}u^2-6u_xu_{xx}u+3u_x^3)F(\omega_1,\omega_2)+\mathrm{e}^\frac{xu_x}{u}G(\omega_1,\omega_2),\\[2mm]
& \qquad\qquad\qquad      \omega_1=\frac{x^2(u_{xx}u-u_x^2)}{u^2},\ \omega_2=\frac{x^3(u_{xxx}u^2-3u_xu_{xx}u+2u_x^3)}{u^3},\\[2mm]
& 2A^6_{2.2}=\langle -t\partial_t-x\partial_x,\partial_x,t\partial_t,tu\partial_x \rangle:\\[2mm]
& \qquad\qquad          F=\frac 1{tu_x^4}F(u,\omega),\\[2mm]
& \qquad\qquad          G=-5\frac{2u_xu_{xx}u_{xxx}-3u_{xx}^3}{tu_x^6}F(u,\omega)+\frac{u_{xx}}{tu_x^2}G(u,\omega)-\frac ut,\\[2mm]
& \qquad\qquad\qquad      \omega=\frac{u_xu_{xxx}-3u_{xx}^2}{u_x^2u_{xx}},\\[2mm]
& 2A^7_{2.2}=\langle -t\partial_t-x\partial_x,\partial_x,\alpha t\partial_t+(1-\alpha)u\partial_u,tu\partial_x\rangle,  \alpha\neq 1 :\\[2mm]
& \qquad\qquad          F=\frac {u^4}{tu_x^4}F(\omega_1,\omega_2),\\[2mm]
& \qquad\qquad          G=-5\frac{u^4(2u_xu_{xx}u_{xxx}-3u_{xx}^3)}{tu_x^6}F(\omega_1,\omega_2)+u^\frac{\alpha}{\alpha-1}u_xG(\omega_1,\omega_2)-\frac ut,\\[2mm]
& \qquad\qquad\qquad      \omega_1=\frac{u_{xx}}{tu_x^3}u^\frac{\alpha-2}{\alpha-1},\ \omega_2=\frac{u_xu_{xxx}-3u_{xx}^2}{tu_x^5}u^\frac{2\alpha-3}{\alpha-1},\\[2mm]
& 2A^8_{2.2}=\langle -t\partial_t-x\partial_x,\partial_x,\alpha t\partial_x-u\partial_u,\partial_u\rangle,  \alpha\in \mathbb{R} :\\[2mm]
& \qquad\qquad          F=t^3F(\omega_1,\omega_2),\quad G=u_xG(\omega_1,\omega_2)+\alpha u_x\ln(tu_x),\\[2mm]
& \qquad\qquad\qquad      \omega_1=\frac{tu_{xx}}{u_x},\ \omega_2=\frac{t^2u_{xxx}}{u_x},\\[2mm]
& 2A^9_{2.2}=\langle -t\partial_t-x\partial_x,\partial_x,t\partial_t-u\partial_u,\partial_u \rangle:\\[2mm]
& \qquad\qquad          F=\frac{1}{t^5u_{xx}^4}F(\omega_1,\omega_2),\quad G=\frac{1}{t^3u_{xx}}G(\omega_1,\omega_2),\\[2mm]
& \qquad\qquad\qquad      \omega_1=tu_x,\ \omega_2=\frac{u_{xxx}}{tu_{xx}^2},\\[2mm]
& 2A^{10}_{2.2}=\langle -t\partial_t-x\partial_x,\partial_x,\alpha t\partial_t-u\partial_u,\partial_u \rangle,\ \alpha\neq 0,1:\\[2mm]
& \qquad\qquad          F=t^{-\frac{\alpha+3}{\alpha-1}}u_x^{-\frac{4\alpha}{\alpha-1}}F(\omega_1,\omega_2),\quad G=t^{-\frac{\alpha}{\alpha-1}}u_x^{-\frac1{\alpha-1}}G(\omega_1,\omega_2),\\[2mm]
& \qquad\qquad\qquad      \omega_1=t^{-\frac1{\alpha-1}}u_x^{-\frac{2\alpha-1}{\alpha-1}}u_{xx},\ \omega_2=t^{-\frac2{\alpha-1}}u_x^{-\frac{3\alpha-1}{\alpha-1}}u_{xxx},\\[2mm]
& 2A^{11}_{2.2}=\langle -t\partial_t-x\partial_x,\partial_x,\partial_u,\mathrm{e}^u\partial_u \rangle:\\[2mm]
& \qquad\qquad          F=t^3F(\omega_1,\omega_2),\\[2mm]
& \qquad\qquad          G=-t^3(u_x^4-6u_x^2u_{xx}+4u_xu_{xxx}+3u_{xx}^2)F(\omega_1,\omega_2)+u_xG(\omega_1,\omega_2),\\[2mm]
& \qquad\qquad\qquad      \omega_1=\frac{t(u_x^2-u_{xx})}{u_x},\ \omega_2=\frac{t^2(u_x^3-3u_xu_{xx}+u_{xxx})}{u_x},\\[2mm]
& 2A^{12}_{2.2}=\langle -t\partial_t-u\partial_u,\partial_u,\partial_x,\alpha \mathrm{e}^x\partial_x+t\mathrm{e}^x\partial_u \rangle,\ \alpha\neq 0:\\[2mm]
& \qquad\qquad          F=\frac{t^3}{(\alpha u_x-t)^4}F(\omega_1,\omega_2),\\[2mm]
& \qquad\qquad          G=\frac{t^3}{\alpha(\alpha u_x-t)^4}(6\alpha u_x+11\alpha u_{xx}+6\alpha u_{xxx}-6t)F(\omega_1,\omega_2)\\[2mm]
& \qquad\qquad\qquad      +G(\omega_1,\omega_2)+\frac{1}{\alpha}\ln{\frac{t}{\alpha u_x-t}},\\[2mm]
& \qquad\qquad\qquad      \omega_1=\frac{t(\alpha u_x+\alpha u_{xx}-t)}{(\alpha u_x-t)^2},\ \omega_2=\frac{t^2(2\alpha u_x+3\alpha u_{xx}+\alpha u_{xxx}-2t)}{(\alpha u_x-t)^3},\\[2mm]
& 2A^{13}_{2.2}=\langle -t\partial_t-x\partial_x,\partial_x,\partial_u,t\mathrm{e}^u\partial_x \rangle:\\[2mm]
& \qquad\qquad          F=\frac{1}{tu_x^4}F(\omega_1,\omega_2),\\[2mm]
& \qquad\qquad          G=\frac{26u_x^6+35u_x^4u_{xx}-10u_x^3u_{xxx}+30u_x^2u_{xx}^2-10u_xu_{xx}u_{xxx}+15u_{xx}^3}{tu_x^6}F(\omega_1,\omega_2)\\[2mm]
& \qquad\qquad\qquad      +u_xG(\omega_1,\omega_2)-\frac1t,\\[2mm]
& \qquad\qquad\qquad      \omega_1=\frac{u_x^2+u_{xx}}{tu_x^3},\ \omega_2=\frac{5u_x^4+6u_x^2u_{xx}-u_xu_{xxx}+3u_{xx}^2}{tu_x^5},\\[2mm]
& 2A^{14}_{2.2}=\langle -t\partial_t-x\partial_x,\partial_x,t\partial_t+\partial_u,\mathrm{e}^u\partial_u \rangle:\\[2mm]
& \qquad\qquad          F=\frac{\mathrm{e}^{4u}}{t^5(u_x^2-u_{xx})^4}F(\omega_1,\omega_2),\\[2mm]
& \qquad\qquad          G=-\frac{\mathrm{e}^{4u}(u_x^4-6u_x^2u_{xx}+4u_xu_{xxx}+3u_{xx}^2)}{t^5(u_x^2-u_{xx})^4}F(\omega_1,\omega_2)-\frac{\mathrm{e}^{2u}}{t^3(u_x^2-u_{xx})}G(\omega_1,\omega_2),\\[2mm]
& \qquad\qquad\qquad      \omega_1=\frac{tu_x}{\mathrm{e}^u},\ \omega_2=\frac{\mathrm{e}^u(u_x^3-3u_xu_{xx}+u_{xxx})}{t(u_x^2-u_{xx})^2},\\[2mm]
& 2A^{15}_{2.2}=\langle -x\partial_x-u\partial_u,\partial_x,u\partial_u,u\partial_x \rangle:\\[2mm]
& \qquad\qquad          F=\frac{u^4}{u_x^4}F(t,\omega),\quad G=-\frac{5u^4u_{xx}(2u_xu_{xxx}-3u_{xx}^2)}{u_x^6}F(t,\omega)-\frac{u^2u_{xx}}{u_x^2}G(t,\omega),\\[2mm]
& \qquad\qquad\qquad      \omega=\frac{u(u_xu_{xxx}-3u_{xx}^2)}{u_x^2u_{xx}},\\[2mm]
& 2A^{16}_{2.2}=\langle -x\partial_x-u\partial_u,\partial_x,\partial_t+u\partial_u,u\partial_x \rangle:\\[2mm]
& \qquad\qquad          F=\frac{u^4}{u_x^4}F(\omega_1,\omega_2),\quad G=-\frac{5u^4u_{xx}(2u_xu_{xxx}-3u_{xx}^2)}{u_x^6}F(\omega_1,\omega_2)-\frac{uu_x}{\mathrm{e}^t}G(\omega_1,\omega_2),\\[2mm]
& \qquad\qquad\qquad      \omega_1=\frac{u\mathrm{e}^tu_{xx}}{u_x^3},\ \omega_2=\frac{u^2\mathrm{e}^t(u_xu_{xxx}-3u_{xx}^2)}{u_x^5},\\[2mm]
& 2A^{17}_{2.2}=\langle -x\partial_x-u\partial_u,\partial_x,-t\partial_t+\alpha u\partial_x,\partial_t \rangle,\ \alpha\neq 0:\\[2mm]
& \qquad\qquad          F=\frac{u^4\mathrm{e}^\frac{1}{\alpha u_x}}{u_x^4}F(\omega_1,\omega_2),\\[2mm]
& \qquad\qquad          G=-\frac{ 5u^4\mathrm{e}^\frac{1}{\alpha u_x}u_{xx}(2u_xu_{xxx}-3u_{xx}^2) }{u_x^6}F(\omega_1,\omega_2)+uu_x\mathrm{e}^\frac{1}{\alpha u_x}G(\omega_1,\omega_2),\\[2mm]
& \qquad\qquad\qquad      \omega_1=\frac{uu_{xx}}{u_x^3},\ \omega_2=\frac{u^2(u_xu_{xxx}-3u_{xx}^2)}{u_x^5},\\[2mm]
& 2A^{18}_{2.2}=\langle -x\partial_x-u\partial_u,\partial_x,-t\partial_t+\alpha u\partial_u,\partial_t \rangle,\ \alpha\neq 0:\\[2mm]
& \qquad\qquad          F=u^4u_x^{\frac{1-4\alpha}{\alpha}}F(\omega_1,\omega_2),\quad G=uu_x^{\frac{1}{\alpha}}G(\omega_1,\omega_2),\\[2mm]
& \qquad\qquad\qquad      \omega_1=\frac{uu_{xx}}{u_x^2},\ \omega_2=\frac{u^2u_{xxx}}{u_x^3},\\[2mm]
& 2A^{19}_{2.2}=\langle -x\partial_x-u\partial_u,\partial_x,\partial_t,u\mathrm{e}^t\partial_x \rangle:\\[2mm]
& \qquad\qquad          F=\frac{u^4}{u_x^4}F(\omega_1,\omega_2),\\[2mm]
& \qquad\qquad          G=-\frac{5u^4u_{xx}(2u_xu_{xxx}-3u_{xx}^2)}{u_x^6}F(\omega_1,\omega_2)+uu_xG(\omega_1,\omega_2)-u,\\[2mm]
& \qquad\qquad\qquad      \omega_1=\frac{uu_{xx}}{u_x^3},\ \omega_2=\frac{u^2(u_xu_{xxx}-3u_{xx}^2)}{u_x^5},\\[2mm]
& 2A^{20}_{2.2}=\langle -x\partial_x-u\partial_u,\partial_x,\partial_t,\mathrm{e}^t\partial_t+\alpha u\mathrm{e}^t\partial_x \rangle,\ \alpha\neq 0:\\[2mm]
& \qquad\qquad          F=\frac{u^4}{u_x^4\mathrm{e}^\frac{1}{\alpha u_x}}F(\omega_1,\omega_2),\\[2mm]
& \qquad\qquad          G=-\frac{ 5u^4u_{xx}(2u_xu_{xxx}-3u_{xx}^2) }{u_x^6\mathrm{e}^\frac{1}{\alpha u_x}}F(\omega_1,\omega_2)+\frac{uu_x}{\mathrm{e}^\frac{1}{\alpha u_x}}G(\omega_1,\omega_2)-\alpha uu_x,\\[2mm]
& \qquad\qquad\qquad      \omega_1=\frac{uu_{xx}}{u_x^3},\ \omega_2=\frac{u^2(u_xu_{xxx}-3u_{xx}^2)}{u_x^5},\\[2mm]
& 2A^{21}_{2.2}=\langle -x\partial_x-u\partial_u,\partial_x,\partial_t,u\mathrm{e}^t\partial_u \rangle:\\[2mm]
& \qquad\qquad          F=\frac{u^4}{u_x^4}F(\omega_1,\omega_2),\quad G=uG(\omega_1,\omega_2)+u\ln u_x,\\[2mm]
& \qquad\qquad\qquad      \omega_1=\frac{uu_{xx}}{u_x^2},\ \omega_2=\frac{u^2u_{xxx}}{u_x^3},\\[2mm]
& 2A^{22}_{2.2}=\langle -x\partial_x-u\partial_u,\partial_x,\partial_t,\mathrm{e}^t\partial_t+\alpha u\mathrm{e}^t\partial_u \rangle,\ \alpha\neq 0:\\[2mm]
& \qquad\qquad          F=\frac{u^4}{u_x^\frac{4\alpha+1}{\alpha}}F(\omega_1,\omega_2),\quad G=\frac{u}{u_x^\frac{1}{\alpha}}G(\omega_1,\omega_2)+\alpha u,\\[2mm]
& \qquad\qquad\qquad      \omega_1=\frac{uu_{xx}}{u_x^2},\ \omega_2=\frac{u^2u_{xxx}}{u_x^3},\\[2mm]
& 2A^{23}_{2.2}=\langle -x\partial_x-u\partial_u,\partial_x,t\partial_t,tu\partial_u \rangle:\\[2mm]
& \qquad\qquad          F=\frac{u^4}{tu_x^4}F(\omega_1,\omega_2),\quad G=\frac{u}{t}G(\omega_1,\omega_2)+\frac{u}{t}\ln u_x,\\[2mm]
& \qquad\qquad\qquad      \omega_1=\frac{uu_{xx}}{u_x^2},\ \omega_2=\frac{u^2u_{xxx}}{u_x^3},\\[2mm]
& 2A^{24}_{2.2}=\langle -x\partial_x,\partial_x,-t\partial_t+\partial_u,\partial_t \rangle:\\[2mm]
& \qquad\qquad          F=\frac{\mathrm{e}^u}{u_x^4}F(\omega_1,\omega_2),\quad G=\mathrm{e}^uG(\omega_1,\omega_2),\\[2mm]
& \qquad\qquad\qquad      \omega_1=\frac{u_{xx}}{u_x^2},\ \omega_2=\frac{u_{xxx}}{u_x^3},\\[2mm]
& 2A^{25}_{2.2}=\langle -x\partial_x,\partial_x,\partial_t,\alpha \mathrm{e}^t\partial_t+\mathrm{e}^t\partial_u \rangle,\ \alpha\neq 0:\\[2mm]
& \qquad\qquad          F=\frac{1}{u_x^4\mathrm{e}^{\alpha u}}F(\omega_1,\omega_2),\quad G=\mathrm{e}^{(1-\alpha)u}G(\omega_1,\omega_2)+\frac1\alpha ,\\[2mm]
& \qquad\qquad\qquad      \omega_1=\frac{u_{xx}}{u_x^2},\ \omega_2=\frac{u_{xxx}}{u_x^3},\\[2mm]
& 2A^{26}_{2.2}=\langle -x\partial_x,\partial_x,\partial_t,\mathrm{e}^t\partial_u \rangle:\\[2mm]
& \qquad\qquad          F=\frac{1}{u_x^4}F(\omega_1,\omega_2),\quad G=G(\omega_1,\omega_2)+u,\\[2mm]
& \qquad\qquad\qquad      \omega_1=\frac{u_{xx}}{u_x^2},\ \omega_2=\frac{u_{xxx}}{u_x^3},\\[2mm]
& 2A^{27}_{2.2}=\langle -x\partial_x,\partial_x,-u\partial_u,\partial_u \rangle:\\[2mm]
& \qquad\qquad          F=\frac{u_x^4}{u_{xx}^4}F(t,\omega),\quad G=\frac{u_x^2}{u_{xx}}G(t,\omega),\\[2mm]
& \qquad\qquad\qquad      \omega=\frac{u_xu_{xxx}}{u_{xx}^2},\\[2mm]
& 2A^{28}_{2.2}=\langle -x\partial_x,\partial_x,\partial_t-u\partial_u,\partial_u \rangle:\\[2mm]
& \qquad\qquad          F=\frac{1}{\mathrm{e}^{4t}u_x^4}F(\omega_1,\omega_2),\quad G=\frac1{\mathrm{e}^t}G(\omega_1,\omega_2)+u,\\[2mm]
& \qquad\qquad\qquad      \omega_1=\frac{u_{xx}}{\mathrm{e}^tu_x^2},\ \omega_2=\frac{u_{xxx}}{\mathrm{e}^{2t}u_x^3},\\[2mm]
& 2A^{29}_{2.2}=\langle -x\partial_x,\partial_x,\partial_u,t\mathrm{e}^u\partial_u \rangle:\\[2mm]
& \qquad\qquad          F=\frac{u_x^4}{(u_x^2-u_{xx})^4}F(t,\omega),\\[2mm]
& \qquad\qquad          G=-\frac{u_x^4(u_x^4-6u_x^2u_{xx}+4u_xu_{xxx}+3u_{xx}^2)}{(u_x^2-u_{xx})^4}F(t,\omega)-\frac{u_x^2}{u_x^2-u_{xx}}G(t,\omega)\\[2mm]
& \qquad\qquad\qquad      +\frac{u_{xx}}{t(u_x^2-u_{xx})},\\[2mm]
& \qquad\qquad\qquad      \omega=\frac{u_x(u_x^3-3u_xu_{xx}+u_{xxx})}{(u_x^2-u_{xx})^2}.\\[2mm]
\end{align*}

$A_{3.1}\oplus A_1-$ invariant equations,
\begin{align*}
& A^2_{3.1}\oplus A_1=\langle \partial_t,\partial_u,x\partial_u,f(x)\partial_u\rangle,\ f''\neq 0:\\[2mm]
& \qquad\qquad          F=F(x,\omega),\quad G=-\frac{f^{(4)}}{f''}u_{xx}F(x,\omega)+G(x,\omega),\\[2mm]
& \qquad\qquad\qquad      \omega=\frac{f''u_{xxx}-f^{(3)}u_{xx}}{f''}.\\[2mm]
& A^3_{3.1}\oplus A_1=\langle \partial_u,x\partial_u,f(t,x)\partial_u,g(t,x)\partial_u \rangle,\ f_{xx}g_{xxx}-g_{xx}f_{xxx}\neq0:\\[2mm]
& \qquad\qquad          F=F(t,x),\\[2mm]
& \qquad\qquad          G=\frac{(f_{xx}u_{xxx}-f_{xxx}u_{xx})g_{xxxx}+(g_{xxx}u_{xx}-g_{xx}u_{xxx})f_{xxxx}}{f_{xxx}g_{xx}-g_{xxx}f_{xx}}F(t,x)+G(t,x)\\[2mm]
& \qquad\qquad\qquad       +\frac{(f_tg_{xx}-g_tf_{xx})u_{xxx}+(g_tf_{xxx}-f_tg_{xxx})u_{xx}}{f_{xxx}g_{xx}-g_{xxx}f_{xx}}.\\[2mm]
\end{align*}

$A_{3.2}\oplus A_1-$ invariant equations,
\begin{align*}
& A^2_{3.2}\oplus A_1=\langle -t\partial_t-u\partial_u,\partial_t,xu\partial_u,uf(x)\partial_u\rangle,\ f''\neq 0:\\[2mm]
& \qquad\qquad          F=\frac{\mathrm{e}^\sigma}{u}F(x,\omega),\\[2mm]
& \qquad\qquad          G=-\mathrm{e}^\sigma\frac{(u_{xx}u^3-u_x^2u^2)f^{(4)}+[4(u_xu_{xxx}+3u_{xx}^2)u^2-12u_x^2u_{xx}u+6u_x^4]f''}{u^4f''}F(x,\omega)\\[2mm]
& \qquad\qquad\qquad      +\mathrm{e}^\sigma G(x,\omega),\\[2mm]
& \qquad\qquad\qquad      \sigma=\frac{xuu_xf''-x(u_{xx}u-u_x^2)f'+(u_{xx}u-u_x^2)f}{f''u^2},\\[2mm]
& \qquad\qquad\qquad      \omega=\frac{(-u_{xx}u^2+u_x^2u)f^{(3)}+(u_{xxx}u^2-3u_xu_{xx}u+2u_x^3)f''}{u^3f''}.\\[2mm]
& A^3_{3.2}\oplus A_1=\langle -t\partial_t-x\partial_x,\partial_x,tu\partial_x,tf(u)\partial_x\rangle,\ f''\neq 0:\\[2mm]
& \qquad\qquad          F=\frac{1}{tu_x^4}F(u,\omega),\\[2mm]
& \qquad\qquad          G=-\frac{u_{xx}[u_x^4f^{(4)}+5(2u_xu_{xxx}-3u_{xx}^2)f'']}{tu_x^6f''}F(u,\omega)+u_xG(u,\omega)\\[2mm]
& \qquad\qquad\qquad      -\frac{u_{xx}(uf'-f)}{tu_x^2f''}-\frac ut,\\[2mm]
& \qquad\qquad\qquad      \omega=\frac{-u_x^2u_{xx}f^{(3)}+(u_xu_{xxx}-3u_{xx}^2)f''}{tu_x^5f''}.\\[2mm]
& A^3_{3.2}\oplus A_1=\langle -t\partial_t-x\partial_x,\partial_x,tu\partial_x,-u\partial_u\rangle:\\[2mm]
& \qquad\qquad          F=\frac{u^4}{tu_x^4}F(\omega_1,\omega_2),\\[2mm]
& \qquad\qquad          G=-\frac{5u^4u_{xx}(2u_xu_{xxx}-3u_{xx}^2)}{tu_x^6}F(\omega_1,\omega_2)+uu_xG(\omega_1,\omega_2)-\frac ut,\\[2mm]
& \qquad\qquad\qquad      \omega_1=\frac{uu_{xx}}{tu_x^3},\ \omega_2=\frac{u^2(u_xu_{xxx}-3u_{xx}^2)}{tu_x^5},\\[2mm]
& A^4_{3.2}\oplus A_1=\langle -t\partial_t-x\partial_x,\partial_x,\partial_u,t\partial_t\rangle:\\[2mm]
& \qquad\qquad          F=\frac{1}{tu_x^4}F(\omega_1,\omega_2),\quad G=\frac1t G(\omega_1,\omega_2),\\[2mm]
& \qquad\qquad\qquad      \omega_1=\frac{u_{xx}}{u_x^2},\ \omega_2=\frac{u_{xxx}}{u_x^3},\\[2mm]
& A^6_{3.2}\oplus A_1=\langle -t\partial_t-x\partial_x,\partial_x,t\partial_t+\partial_u,t\mathrm{e}^{-u}\partial_x\rangle:\\[2mm]
& \qquad\qquad          F=\frac{1}{tu_x^4}F(\omega_1,\omega_2),\\[2mm]
& \qquad\qquad          G=\frac{ -26u_x^6+35u_x^4u_{xx}+10u_x^3u_{xxx}-30u_x^2u_{xx}^2-10u_xu_{xx}u_{xxx}+15u_{xx}^3 }{tu_x^6} F(\omega_1,\omega_2)\\[2mm]
& \qquad\qquad\qquad      +\frac{u_x}{\mathrm{e}^u}G(\omega_1,\omega_2)+\frac1t,\\[2mm]
& \qquad\qquad\qquad      \omega_1=\frac{\mathrm{e}^u(u_x^2-u_{xx})}{tu_x^3},\ \omega_2=\frac{\mathrm{e}^u(5u_x^4-6u_x^2u{xx}-u_xu_{xxx}+3u_{xx}^2)}{tu_x^5},\\[2mm]
& A^7_{3.2}\oplus A_1=\langle -x\partial_x-u\partial_u,\partial_x,u\partial_x,\partial_t \rangle:\\[2mm]
& \qquad\qquad          F=\frac{u^4}{u_x^4}F(\omega_1,\omega_2),\\[2mm]
& \qquad\qquad          G=-\frac{5u^4u_{xx}(2u_xu_{xxx}-3u_{xx}^2)}{u_x^6}F(\omega_1,\omega_2)+uu_xG(\omega_1,\omega_2),\\[2mm]
& \qquad\qquad\qquad      \omega_1=\frac{uu_{xx}}{u_x^3},\ \omega_2=\frac{u^2(u_xu_{xxx}-3u_{xx}^2)}{u_x^5},\\[2mm]
& A^8_{3.2}\oplus A_1=\langle -x\partial_x-u\partial_u,\partial_x,\partial_t,u\partial_u \rangle:\\[2mm]
& \qquad\qquad          F=\frac{u^4}{u_x^4}F(\frac{uu_{xx}}{u_x^2},\frac{u^2u_{xxx}}{u_x^3}),\quad G=u G(\frac{uu_{xx}}{u_x^2},\frac{u^2u_{xxx}}{u_x^3}),\\[2mm]
& A^{10}_{3.2}\oplus A_1=\langle -x\partial_x,\partial_x,\partial_t,\partial_u \rangle:\\[2mm]
& \qquad\qquad          F=\frac{1}{u_x^4}F(\frac{u_{xx}}{u_x^2},\frac{u_{xxx}}{u_x^3}),\quad G=G(\frac{u_{xx}}{u_x^2},\frac{u_{xxx}}{u_x^3}).\\[2mm]
\end{align*}

$A_{3.3}\oplus A_1-$ invariant equations,
\begin{align*}
& A^2_{3.3}\oplus A_1=\langle \partial_u,\partial_t,(t+x)\partial_u,f(x)\partial_u \rangle,\ f''\neq 0:\\[2mm]
& \qquad\qquad          F=F(x,\omega),\\[2mm]
& \qquad\qquad          G=-\frac{u_{xx}f^{(4)}}{f''}F(x,\omega)+G(x,\omega)-\frac{f'}{f''}u_{xx}+u_x,\\[2mm]
& \qquad\qquad\qquad      \omega=\frac{u_{xx}f^{(3)}-u_{xxx}f''}{f''},\\[2mm]
& A^2_{3.3}\oplus A_1=\langle \partial_u,\partial_t,(t+x)\partial_u,\partial_t-\partial_x \rangle:\\[2mm]
& \qquad\qquad          F=F(u_{xx},u_{xxx}),\quad G=F(u_{xx},u_{xxx})+u_x, \\[2mm]
& A^4_{3.3}\oplus A_1=\langle \partial_x,\partial_u,\partial_t+u\partial_x,\partial_t+\alpha t\partial_x+\alpha \partial_u \rangle,\ \alpha\in\mathbb{R}:\\[2mm]
& \qquad\qquad          F=\frac{1}{u_x^4}F(\omega_1,\omega_2),\\[2mm]
& \qquad\qquad          G=-\frac{5u_{xx}(2u_xu_{xxx}-3u_{xx}^2)}{u_x^6}F(\omega_1,\omega_2)+u_xG(\omega_1,\omega_2)-\alpha tu_x+\alpha,\\[2mm]
& \qquad\qquad\qquad      \omega_1=\frac{u_{xx}}{u_x^3},\ \omega_2=\frac{u_xu_{xxx}-3u_{xx}^2}{u_x^5},\\[2mm]
& A^5_{3.3}\oplus A_1=\langle \partial_x,u\partial_x,-\partial_u,\partial_t \rangle:\\[2mm]
& \qquad\qquad          F=\frac{1}{u_x^4}F(\omega_1,\omega_2),\\[2mm]
& \qquad\qquad          G=-\frac{5u_{xx}(2u_xu_{xxx}-3u_{xx}^2)}{u_x^6}F(\omega_1,\omega_2)+u_xG(\omega_1,\omega_2),\\[2mm]
& \qquad\qquad\qquad      \omega_1=\frac{u_{xx}}{u_x^3},\ \omega_2=\frac{u_xu_{xxx}-3u_{xx}^2}{u_x^5},\\[2mm]
& A^6_{3.3}\oplus A_1=\langle \partial_x,u\partial_x,\partial_t-\partial_u,f(t+u)\partial_x \rangle,\ f''\neq 0:\\[2mm]
& \qquad\qquad          F=\frac{1}{u_x^4}F(\omega_1,\omega_2),\\[2mm]
& \qquad\qquad          G=[\frac{15u_{xx}^3}{u_x^6}-\frac{10u_{xx}u_{xxx}}{u_x^5}-\frac{u_{xx}f^{(4)}}{u_x^2f''}]F(\omega_1,\omega_2)+u_xG(\omega_1,\omega_2)+\frac{u_{xx}f'}{u_x^2f''},\\[2mm]
& \qquad\qquad\qquad      \omega_1=t+u,\ \omega_2=\frac{(u_xu_{xxx}-3u_{xx}^2)f''-u_x^2u_{xx}f^{(3)}}{u_x^5f''},\\[2mm]
& A^7_{3.3}\oplus A_1=\langle x\partial_u,\partial_u,x^2\partial_x+xu\partial_u,\partial_t \rangle:\\[2mm]
& \qquad\qquad          F=x^8F(\omega_1,\omega_2),\\[2mm]
& \qquad\qquad          G=4x^6(3u_{xx}+2xu_{xxx})F(\omega_1,\omega_2)+x G(\omega_1,\omega_2),\\[2mm]
& \qquad\qquad\qquad      \omega_1=x^3u_{xx},\ \omega_2=x^4(3u_{xx}+xu_{xxx}).\\[2mm]
\end{align*}

$A_{3.4}\oplus A_1-$ invariant equations,
\begin{align*}
& A^1_{3.4}\oplus A_1=\langle \partial_u,\partial_t,t\partial_t+\partial_x+(t+u)\partial_u,\partial_x \rangle:\\[2mm]
& \qquad\qquad          F=\frac{1}{u_x}F(\omega_1,\omega_2),\quad G=G(\omega_1,\omega_2)+\ln{u_x},\\[2mm]
& \qquad\qquad\qquad      \omega_1=\frac{u_{xx}}{u_x},\  \omega_2=\frac{u_{xxx}}{u_x},\\[2mm]
& A^1_{3.4}\oplus A_1=\langle \partial_u,\partial_t,t\partial_t+\partial_x+(t+u)\partial_u,\mathrm{e}^x\partial_x \rangle:\\[2mm]
& \qquad\qquad          F=\mathrm{e}^{-x}F(\omega_1,\omega_2),\quad G=-\mathrm{e}^{-x}u_xF(\omega_1,\omega_2)+G(\omega_1,\omega_2)+x,\\[2mm]
& \qquad\qquad\qquad      \omega_1=\mathrm{e}^{-x}(u_x-u_{xx}),\  \omega_2=\mathrm{e}^{-x}(u_x-u_{xxx}),\\[2mm]
& A^4_{3.4}\oplus A_1=\langle \partial_x,\partial_u,(x+u)\partial_x+u\partial_u,\partial_t \rangle:\\[2mm]
& \qquad\qquad          F=\frac{\mathrm{e}^\frac{4}{u_x}}{u_x^4}F(\omega_1,\omega_2),\\[2mm]
& \qquad\qquad          G=\frac{5u_{xx}(3u_{xx}^2-2u_xu_{xxx})\mathrm{e}^\frac{4}{u_x}}{u_x^6}F(\omega_1,\omega_2)+u_x\mathrm{e}^\frac1{u_x}G(\omega_1,\omega_2),\\[2mm]
& \qquad\qquad\qquad      \omega_1=\frac{u_{xx}\mathrm{e}^\frac1{u_x}}{u_x^3},\  \omega_2=\frac{(u_xu_{xxx}-3u_{xx}^2)\mathrm{e}^\frac2{u_x}}{u_x^3},\\[2mm]
& A^5_{3.4}\oplus A_1=\langle \partial_x,u\partial_x,x\partial_x-\partial_u,\mathrm{e}^{-u}\partial_x \rangle:\\[2mm]
& \qquad\qquad          F=\frac1{u_x^4}F(t,\omega),\\[2mm]
& \qquad\qquad          G=-\frac{u_{xx}(u_x^4+10u_xu_{xxx}-15u_{xx}^2)}{u_x^6}F(t,\omega)+u_x\mathrm{e}^{-u}G(t,\omega),\\[2mm]
& \qquad\qquad\qquad      \omega=\frac{(u_x^2u_{xx}+u_xu_{xxx}-3u_{xx}^2)\mathrm{e}^u}{u_x^5},\\[2mm]
& A^5_{3.4}\oplus A_1=\langle \partial_x,u\partial_x,x\partial_x-\partial_u,t\mathrm{e}^{-u}\partial_x \rangle:\\[2mm]
& \qquad\qquad          F=\frac1{u_x^4}F(t,\omega),\\[2mm]
& \qquad\qquad          G=-\frac{u_{xx}(u_x^4+10u_xu_{xxx}-15u_{xx}^2)}{u_x^6}F(t,\omega)+u_x\mathrm{e}^{-u}G(t,\omega)+\frac{u_{xx}}{tu_x^2},\\[2mm]
& \qquad\qquad\qquad      \omega=\frac{(u_x^2u_{xx}+u_xu_{xxx}-3u_{xx}^2)\mathrm{e}^u}{u_x^5},\\[2mm]
& A^5_{3.4}\oplus A_1=\langle \partial_x,u\partial_x,x\partial_x-\partial_u,\partial_t \rangle:\\[2mm]
& \qquad\qquad          F=\frac1{u_x^4}F(\omega_1,\omega_2),\\[2mm]
& \qquad\qquad          G=-\frac{5u_{xx}(2u_xu_{xxx}-3u_{xx}^2)}{u_x^6}F(\omega_1,\omega_2)+\mathrm{e}^{-u}u_x(\omega_1,\omega_2),\\[2mm]
& \qquad\qquad\qquad      \omega_1=\frac{\mathrm{e}^uu_{xx}}{u_x^3},\ \omega_2=\frac{(u_xu_{xxx}-3u_{xx}^2)\mathrm{e}^u}{u_x^5},\\[2mm]
& A^6_{3.4}\oplus A_1=\langle \partial_x,u\partial_x,\partial_t+x\partial_x-\partial_u,f(t+u)\mathrm{e}^t\partial_x\rangle,\ f''\neq 0 :\\[2mm]
& \qquad\qquad          F=\frac1{u_x^4}F(\omega_1,\omega_2),\\[2mm]
& \qquad\qquad          G=-[\frac{u_{xx}f^{(4)}}{u_x^2f''}+\frac{5u_{xx}(2u_xu_{xxx}-3u_{xx}^2)}{u_x^6}]F(\omega_1,\omega_2)+u_x\mathrm{e}^tG(\omega_1,\omega_2)\\[2mm]
& \qquad\qquad\qquad       +\frac{(f+f')u_{xx}}{f''u_x^2},\\[2mm]
& \qquad\qquad\qquad      \omega_1=t+u,\ \omega_2=\frac{(u_xu_{xxx}-3u_{xx}^2)f''-u_x^2u_{xx}f^{(3)}}{u_x^5\mathrm{e}^tf''},\\[2mm]
& A^7_{3.4}\oplus A_1=\langle u\partial_x,\partial_x,x(1+u)\partial_x+u^2\partial_u,u\mathrm{e}^{-\frac1u}\partial_x \rangle:\\[2mm]
& \qquad\qquad          F=\frac{u^8}{u_x^4}F(t,\omega),\\[2mm]
& \qquad\qquad          G=-\frac{u^4[5u^4u_{xx}(2u_xu_{xxx}-3u_{xx}^2)+8u^3u_x^2(3u_{xx}^2-u_xu_{xxx})+u_x^4u_{xx}(1-12u^2)]}{u_x^6}F(t,\omega)\\[2mm]
& \qquad\qquad\qquad       +uu_x\mathrm{e}^{-\frac1u}G(t,\omega),\\[2mm]
& \qquad\qquad\qquad      \omega=\frac{u^3[u^2(u_xu_{xxx}-3u_{xx}^2)-3u_x^2u_{xx}u-u_x^2u_{xx}]\mathrm{e}^\frac1u}{u_x^5},\\[2mm]
& A^7_{3.4}\oplus A_1=\langle u\partial_x,\partial_x,x(1+u)\partial_x+u^2\partial_u,tu\mathrm{e}^{-\frac1u}\partial_x \rangle:\\[2mm]
& \qquad\qquad          F=\frac{u^8}{u_x^4}F(t,\omega),\\[2mm]
& \qquad\qquad          G=-\frac{u^4[5u^4u_{xx}(2u_xu_{xxx}-3u_{xx}^2)+8u^3u_x^2(3u_{xx}^2-u_xu_{xxx})+u_x^4u_{xx}(1-12u^2)]}{u_x^6}F(t,\omega)\\[2mm]
& \qquad\qquad\qquad       +uu_x\mathrm{e}^{-\frac1u}G(t,\omega)+\frac{u^4u_{xx}}{tu_x^2},\\[2mm]
& \qquad\qquad\qquad      \omega=\frac{u^3[u^2(u_xu_{xxx}-3u_{xx}^2)-3u_x^2u_{xx}u-u_x^2u_{xx}]\mathrm{e}^\frac1u}{u_x^5},\\[2mm]
& A^7_{3.4}\oplus A_1=\langle u\partial_x,\partial_x,x(1+u)\partial_x+u^2\partial_u,\partial_t \rangle:\\[2mm]
& \qquad\qquad          F=\frac{u^8}{u_x^4}F(\omega_1,\omega_2),\\[2mm]
& \qquad\qquad          G=-\frac{u^4[5u^4u_{xx}(2u_xu_{xxx}-3u_{xx}^2)+8u^3u_x^2(3u_{xx}^2-u_xu_{xxx})-12u^2u_x^4u_{xx}]}{u_x^6}F(\omega_1,\omega_2)\\[2mm]
& \qquad\qquad\qquad       +uu_x\mathrm{e}^{-\frac1u}G(\omega_1,\omega_2),\\[2mm]
& \qquad\qquad\qquad      \omega_1=\frac{u^3\mathrm{e}^{\frac1u}u_{xx}}{u_x^3},\ \omega_2=\frac{u^4\mathrm{e}^\frac1u[u(u_xu_{xxx}-3u_{xx}^2)+3u_x^2u_{xx}]}{u_x^5},\\[2mm]
& A^8_{3.4}\oplus A_1=\langle u\partial_x,\partial_x,\partial_t+x(1+u)\partial_x+u^2\partial_u,u\mathrm{e}^tf(t+\frac1u)\partial_x \rangle,\ f''\neq 0:\\[2mm]
& \qquad\qquad          F=\frac{u^8}{u_x^4}F(\omega_1,\omega_2),\\[2mm]
& \qquad\qquad          G=\{\frac{u^6[5u^2u_{xx}(3u_{xx}^2-2u_xu_{xxx})+8uu_x^2(u_xu_{xxx}-3u_{xx}^2)-12u_x^4u_{xx}]}{u_x^6}\\[2mm]
& \qquad\qquad\qquad       -\frac{u^4f^{(4)}u_{xx}}{f''u_x^2}\}F(\omega_1,\omega_2)-uu_x\mathrm{e}^tG(\omega_1,\omega_2)+\frac{(f'+f)u^4u_{xx}}{f''u_x^2},\\[2mm]
& \qquad\qquad\qquad      \omega_1=t+\frac1u,\ \omega_2=\frac{u^4[u(u_xu_{xxx}-3u_{xx}^2)+3u_x^2u_{xx}]f''}{u_x^5\mathrm{e}^t}+\frac{u^3u_{xx}f^{(3)}}{\mathrm{e}^tu_x^3}.\\[2mm]
\end{align*}

$A_{3.5}\oplus A_1-$ invariant equations,
\begin{align*}
& A^1_{3.5}\oplus A_1=\langle \partial_u,\partial_t,t\partial_t+u\partial_u,\partial_x \rangle:\\[2mm]
& \qquad\qquad          F=\frac{1}{u_x}F(\frac{u_{xx}}{u_x},\frac{u_{xxx}}{u_x}),\quad G=G(\frac{u_{xx}}{u_x},\frac{u_{xxx}}{u_x}),\\[2mm]
& A^2_{3.5}\oplus A_1=\langle \partial_u,\partial_t,t\partial_t+\partial_x+u\partial_u,\mathrm{e}^x\partial_u \rangle:\\[2mm]
& \qquad\qquad          F=\mathrm{e}^{-x}F(\omega_1,\omega_2),\quad G=-\mathrm{e}^{-x}u_xF(\omega_1,\omega_2)+G(\omega_1,\omega_2),\\[2mm]
& \qquad\qquad\qquad      \omega_1=\mathrm{e}^{-x}(u_{xx}-u_x),\  \omega_2=\mathrm{e}^{-x}(u_{xxx}-u_x),\\[2mm]
& A^3_{3.5}\oplus A_1=\langle \partial_x,\partial_u,x\partial_x+u\partial_u,\partial_t \rangle:\\[2mm]
& \qquad\qquad          F=\frac{1}{u_{xx}^4}F(u_x,\frac{u_{xxx}}{u_{xx}^2}),\quad G=\frac1{u_{xx}}G(u_x,\frac{u_{xxx}}{u_{xx}^2}),\\[2mm]
& A^5_{3.5}\oplus A_1=\langle \partial_x,u\partial_x,x\partial_x,\partial_t \rangle:\\[2mm]
& \qquad\qquad          F=\frac{1}{u_x^4}F(u,\omega),\quad G=-5\frac{u_{xx}(2u_xu_{xxx}-3u_{xx}^2)}{u_x^6}F(u,\omega)+\frac{u_{xx}}{u_x^2}G(u,\omega),\\[2mm]
& \qquad\qquad\qquad      \omega=\frac{u_xu_{xxx}-3u_{xx}^2}{u_x^2u_{xx}},\\[2mm]
& A^6_{3.5}\oplus A_1=\langle \partial_x,u\partial_x,\partial_t+x\partial_x,f(u)\mathrm{e}^t\partial_x \rangle\ f''\neq0:\\[2mm]
& \qquad\qquad          F=\frac{1}{u_x^4}F(u,\omega),\\[2mm]
& \qquad\qquad          G=-[\frac{5u_{xx}(2u_xu_{xxx}-3u_{xx}^2)}{u_x^6}+\frac{u_{xx}f^{(4)}}{u_x^2f''}]F(u,\omega)+u_x\mathrm{e}^tG(u,\omega)+\frac{u_{xx}f}{u_x^2f''},\\[2mm]
& \qquad\qquad\qquad      \omega=\frac{(3u_{xx}^2-u_xu_{xxx})f''+u_x^2u_{xx}f^{(3)}}{\mathrm{e}^tu_x^5f''}.\\[2mm]
\end{align*}

$A_{3.6}\oplus A_1-$ invariant equations,
\begin{align*}
& A^1_{3.6}\oplus A_1=\langle \partial_u,\partial_t,t\partial_t-u\partial_u,\partial_x \rangle:\\[2mm]
& \qquad\qquad          F=u_xF(\omega_1,\omega_2),\quad G=u_x^2G(\omega_1,\omega_2),\\[2mm]
& \qquad\qquad\qquad      \omega_1=\frac{u_{xx}}{u_x},\  \omega_2=\frac{u_{xxx}}{u_x},\\[2mm]
& A^2_{3.6}\oplus A_1=\langle \partial_u,\partial_t,t\partial_t+\partial_x-u\partial_u,\mathrm{e}^{-x}\partial_u \rangle:\\[2mm]
& \qquad\qquad          F=\mathrm{e}^{-x}F(\omega_1,\omega_2),\quad G=u_x\mathrm{e}^{-x}F(\omega_1,\omega_2)+\mathrm{e}^{-2x}G(\omega_1,\omega_2),\\[2mm]
& \qquad\qquad\qquad      \omega_1=\mathrm{e}^x(u_x+u_{xx}),\  \omega_2=\mathrm{e}^x(u_x-u_{xxx}),\\[2mm]
& A^3_{3.6}\oplus A_1=\langle \partial_x,\partial_u,x\partial_x-u\partial_u,\partial_t \rangle:\\[2mm]
& \qquad\qquad          F=\frac1{u_x^2}F(\omega_1,\omega_2),\quad G=u_x^\frac12G(\omega_1,\omega_2),\\[2mm]
& \qquad\qquad\qquad      \omega_1=\frac{u_{xx}}{u_x^\frac32},\  \omega_2=\frac{u_{xxx}}{u_x^2},\\[2mm]
& A^5_{3.6}\oplus A_1=\langle \partial_x,u\partial_x,x\partial_x+2u\partial_u,u^\frac12f(t)\partial_x \rangle:\\[2mm]
& \qquad\qquad          F=\frac{u^4}{u_x^4}F(t,\omega),\\[2mm]
& \qquad\qquad          G=\frac{5u^2u_{xx}}{4u_x^6}[4u^2(3u_{xx}^2-2u_xu_{xxx})-3u_x^4]F(t,\omega)+u^\frac12u_xG(t,\omega)-\frac{4u^2u_{xx}f'}{u_x^2f},\\[2mm]
& \qquad\qquad\qquad      \omega=\frac{u^\frac32}{u_x^5}[2u(u_xu_{xxx}-3u_{xx}^2)+3u_x^2u_{xx}],\\[2mm]
& A^5_{3.6}\oplus A_1=\langle \partial_x,u\partial_x,x\partial_x+2u\partial_u,\partial_t \rangle:\\[2mm]
& \qquad\qquad          F=\frac{u^4}{u_x^4}F(\omega_1,\omega_2),\\[2mm]
& \qquad\qquad          G=-\frac{5u^4u_{xx}(2u_xu_{xxx}-3u_{xx}^2)}{u_x^6}F(\omega_1,\omega_2)+u^\frac12u_xG(\omega_1,\omega_2),\\[2mm]
& \qquad\qquad\qquad      \omega_1=\frac{u^\frac32u_{xx}}{u_x^3},\ \omega_2=\frac{u^\frac52(u_xu_{xxx}-3u_{xx}^2)}{u_x^5}.\\[2mm]
\end{align*}

$A_{3.7}\oplus A_1-$ invariant equations,
\begin{align*}
& A^1_{3.7}\oplus A_1=\langle \partial_t,\partial_u,t\partial_t+qu\partial_u,\partial_x \rangle:\\[2mm]
& \qquad\qquad          F=u_x^{-\frac1q}F(\omega_1,\omega_2),\quad G=u_x^{1-\frac1q}G(\omega_1,\omega_2),\\[2mm]
& \qquad\qquad\qquad      \omega_1=\frac{u_{xx}}{u_x},\  \omega_2=\frac{u_{xxx}}{u_x},\\[2mm]
& A^2_{3.7}\oplus A_1=\langle \partial_t,\partial_u,t\partial_t+\partial_x+qu\partial_u,\mathrm{e}^{qx}\partial_u \rangle:\\[2mm]
& \qquad\qquad          F=q^{-3}\mathrm{e}^{-x}F(\omega_1,\omega_2),\quad G=-u_x\mathrm{e}^{-x}F(\omega_1,\omega_2)+\mathrm{e}^{(q-1)x}G(\omega_1,\omega_2),\\[2mm]
& \qquad\qquad\qquad      \omega_1=\mathrm{e}^{-qx}{qu_x-u_{xx}},\  \omega_2=\mathrm{e}^{-qx}{q^2u_x-u_{xxx}},\\[2mm]
& A^3_{3.7}\oplus A_1=\langle \partial_x,\partial_u,x\partial_x+qu\partial_u,\partial_t \rangle:\\[2mm]
& \qquad\qquad                        F=u_x^{\frac4{q-1}}F(\omega_1,\omega_2),\quad G=u_x^{\frac q{q-1}}G(\omega_1,\omega_2),\\[2mm]
& \qquad\qquad\qquad                      \omega_1=u_x^{\frac{2-q}{q-1}}u_{xx},\ \omega_2=u_x^{\frac{3-q}{q-1}}u_{xxx},\\[2mm]
& A^5_{3.7}\oplus A_1=\langle \partial_u,x\partial_u,(1-q)x\partial_x+u\partial_u,x^{\frac1{1-q}}\partial_u \rangle:\\[2mm]
& \qquad\qquad                        F=x^4F(t,\omega),\\[2mm]
& \qquad\qquad                        G=-\frac{(2q-1)(3q-2)}{(q-1)^2}x^2u_{xx}F(t,\omega)+x^{\frac 1{1-q}}G(t,\omega),\\[2mm]
& \qquad\qquad\qquad                      \omega=x^{\frac{2q-1}{q-1}}[(q-1)xu_{xxx}+(2q-1)u_{xx}],\\[2mm]
& A^5_{3.7}\oplus A_1=\langle \partial_u,x\partial_u,(1-q)x\partial_x+u\partial_u,tx^{\frac1{1-q}}\partial_u \rangle:\\[2mm]
& \qquad\qquad                        F=x^4F(t,\omega),\\[2mm]
& \qquad\qquad                        G=-\frac{(2q-1)(3q-2)}{(q-1)^2}x^2u_{xx}F(t,\omega)+x^{\frac 1{1-q}}G(t,\omega)+\frac{(q-1)^2x^2u_{xx}}{qt},\\[2mm]
& \qquad\qquad\qquad                      \omega=x^{\frac{2q-1}{q-1}}[(q-1)xu_{xxx}+(2q-1)u_{xx}],\\[2mm]
& A^5_{3.7}\oplus A_1=\langle \partial_u,x\partial_u,(1-q)x\partial_x+u\partial_u,\partial_t \rangle:\\[2mm]
& \qquad\qquad                        F=x^4F(\omega_1,\omega_2),\quad G=x^{\frac 1{1-q}}G(\omega_1,\omega_2),\\[2mm]
& \qquad\qquad\qquad                      \omega_1=x^{\frac{2q-1}{q-1}}u_{xx},\ \omega_2=x^{\frac{3q-2}{q-1}}u_{xxx},\\[2mm]
& A^6_{3.7}\oplus A_1=\langle \partial_u,x\partial_u,\partial_t+(1-q)x\partial_x+u\partial_u,x^{\frac 1{1-q}}\partial_u \rangle:\\[2mm]
& \qquad\qquad                        F=\mathrm{e}^{4(1-q)t}F(\omega_1,\omega_2),\\[2mm]
& \qquad\qquad                        G=-\frac{(2q-1)(3q-2)u_{xx}\mathrm{e}^{4(1-q)t}}{(q-1)^2x^2}F(\omega_1,\omega_2)+\mathrm{e}^tG(\omega_1,\omega_2),\\[2mm]
& \qquad\qquad\qquad                      \omega_1=\mathrm{e}^{(q-1)t}x,\ \omega_2=\frac{\mathrm{e}^{(2-3q)t}[(q-1)xu_{xxx}+(2q-1)u_{xx}]}x,\\[2mm]
& A^6_{3.7}\oplus A_1=\langle \partial_u,x\partial_u,\partial_t+(1-q)x\partial_x+u\partial_u,\partial_t+\alpha x^{\frac 1{1-q}}\partial_u \rangle,\ \alpha\neq 0:\\[2mm]
& \qquad\qquad                        F=x^4F(\omega_1,\omega_2),\\[2mm]
& \qquad\qquad                        G=-\frac{\alpha q(2q-1)(3q-2)x^{\frac1{1-q}}[\ln{x}+(q-1)t]}{(q-1)^5}F(\omega_1,\omega_2)+x^{\frac1{1-q}}G(\omega_1,\omega_2),\\[2mm]
& \qquad\qquad\qquad                      \omega_1=(q-1)^3x^{\frac{2q-1}{q-1}}u_{xx}-\alpha q[\ln{x}+(q-1)t],\\[2mm]
& \qquad\qquad\qquad                      \omega_2=(q-1)^4x^{\frac{3q-2}{q-1}}u_{xxx}+\alpha q(2q-1)[\ln{x}+(q-1)t].\\[2mm]
\end{align*}

$A_{3.8}\oplus A_1-$ invariant equations,
\begin{align*}
& A^1_{3.8}\oplus A_1=\langle \partial_x,\partial_u,u\partial_x-x\partial_u,\partial_t \rangle:\\[2mm]
& \qquad\qquad                        F=\frac1{(1+u_x^2)^2}F(\omega_1,\omega_2),\\[2mm]
& \qquad\qquad                        G=-5\frac{u_xu_{xx}(2u_x^2u_{xxx}-3u_xu_{xx}^2+2u_{xxx})}{(1+u_x^2)^4}F(\omega_1,\omega_2)+(1+u_x^2)^{\frac12}G(\omega_1,\omega_2),\\[2mm]
& \qquad\qquad\qquad                      \omega_1=\frac{u_{xx}}{(1+u_x^2)^{\frac32}},\ \omega_2=\frac{u_x^2u_{xxx}-3u_xu_{xx}^2+u_{xxx}}{(1+u_x^2)^3},\\[2mm]
& A^3_{3.8}\oplus A_1=\langle \partial_u,x\partial_u,-(x^2+1)\partial_x-xu\partial_u,\partial_t \rangle:\\[2mm]
& \qquad\qquad                        F=(1+x^2)^4F(\omega_1,\omega_2),\\[2mm]
& \qquad\qquad                        G=4x(1+x^2)^2(2x^2u_{xxx}+3xu_{xx}+2u_{xxx})F(\omega_1,\omega_2)+(1+x^2)^{\frac12}G(\omega_1,\omega_2),\\[2mm]
& \qquad\qquad\qquad                      \omega_1=(1+x^2)^{\frac32}u_{xx},\ \omega_2=(1+x^2)^{\frac32}(x^2u_{xxx}+3xu_{xx}+u_{xxx}),\\[2mm]
& A^3_{3.8}\oplus A_1=\langle \partial_u,x\partial_u,-(x^2+1)\partial_x-xu\partial_u,(x^2+1)^\frac12\partial_u \rangle:\\[2mm]
& \qquad\qquad                        F=(1+x^2)^4F(t,\omega),\\[2mm]
& \qquad\qquad                        G=x(1+x^2)^2(8x^3u_{xxx}+12x^2u_{xx}+8xu_{xxx}+3u_{xx})F(t,\omega)+(1+x^2)^{\frac12}G(t,\omega),\\[2mm]
& \qquad\qquad\qquad                      \omega=(1+x^2)^{\frac32}(x^2u_{xxx}+3xu_{xx}+u_{xxx}),\\[2mm]
& A^3_{3.8}\oplus A_1=\langle \partial_u,x\partial_u,-(x^2+1)\partial_x-xu\partial_u,t(x^2+1)^\frac12\partial_u \rangle:\\[2mm]
& \qquad\qquad                        F=(1+x^2)^4F(t,\omega),\\[2mm]
& \qquad\qquad                        G=(1+x^2)^2(8x^3u_{xxx}+12x^2u_{xx}+8xu_{xxx}+3u_{xx})F(t,\omega)+(1+x^2)^{\frac12}G(t,\omega)\\[2mm]
& \qquad\qquad\qquad                    +\frac{u_{xx}}t(1+x^2)^2,\\[2mm]
& \qquad\qquad\qquad                      \omega=(1+x^2)^{\frac32}(x^2u_{xxx}+3xu_{xx}+u_{xxx}),\\[2mm]
& A^4_{3.8}\oplus A_1=\langle \partial_x,u\partial_x,\partial_t-xu\partial_x-(u^2+1)\partial_u,(1+u^2)^\frac12f(t+\arctan{u})\partial_x \rangle\ f+f''\neq0:\\[2mm]
& \qquad\qquad                        F=\frac{(1+u^2)^4}{u_x^4}F(\omega_1,\omega_2),\\[2mm]
& \qquad\qquad                        G=\{\frac{(1+u^2)^2}{u_x^6}[24u_x^4u_{xx}\arctan^2{u}-8u_x^2(3u_x^2u_{xx}(u-2t)+(1+u^2)(u_xu_{xxx}-3u_{xx}^2))\\[2mm]
& \qquad\qquad\qquad                    \arctan{u}+12u_x^4u_{xx}(u^2-2tu+2t^2)+8u_x^2(1+u^2)(u_xu_{xxx}-3u_{xx}^2)(u-t)\\[2mm]
& \qquad\qquad\qquad                    -5(1+u^2)^2u_{xx}(2u_xu_{xxx}-3u_{xx}^2)]+\frac{(1+u^2)^2u_{xx}}{(f+f'')u_x^2}[3(1-8\omega_1^2)f+8\omega_1f'\\[2mm]
& \qquad\qquad\qquad                    +2(1-12\omega_1^2)f''+8\omega_1f^{(3)}-f^{(4)}]\}F(\omega_1,\omega_2)+(1+u^2)^\frac12u_xG(\omega_1,\omega_2)\\[2mm]
& \qquad\qquad\qquad                    +\frac{f'(1+u^2)^2u_{xx}}{(f+f'')u_x^2},\\[2mm]
& \qquad\qquad\qquad                      \omega_1=t+\arctan u,\\[2mm]
& \qquad\qquad\qquad                      \omega_2=(1+u^2)^{\frac32}[\frac{u_{xx}}{u_x^3}(3u-\frac{f'+f^{(3)}}{f+f''})+(1+u^2)\frac{u_xu_{xxx}-3u_{xx}^2}{u_x^5}].\\[2mm]
\end{align*}

$A_{3.9}\oplus A_1-$ invariant equations,
\begin{align*}
& A^1_{3.9}\oplus A_1=\langle \partial_x,\partial_u,(qx+u)\partial_x+(qu-x)\partial_u,\partial_t \rangle:\\[2mm]
& \qquad\qquad                        F=\frac1{(1+u_x^2)^2\mathrm{e}^{4q\arctan u_x}}F(\omega_1,\omega_2),\\[2mm]
& \qquad\qquad                        G=-5\frac{u_xu_{xx}(2u_x^2u_{xxx}-3u_xu_{xx}^2+2u_{xxx})}{(1+u_x^2)^4\mathrm{e}^{4q\arctan u_x}}F(\omega_1,\omega_2)+\frac{(1+u_x^2)^{\frac12}}{\mathrm{e}^{q\arctan u_x}}G(\omega_1,\omega_2),\\[2mm]
& \qquad\qquad\qquad                      \omega_1=\frac{u_{xx}}{(1+u_x^2)^{\frac32}\mathrm{e}^{q\arctan u_x}},\ \omega_2=\frac{u_x^2u_{xxx}-3u_xu_{xx}^2+u_{xxx}}{(1+u_x^2)^3\mathrm{e}^{2q\arctan u_x}},\\[2mm]
& A^3_{3.9}\oplus A_1=\langle \partial_u,x\partial_u,-(x^2+1)\partial_x+(q-x)u\partial_u,\partial_t \rangle:\\[2mm]
& \qquad\qquad                        F=(x^2+1)^4F(\omega_1,\omega_2),\\[2mm]
& \qquad\qquad                        G=4x(x^2+1)^2(2x^2u_{xxx}+3xu_{xx}+2u_{xxx})F(\omega_1,\omega_2)+\frac{(x^2+1)^{\frac12}}{\mathrm{e}^{q\arctan x}}G(\omega_1,\omega_2),\\[2mm]
& \qquad\qquad\qquad                      \omega_1=(x^2+1)^{\frac32}\mathrm{e}^{q\arctan x}u_{xx},\\[2mm]
& \qquad\qquad\qquad                      \omega_2=(1+u_x^2)^{\frac32}\mathrm{e}^{q\arctan x}(x^2u_{xxx}+3xu_{xx}+u_{xxx}),\\[2mm]
& A^3_{3.9}\oplus A_1=\langle \partial_u,x\partial_u,-(x^2+1)\partial_x+(q-x)u\partial_u,(x^2+1)^\frac12\mathrm{e}^{-q\arctan{x}}\partial_u \rangle:\\[2mm]
& \qquad\qquad                        F=(x^2+1)^4F(t,\omega),\\[2mm]
& \qquad\qquad                        G=(x^2+1)^2[8x^3u_{xxx}+12x^2u_{xx}+8xu_{xxx}+(3-q^2)u_{xx}]F(t,\omega)+\frac{(1+x^2)^{\frac12}}{\mathrm{e}^{q\arctan{x}}}G(t,\omega),\\[2mm]
& \qquad\qquad\qquad                      \omega=(1+x^2)^{\frac32}(x^2u_{xxx}+3xu_{xx}+qu_{xx}+u_{xxx}),\\[2mm]
& A^3_{3.9}\oplus A_1=\langle \partial_u,x\partial_u,-(x^2+1)\partial_x+(q-x)u\partial_u,t(x^2+1)^\frac12\mathrm{e}^{-q\arctan{x}}\partial_u \rangle:\\[2mm]
& \qquad\qquad                        F=(x^2+1)^4F(t,\omega),\\[2mm]
& \qquad\qquad                        G=(x^2+1)^2(8x^3u_{xxx}+12x^2u_{xx}+8xu_{xxx}+(3-q^2)u_{xx})F(t,\omega)+\frac{(1+x^2)^{\frac12}}{\mathrm{e}^{q\arctan{x}}}G(t,\omega)\\[2mm]
& \qquad\qquad\qquad                    +\frac{(1+x^2)^2u_{xx}}{(1+q^2)t},\\[2mm]
& \qquad\qquad\qquad                      \omega=(1+x^2)^{\frac32}(x^2u_{xxx}+3xu_{xx}+qu_{xx}+u_{xxx})\mathrm{e}^{q\arctan{x}},\\[2mm]
& A^4_{3.9}\oplus A_1=\langle \partial_x,u\partial_x,\partial_t+(q-u)x\partial_x-(u^2+1)\partial_u,(u^2+1)^\frac12\mathrm{e}^{qt}\partial_x \rangle:\\[2mm]
& \qquad\qquad                        F=\frac{(u^2+1)^4}{u_x^4}F(\omega_1,\omega_2),\\[2mm]
& \qquad\qquad                        G=(u^2+1)^2[-\frac{5(u^2+1)^2u_{xx}}{u_x^6}(2u_xu_{xxx}-3u_{xx}^2)+\frac{8u(u^2+1)}{u_x^4}(u_xu_{xxx}-3u_{xx}^2)\\[2mm]
& \qquad\qquad\qquad                    +\frac{3u_{xx}}{u_x^2}(4u^2+1)]F(\omega_1,\omega_2)+(u^2+1)^{\frac12}\mathrm{e}^{qt}u_xG(\omega_1,\omega_2)+\frac{q(u^2+1)^2u_{xx}}{u_x^2},\\[2mm]
& \qquad\qquad\qquad                      \omega_1=t+\arctan u,\\[2mm]
& \qquad\qquad\qquad                      \omega_2=\frac{(u^2+1)^{\frac32}}{u_x^5\mathrm{e}^{qt}}[(u^2+1)(u_xu_{xxx}-3u_{xx}^2)+3uu_x^2u_{xx}],\\[2mm]
& A^4_{3.9}\oplus A_1=\langle \partial_x,u\partial_x,\partial_t+(q-u)x\partial_x-(u^2+1)\partial_u,f(t+\arctan u)(u^2+1)^\frac12\mathrm{e}^{qt}\partial_x  \rangle,\\[2mm]
& \qquad\qquad\quad                      \ f+f''\neq 0:\\[2mm]
& \qquad\qquad                        F=\frac{(u^2+1)^4}{u_x^4}F(\omega_1,\omega_2),\\[2mm]
& \qquad\qquad                        G=(u^2+1)^2[-\frac{5(u^2+1)^2u_{xx}}{u_x^6}(2u_xu_{xxx}-3u_{xx}^2)+\frac{8u(u^2+1)}{u_x^4}(u_xu_{xxx}-3u_{xx}^2)\\[2mm]
& \qquad\qquad\qquad                    +\frac{u_{xx}}{u_x^2}(2(6u^2+1)+\frac{f-f^{(4)}}{f+f''})]F(\omega_1,\omega_2)+(u^2+1)^{\frac12}\mathrm{e}^{qt}u_xG(\omega_1,\omega_2)\\[2mm]
& \qquad\qquad\qquad                    +\frac{(u^2+1)^2u_{xx}(qf+f')}{u_x^2(f+f'')},\\[2mm]
& \qquad\qquad\qquad                      \omega_1=t+\arctan u,\\[2mm]
& \qquad\qquad\qquad                      \omega_2=\frac{(u^2+1)^{\frac32}}{u_x^5\mathrm{e}^{qt}}[(u^2+1)(u_xu_{xxx}-3u_{xx}^2)+3uu_x^2u_{xx}-\frac{f+f^{(3)}}{f+f''}u_x^2u_{xx}],\\[2mm]
\end{align*}

\section*{Appendix A. Equations admitting non-decompos\-able four-dimensional solvable Lie algebras}

$A_{4.1}-$ invariant equations,
\begin{align*}
& A^1_{4.1}=\langle \partial_u,\partial_x,\partial_t,t\partial_x+x\partial_u \rangle:\\[2mm]
& \qquad\qquad        F=F(u_{xx},u_{xxx}),\quad G=G(u_{xx},u_{xxx})-\frac{u_x^2}2,\\[2mm]
& A^2_{4.1}=\langle \partial_u,x\partial_u,\partial_t,-\partial_x+tx\partial_u \rangle:\\[2mm]
& \qquad\qquad        F=F(u_{xx},u_{xxx}),\quad G=G(u_{xx},u_{xxx})-\frac{x^2}2,\\[2mm]
& A^3_{4.1}=\langle x\partial_u,\partial_u,\partial_t,x^2\partial_x+(t+xu)\partial_u \rangle:\\[2mm]
& \qquad\qquad        F=x^8F(\omega_1,\omega_2),\quad G=4x^2(2\omega_2-3x\omega_1)F(\omega_1,\omega_2)+xG(\omega_1,\omega_2)-\frac1{2x},\\[2mm]
& \qquad\qquad\qquad  \omega_1=x^3u_{xx},\ \omega_2=x^4(xu_{xxx}+3u_{xx}),\\[2mm]
\end{align*}

$A_{4.2}-$ invariant equations,
\begin{align*}
& A^1_{4.2}=\langle \partial_t,\partial_u,\partial_x,qt\partial_t+x\partial_x+(x+u)\partial_u \rangle:\\[2mm]
& \qquad\qquad        F=\mathrm{e}^{(4-q)u_x}F(\omega_1,\omega_2),\quad G=\mathrm{e}^{(1-q)u_x}G(\omega_1,\omega_2),\\[2mm]
& \qquad\qquad\qquad       \omega_1=\mathrm{e}^{u_x}u_{xx},\ \omega_2=\mathrm{e}^{2u_x}u_{xxx},\\[2mm]
& A^2_{4.2}=\langle \partial_x,\partial_u,\partial_t,t\partial_t+qx\partial_x+(t+u)\partial_u \rangle\ q\neq 1:\\[2mm]
& \qquad\qquad        F=u_x^{\frac{1-4q}{q-1}}F(\omega_1,\omega_2),\quad G=G(\omega_1,\omega_2)-\frac{\ln{|u_x|}}{q-1},\\[2mm]
& \qquad\qquad\qquad       \omega_1=u_x^{\frac{1-2q}{q-1}}u_{xx},\ \omega_2=u_x^{\frac{1-3q}{q-1}}u_{xxx},\\[2mm]
& A^3_{4.2}=\langle \partial_x,\partial_u,\partial_t,t\partial_t+x\partial_x+(t+u)\partial_u \rangle:\\[2mm]
& \qquad\qquad        F=\frac1{u_{xx}^3}F(u_x,\frac{u_{xxx}}{u_{xx}^2}),\quad G=G(u_x,\frac{u_{xxx}}{u_{xx}^2})-\ln{|u_{xx}|},\\[2mm]
& A^4_{4.2}=\langle \partial_t,\partial_u,x\partial_u,qt\partial_t-\partial_x+u\partial_u \rangle:\\[2mm]
& \qquad\qquad        F=\mathrm{e}^{qx}F(\omega_1,\omega_2),\quad G=\mathrm{e}^{(q-1)x}G(\omega_1,\omega_2),\\[2mm]
& \qquad\qquad\qquad       \omega_1=\mathrm{e}^xu_{xx},\ \omega_2=\mathrm{e}^xu_{xxx},\\[2mm]
& A^5_{4.2}=\langle \partial_t,x\partial_u,\partial_u,qt\partial_t+x^2\partial_x+(x+1)u\partial_u \rangle:\\[2mm]
& \qquad\qquad        F=x^8\mathrm{e}^{\frac qx}F(\omega_1,\omega_2),\\[2mm]
& \qquad\qquad        G=4x^6\mathrm{e}^{\frac qx}(3u_{xx}+2xu_{xxx})F(\omega_1,\omega_2)+x\mathrm{e}^{\frac{q-1}x}G(\omega_1,\omega_2),\\[2mm]
& \qquad\qquad\qquad       \omega_1=x^3\mathrm{e}^{\frac1x}u_{xx},\ \omega_2=x^4\mathrm{e}^{\frac1x}(3u_{xx}+xu_{xxx}),\\[2mm]
& A^6_{4.2}=\langle \partial_u,x\partial_u,\partial_t,t\partial_t+(q-1)x\partial_x+(tx+qu)\partial_u \rangle\ q\neq1:\\[2mm]
& \qquad\qquad        F=x^{\frac{4q-5}{q-1}}F(\omega_1,\omega_2),\quad G=xG(\omega_1,\omega_2)+\frac{x\ln{|x|}}{q-1},\\[2mm]
& \qquad\qquad\qquad       \omega_1=x^{\frac{q-2}{q-1}}u_{xx},\ \omega_2=x^{\frac{2q-3}{q-1}}u_{xxx},\\[2mm]
& A^7_{4.2}=\langle \partial_u,x\partial_u,\partial_t,t\partial_t+(tx+u)\partial_u \rangle:\\[2mm]
& \qquad\qquad        F=\frac1{u_{xx}}F(x,\frac{u_{xxx}}{u_{xx}}),\quad G=G(x,\frac{u_{xxx}}{u_{xx}})+x\ln {|u_{xx}|},\\[2mm]
& A^8_{4.2}=\langle x\partial_u,\partial_u,\partial_t,t\partial_t+(1-q)x\partial_x+(t+u)\partial_u \rangle\ q\neq1:\\[2mm]
& \qquad\qquad        F=x^{\frac{4q-3}{q-1}}F(\omega_1,\omega_2),\quad G=G(\omega_1,\omega_2)-\frac{\ln x}{q-1},\\[2mm]
& \qquad\qquad\qquad       \omega_1=x^{\frac{2q-1}{q-1}}u_{xx},\ \omega_2=x^{\frac{3q-2}{q-1}}u_{xxx},\\[2mm]
& A^9_{4.2}=\langle x\partial_u,\partial_u,\partial_t,t\partial_t+(t+u)\partial_u \rangle:\\[2mm]
& \qquad\qquad        F=\frac1{u_{xx}}F(x,\frac{u_{xxx}}{u_{xx}}),\quad G=G(x,\frac{u_{xxx}}{u_{xx}})+\ln {|u_{xx}|}.\\[2mm]
\end{align*}

$A_{4.3}-$ invariant equations,
\begin{align*}
& A^1_{4.3}=\langle \partial_t,\partial_u,\partial_x,t\partial_t+x\partial_u \rangle:\\[2mm]
& \qquad\qquad        F=\mathrm{e}^{-u_x}F(u_{xx},u_{xxx}),\quad G=\mathrm{e}^{-u_x}G(u_{xx},u_{xxx}),\\[2mm]
& A^2_{4.3}=\langle \partial_x,\partial_u,\partial_t,x\partial_x+t\partial_u \rangle:\\[2mm]
& \qquad\qquad        F=\frac1{u_x^4}F(\omega_1,\omega_2),\quad G=G(\omega_1,\omega_2)-\ln{|u_x|},\\[2mm]
& \qquad\qquad\qquad       \omega_1=\frac{u_{xx}}{u_x^2},\ \omega_2=\frac{u_{xxx}}{u_x^3},\\[2mm]
& A^3_{4.3}=\langle \partial_t,\partial_u,x\partial_u,t\partial_t-\partial_x \rangle:\\[2mm]
& \qquad\qquad        F=\mathrm{e}^xF(u_{xx},u_{xxx}),\quad G=\mathrm{e}^xG(u_{xx},u_{xxx}),\\[2mm]
& A^4_{4.3}=\langle \partial_t,x\partial_u,\partial_u,t\partial_t+x^2\partial_x+xu\partial_u \rangle:\\[2mm]
& \qquad\qquad        F=x^8\mathrm{e}^{\frac1x}F(\omega_1,\omega_2),\\[2mm]
& \qquad\qquad        G=4x^6(3u_{xx}+2xu_{xxx})\mathrm{e}^{\frac1x}F(\omega_1,\omega_2)+x\mathrm{e}^{\frac1x}G(\omega_1,\omega_2),\\[2mm]
& \qquad\qquad\qquad       \omega_1=x^3u_{xx},\ \omega_2=x^4(3u_{xx}+xu_{xxx}),\\[2mm]
& A^5_{4.3}=\langle \partial_u,x\partial_u,\partial_t,x\partial_x+(tx+u)\partial_u \rangle:\\[2mm]
& \qquad\qquad        F=x^4F(\omega_1,\omega_2),\quad G=xG(\omega_1,\omega_2)+x\ln{|x|},\\[2mm]
& \qquad\qquad\qquad       \omega_1=xu_{xx},\ \omega_2=x^2u_{xxx},\\[2mm]
& A^6_{4.3}=\langle x\partial_u,\partial_u,\partial_t,-x\partial_x+t\partial_u \rangle:\\[2mm]
& \qquad\qquad        F=x^4F(\omega_1,\omega_2),\quad G=G(\omega_1,\omega_2)-\ln{|x|},\\[2mm]
& \qquad\qquad\qquad       \omega_1=x^2u_{xx},\ \omega_2=x^3u_{xxx}.\\[2mm]
\end{align*}

$A_{4.4}-$ invariant equations,
\begin{align*}
& A^1_{4.4}=\langle \partial_u,\partial_x,\partial_t,t\partial_t+(t+x)\partial_x+(x+u)\partial_u \rangle:\\[2mm]
& \qquad\qquad        F=\mathrm{e}^{3u_x}F(\omega_1,\omega_2),\quad G=G(\omega_1,\omega_2)-\frac{u_x^2}2,\\[2mm]
& \qquad\qquad\qquad       \omega_1=\mathrm{e}^{u_x}u_{xx},\ \omega_2=\mathrm{e}^{2u_x}u_{xxx},\\[2mm]
& A^2_{4.4}=\langle \partial_u,x\partial_u,\partial_t,t\partial_t-\partial_x +(tx+u)\partial_u \rangle:\\[2mm]
& \qquad\qquad        F=\mathrm{e}^xF(\omega_1,\omega_2),\quad G=\mathrm{e}^xG(\omega_1,\omega_2)-\frac{x^2}2,\\[2mm]
& \qquad\qquad\qquad      \omega_1=\mathrm{e}^xu_{xx},\omega_2=\mathrm{e}^xu_{xxx},\\[2mm]
& A^3_{4.4}=\langle x\partial_u,\partial_u,\partial_t,t\partial_t+x^2\partial_x+[t+(x+1)u]\partial_u \rangle:\\[2mm]
& \qquad\qquad        F=x^8\mathrm{e}^{\frac1x}F(\omega_1,\omega_2),\\[2mm]
& \qquad\qquad        G=4x^6(3u_{xx}+2xu_{xxx})\mathrm{e}^{\frac1x}F(\omega_1,\omega_2)+xG(\omega_1,\omega_2)-\frac1{2x},\\[2mm]
& \qquad\qquad\qquad       \omega_1=x^3\mathrm{e}^{\frac1x}u_{xx},\ \omega_2=x^4\mathrm{e}^{\frac1x}(3u_{xx}+xu_{xxx}).\\[2mm]
\end{align*}

$A_{4.5}-$ invariant equations,
\begin{align*}
& A^1_{4.5}=\langle \partial_t,\partial_x,\partial_u,t\partial_t+qx\partial_x+pu\partial_u \rangle,\ p\neq q:\\[2mm]
& \qquad\qquad        F=u_x^{\frac{1-4q}{q-p}}F(\omega_1,\omega_2),\quad G=u_x^{\frac{1-p}{q-p}}G(\omega_1,\omega_2),\\[2mm]
& \qquad\qquad\qquad       \omega_1=u_x^{\frac{p-2q}{q-p}}u_{xx},\ \omega_2=u_x^{\frac{p-3q}{q-p}}u_{xxx},\\[2mm]
& A^2_{4.5}=\langle \partial_t,\partial_x,\partial_u,t\partial_t+qx\partial_x+qu\partial_u \rangle:\\[2mm]
& \qquad\qquad        F=u_{xx}^{\frac{1-4q}q}F(u_x,\frac{u_{xxx}}{u_{xx}^2}),\quad G=u_{xx}^{\frac{1-q}q}G(u_x,\frac{u_{xxx}}{u_{xx}^2}),\\[2mm]
& A^3_{4.5}=\langle \partial_x,\partial_t,\partial_u,qt\partial_t+x\partial_x+pu\partial_u \rangle,\ p\neq1:\\[2mm]
& \qquad\qquad        F=u_x^{\frac{4-q}{p-1}}F(\omega_1,\omega_2),\quad G=u_x^{\frac{p-q}{p-1}}G(\omega_1,\omega_2),\\[2mm]
& \qquad\qquad\qquad       \omega_1=u_x^{\frac{2-p}{p-1}}u_{xx},\ \omega_2=u_x^{\frac{3-p}{p-1}}u_{xxx},\\[2mm]
& A^4_{4.5}=\langle \partial_x,\partial_t,\partial_u,t\partial_t+x\partial_x+u\partial_u \rangle,\ p\neq1:\\[2mm]
& \qquad\qquad        F=u_{xx}^{-3}F(u_x,\frac{u_{xxx}}{u_{xx}^2}),\quad G=G(u_x,\frac{u_{xxx}}{u_{xx}^2}),\\[2mm]
& A^5_{4.5}=\langle \partial_x,\partial_u,\partial_t,pt\partial_t+x\partial_x+qu\partial_u \rangle,\ q\neq1:\\[2mm]
& \qquad\qquad        F=u_x^{\frac{4-p}{q-1}}F(\omega_1,\omega_2),\quad G=u_x^{\frac{q-p}{q-1}}G(\omega_1,\omega_2),\\[2mm]
& \qquad\qquad\qquad       \omega_1=u_x^{\frac{2-q}{q-1}}u_{xx},\ \omega_2=u_x^{\frac{3-q}{q-1}}u_{xxx},\\[2mm]
& A^6_{4.5}=\langle \partial_x,\partial_u,\partial_t,pt\partial_t+x\partial_x+u\partial_u \rangle:\\[2mm]
& \qquad\qquad        F=u_{xx}^{p-4}F(u_x,\frac{u_{xxx}}{u_{xx}^2}),\quad G=u_{xx}^{p-1}G(u_x,\frac{u_{xxx}}{u_{xx}^2}),\\[2mm]
& A^7_{4.5}=\langle \partial_t,\partial_u,x\partial_u,t\partial_t+(q-p)x\partial_x+qu\partial_u \rangle,\ p\neq q:\\[2mm]
& \qquad\qquad        F=x^{\frac{-1+4q-4p}{q-p}}F(\omega_1,\omega_2),\quad G=x^{\frac{q-1}{q-p}}G(\omega_1,\omega_2),\\[2mm]
& \qquad\qquad\qquad      \omega_1=x^{\frac{q-2p}{q-p}}u_{xx},\omega_2=x^{\frac{2q-3p}{q-p}}u_{xxx},\\[2mm]
& A^8_{4.5}=\langle \partial_t,\partial_u,x\partial_u,t\partial_t+qu\partial_u \rangle,\ q\neq 0:\\[2mm]
& \qquad\qquad        F=u_{xx}^{\frac{q-1}q}F(x,\frac{u_{xxx}}{u_{xx}}),\quad G=u_{xx}^{-\frac1q}G(x,\frac{u_{xxx}}{u_{xx}}),\\[2mm]
& A^9_{4.5}=\langle \partial_t,x\partial_u,\partial_u,t\partial_t+(p-q)x\partial_x+pu\partial_u \rangle,\ p\neq q:\\[2mm]
& \qquad\qquad        F=x^{\frac{1-4p+4q}{q-p}}F(\omega_1,\omega_2),\quad G=x^{\frac{1-p}{q-p}}G(\omega_1,\omega_2),\\[2mm]
& \qquad\qquad\qquad      \omega_1=x^{\frac{2q-p}{q-p}}u_{xx},\omega_2=x^{\frac{3q-2p}{q-p}}u_{xxx},\\[2mm]
& A^{10}_{4.5}=\langle \partial_u,\partial_t,x\partial_u,qt\partial_t+(1-p)x\partial_x+u\partial_u \rangle,\ p\neq 1:\\[2mm]
& \qquad\qquad        F=x^{\frac{q+4p-4}{p-1}}F(\omega_1,\omega_2),\quad G=x^{\frac{q-1}{p-1}}G(\omega_1,\omega_2),\\[2mm]
& \qquad\qquad\qquad      \omega_1=x^{\frac{2p-1}{p-1}}u_{xx},\omega_2=x^{\frac{3p-2}{p-1}}u_{xxx},\\[2mm]
& A^{11}_{4.5}=\langle \partial_u,\partial_t,x\partial_u,t\partial_t+u\partial_u \rangle:\\[2mm]
& \qquad\qquad        F=\frac1{u_{xx}}F(x,\frac{u_{xxx}}{u_{xx}}),\quad G=G(x,\frac{u_{xxx}}{u_{xx}}),\\[2mm]
& A^{12}_{4.5}=\langle \partial_u,x\partial_u,\partial_t,pt\partial_t+(1-q)x\partial_x+u\partial_u \rangle,\ q\neq 1:\\[2mm]
& \qquad\qquad        F=x^{\frac{p+4q-4}{q-1}}F(\omega_1,\omega_2),\quad G=x^{\frac{p-1}{q-1}}G(\omega_1,\omega_2),\\[2mm]
& \qquad\qquad\qquad      \omega_1=x^{\frac{2q-1}{q-1}}u_{xx},\omega_2=x^{\frac{3q-2}{q-1}}u_{xxx},\\[2mm]
& A^{13}_{4.5}=\langle \partial_u,x\partial_u,\partial_t,pt\partial_t+u\partial_u \rangle:\\[2mm]
& \qquad\qquad        F=\frac1{u_{xx}^p}F(x,\frac{u_{xxx}}{u_{xx}}),\quad G=u_{xx}^{1-p}G(x,\frac{u_{xxx}}{u_{xx}}),\\[2mm]
& A^{14}_{4.5}=\langle x\partial_u,\partial_t,\partial_u,qt\partial_t+(p-1)x\partial_x+pu\partial_u \rangle,\ p\neq 1:\\[2mm]
& \qquad\qquad        F=x^{\frac{4p-q-4}{p-1}}F(\omega_1,\omega_2),\quad G=x^{\frac{p-q}{p-1}}G(\omega_1,\omega_2),\\[2mm]
& \qquad\qquad\qquad      \omega_1=x^{\frac{p-2}{p-1}}u_{xx},\omega_2=x^{\frac{2p-3}{p-1}}u_{xxx},\\[2mm]
& A^{15}_{4.5}=\langle x\partial_u,\partial_u,\partial_t,pt\partial_t+(q-1)x\partial_x+qu\partial_u \rangle,\ q\neq 1:\\[2mm]
& \qquad\qquad        F=x^{\frac{4q-p-4}{q-1}}F(\omega_1,\omega_2),\quad G=x^{\frac{q-p}{q-1}}G(\omega_1,\omega_2),\\[2mm]
& \qquad\qquad\qquad      \omega_1=x^{\frac{q-2}{q-1}}u_{xx},\omega_2=x^{\frac{2q-3}{q-1}}u_{xxx}.\\[2mm]
\end{align*}

$A_{4.6}-$ invariant equations,
\begin{align*}
& A^1_{4.6}=\langle \partial_t,\partial_x,\partial_u,qt\partial_t+(px+u)\partial_x+(pu-x)\partial_u \rangle:\\[2mm]
& \qquad\qquad        F=\frac{\mathrm{e}^{(q-4p)\arctan{u_x}}}{(u_x^2+1)^2}F(\omega_1,\omega_2),\\[2mm]
& \qquad\qquad        G=-\frac{5u_xu_{xx}(2u_x^2u_{xxx}-3u_xu_{xx}^2+2u_{xxx})\mathrm{e}^{(q-4p)\arctan{u_x}}}{(u_x^2+1)^4}F(\omega_1,\omega_2)\\[2mm]
& \qquad\qquad\qquad     +(u_x^2+1)^\frac12\mathrm{e}^{(q-p)\arctan{u_x}}G(\omega_1,\omega_2),\\[2mm]
& \qquad\qquad\qquad       \omega_1=\frac{u_{xx}}{(u_x^2+1)^\frac32}\mathrm{e}^{-p\arctan{u_x}},\ \omega_2=\frac{u_x^2u_{xxx}-3u_xu_{xx}^2+u_{xxx}}{(u_x^2+1)^3}\mathrm{e}^{-2p\arctan{u_x}},\\[2mm]
& A^2_{4.6}=\langle \partial_t,\partial_u,x\partial_u,qt\partial_t-(x^2+1)\partial_x+(p-x)u\partial_u \rangle:\\[2mm]
& \qquad\qquad        F=(x^2+1)^4\mathrm{e}^{q\arctan{x}}F(\omega_1,\omega_2),\\[2mm]
& \qquad\qquad        G=4x(x^2+1)^2(2x^2u_{xxx}+3xu_{xx}+2u_{xxx})\mathrm{e}^{q\arctan{x}}F(\omega_1,\omega_2)\\[2mm]
& \qquad\qquad\qquad      +(x^2+1)^\frac12\mathrm{e}^{(q-p)\arctan{x}}G(\omega_1,\omega_2),\\[2mm]
& \qquad\qquad\qquad       \omega_1=(x^2+1)^\frac32\mathrm{e}^{p\arctan{x}}u_{xx},\ \omega_2=(x^2+1)^\frac32\mathrm{e}^{p\arctan{x}}(x^2u_{xxx}+3xu_{xx}+u_{xxx}).\\[2mm]
\end{align*}

$A_{4.7}-$ invariant equations,
\begin{align*}
& A^1_{4.7}=\langle \partial_u,\partial_x,t\partial_x+x\partial_u,-\partial_t+x\partial_x+2u\partial_u \rangle:\\[2mm]
& \qquad\qquad        F=\mathrm{e}^{-4t}F(u_{xx},\mathrm{e}^{-t}u_{xxx}),\quad G=\mathrm{e}^{-2t}G(u_{xx},\mathrm{e}^{-t}u_{xxx})-\frac{u_x^2}2,\\[2mm]
& A^2_{4.7}=\langle \partial_u,\partial_x,\partial_t+x\partial_u,t\partial_t+(t+x)\partial_x+(\frac {t^2}2+2u)\partial_u \rangle:\\[2mm]
& \qquad\qquad        F=(t-u_x)^3F(u_{xx},\omega),\quad G=(t-u_x)G(u_{xx},\omega)+(t-u_x)\ln{|t-u_x|},\\[2mm]
& \qquad\qquad\qquad      \omega=(t-u_x)u_{xxx},\\[2mm]
& A^3_{4.7}=\langle \partial_u,x\partial_u,-\partial_x,x\partial_x+(u-\frac{x^2}2)\partial_u \rangle:\\[2mm]
& \qquad\qquad        F=\frac{1}{(u_{xx}+1)^4}F(t,\omega),\quad G=\frac{1}{u_{xx}+1}G(t,\omega),\\[2mm]
& \qquad\qquad\qquad      \omega=\frac{u_{xxx}}{(u_{xx}+1)^2},\\[2mm]
& A^4_{4.7}=\langle \partial_u,x\partial_u,-\partial_x,\partial_t+x\partial_x+(u-\frac{x^2}2)\partial_u \rangle:\\[2mm]
& \qquad\qquad        F=\mathrm{e}^{4t}F(\omega_1,\omega_2),\quad G=\mathrm{e}^{t}G(\omega_1,\omega_2),\\[2mm]
& \qquad\qquad\qquad      \omega_1=\mathrm{e}^t(u_{xx}+1),\omega_2=\mathrm{e}^{2t}u_{xxx},\\[2mm]
& A^5_{4.7}=\langle \partial_u,x\partial_u,\partial_t-\partial_x,t\partial_t+x\partial_x+(\frac{t^2}2+tx+2u)\partial_u \rangle:\\[2mm]
& \qquad\qquad        F=(t+x)^3F(u_{xx},\omega),\quad G=(t+x)G(u_{xx},\omega)+(t+x)\ln{|t+x|},\\[2mm]
& \qquad\qquad\qquad    \omega=(t+x)u_{xxx},\\[2mm]
& A^6_{4.7}=\langle x\partial_u,\partial_u,x^2\partial_x+xu\partial_u,-x\partial_x+(u-\frac1{2x})\partial_u \rangle:\\[2mm]
& \qquad\qquad        F=x^8\mathrm{e}^{-4x^3u_{xx}}F(t,\omega),\\[2mm]
& \qquad\qquad        G=4x^6\mathrm{e}^{-4x^3u_{xx}}(2xu_{xxx}+3u_{xx})F(t,\omega)+x\mathrm{e}^{-2x^3u_{xx}}G(t,\omega),\\[2mm]
& \qquad\qquad\qquad    \omega=x^4\mathrm{e}^{-x^3u_{xx}}(xu_{xxx}+3u_{xx}),\\[2mm]
& A^7_{4.7}=\langle x\partial_u,\partial_u,x^2\partial_x+xu\partial_u,\partial_t-x\partial_x+(u-\frac1{2x})\partial_u \rangle:\\[2mm]
& \qquad\qquad        F=x^8\mathrm{e}^{4t}F(\omega_1,\omega_2),\\[2mm]
& \qquad\qquad        G=4x^6\mathrm{e}^{4t}(2xu_{xxx}+3u_{xx})F(\omega_1,\omega_2)+x\mathrm{e}^{2t}G(\omega_1,\omega_2),\\[2mm]
& \qquad\qquad\qquad    \omega_1=x^3u_{xx}+t,\ \omega_2=x^4\mathrm{e}^t(xu_{xxx}+3u_{xx}),\\[2mm]
& A^8_{4.7}=\langle x\partial_u,\partial_u,\partial_t+x^2\partial_x+xu\partial_u,t\partial_t-x\partial_x+(t+u+\frac{t^2x}2)\partial_u \rangle:\\[2mm]
& \qquad\qquad        F=x^5(1+tx)^3F(\omega_1,\omega_2),\\[2mm]
& \qquad\qquad        G=4x^5t(1+tx)[(2xu_{xxx}+3u_{xx})t+2u_{xxx}]F(\omega_1,\omega_2)+(1+tx)G(\omega_1,\omega_2)\\[2mm]
& \qquad\qquad\qquad    +(1+tx)\ln{|\frac{1+tx}x|},\\[2mm]
& \qquad\qquad\qquad    \omega_1=x^3u_{xx},\ \omega_2=x^4[(xu_{xxx}+3u_{xx})t+u_{xxx}].\\[2mm]
\end{align*}

$A_{4.8}-$ invariant equations,
\begin{align*}
& A^1_{4.8}=\langle \partial_u,\partial_t,\partial_x+t\partial_u,t\partial_t+qx\partial_x+(q+1)u\partial_u \rangle:\\[2mm]
& \qquad\qquad        F=u_x^{4q-1}F(\omega_1,\omega_2),\quad G=u_x^qG(\omega_1,\omega_2)+x,\\[2mm]
& \qquad\qquad\qquad      \omega_1=u_x^{q-1}u_{xx},\ \omega_2=u_x^{2q-1}u_{xxx},\\[2mm]
& A^2_{4.8}=\langle \partial_u,\partial_t,(t+x)\partial_u,t\partial_t+x\partial_x+(q+1)u\partial_u \rangle:\\[2mm]
& \qquad\qquad        F=x^3F(\omega_1,\omega_2),\quad G=x^qG(\omega_1,\omega_2)+u_x,\\[2mm]
& \qquad\qquad\qquad      \omega_1=x^{1-q}u_{xx},\ \omega_2=x^{2-q}u_{xxx},\\[2mm]
& A^3_{4.8}=\langle \partial_u,\partial_x,t\partial_x+x\partial_u,(1-q)t\partial_t+x\partial_x+(1+q)u\partial_u \rangle,\ q\neq1:\\[2mm]
& \qquad\qquad        F=t^\frac{3+q}{1-q}F(\omega_1,\omega_2),\quad G=t^\frac{2q}{1-q}G(\omega_1,\omega_2)-\frac{u_x^2}2,\\[2mm]
& \qquad\qquad\qquad      \omega_1=tu_{xx},\ \omega_2=t^\frac{q-2}{q-1}u_{xxx},\\[2mm]
& A^4_{4.8}=\langle \partial_u,\partial_x,t\partial_x+x\partial_u,x\partial_x+2u\partial_u \rangle,\ q\neq1:\\[2mm]
& \qquad\qquad        F=\frac{1}{u_{xxx}^4}F(t,u_{xx}),\quad G=\frac{1}{u_{xxx}^2}G(t,u_{xx})-\frac{u_x^2}2,\\[2mm]
& A^5_{4.8}=\langle \partial_u,\partial_x,\partial_t+x\partial_u,qt\partial_t+x\partial_x+(1+q)u\partial_u \rangle,\ q\neq0:\\[2mm]
& \qquad\qquad        F=(t-u_x)^\frac{4-q}q F(\omega_1,\omega_2),\quad G=(t-u_x)^\frac1q G(\omega_1,\omega_2),\\[2mm]
& \qquad\qquad\qquad      \omega_1=(t-u_x)^\frac{1-q}q u_{xx},\ \omega_2=(t-u_x)^\frac{2-q}q u_{xxx},\\[2mm]
& A^6_{4.8}=\langle \partial_u,\partial_x,\partial_t+x\partial_u,x\partial_x+u\partial_u \rangle:\\[2mm]
& \qquad\qquad        F=\frac{1}{u_{xx}^4}F(\omega_1,\omega_2),\quad G=\frac{1}{u_{xx}}G(\omega_1,\omega_2),\\[2mm]
& \qquad\qquad\qquad      \omega_1=t-u_x,\ \omega_2=\frac{u_{xxx}}{u_{xx}^2},\\[2mm]
& A^7_{4.8}=\langle \partial_u,x\partial_u,-\partial_x,-x\partial_x+t\partial_u \rangle:\\[2mm]
& \qquad\qquad        F=\frac{1}{u_{xx}^2}F(t,\omega),\quad G=G(t,\omega)+\frac{\ln{|u_{xx}|}}2,\\[2mm]
& \qquad\qquad\qquad      \omega=\frac{u_{xxx}}{u_{xx}^\frac32},\\[2mm]
& A^8_{4.8}=\langle \partial_u,x\partial_u,-\partial_x,qx\partial_x+(1+q)u\partial_u \rangle,\ q\neq 1:\\[2mm]
& \qquad\qquad        F=u_{xx}^\frac{4q}{1-q}F(t,\omega),\quad G=u_{xx}^\frac{1+q}{1-q}G(t,\omega),\\[2mm]
& \qquad\qquad\qquad      \omega=u_{xx}^\frac{1-2q}{q-1}u_{xxx},\\[2mm]
& A^9_{4.8}=\langle \partial_u,x\partial_u,-\partial_x,x\partial_x+2u\partial_u \rangle:\\[2mm]
& \qquad\qquad        F=\frac{1}{u_{xxx}^4}F(t,u_{xx}),\quad G=\frac{1}{u_{xxx}^2}G(t,u_{xx}),\\[2mm]
& A^{10}_{4.8}=\langle \partial_u,x\partial_u,-\partial_x,\partial_t+qx\partial_x+(1+q)u\partial_u \rangle:\\[2mm]
& \qquad\qquad        F=\mathrm{e}^{(1+q)t}F(\omega_1,\omega_2),\quad G=\mathrm{e}^{4qt}G(\omega_1,\omega_2),\\[2mm]
& \qquad\qquad\qquad      \omega_1=\mathrm{e}^{(q-1)t}u_{xx},\ \omega_2=\mathrm{e}^{(2q-1)t}u_{xxx},\\[2mm]
& A^{11}_{4.8}=\langle \partial_u,x\partial_u,\partial_t-\partial_x,qt\partial_t+(1+q)x\partial_x+qu\partial_u \rangle,\ q\neq0:\\[2mm]
& \qquad\qquad        F=(t+x)^3F(\omega_1,\omega_2),\quad G=(t+x)^\frac1q G(\omega_1,\omega_2),\\[2mm]
& \qquad\qquad\qquad      \omega_1=(t+x)^\frac{q-1}{q}u_{xx},\ \omega_2=(t+x)^\frac{2q-1}{q}u_{xxx},\\[2mm]
& A^{12}_{4.8}=\langle \partial_u,x\partial_u,\partial_t-\partial_x,x\partial_x \rangle:\\[2mm]
& \qquad\qquad        F=F(t+x,\frac{u_{xxx}}{u_{xx}}),\quad G=u_{xx}G(t+x,\frac{u_{xxx}}{u_{xx}}),\\[2mm]
& A^{13}_{4.8}=\langle \partial_u,x\partial_u,\partial_t-\partial_x,\alpha\partial_t+x\partial_x \rangle,\ \alpha\neq0:\\[2mm]
& \qquad\qquad        F=F(\omega_1,\omega_2),\quad G=\mathrm{e}^\frac{t+x}{\alpha}G(\omega_1,\omega_2),\\[2mm]
& \qquad\qquad\qquad      \omega_1=\mathrm{e}^{-\frac{t+x}{\alpha}}u_{xx},\ \omega_2=\mathrm{e}^{-\frac{t+x}{\alpha}}u_{xxx},\\[2mm]
& A^{14}_{4.8}=\langle x\partial_u,\partial_u,x^2\partial_x+xu\partial_u,-qx\partial_x+u\partial_u \rangle,\ q\neq1:\\[2mm]
& \qquad\qquad        F=x^\frac{4q+8}{1-q}u_{xx}^\frac{4q}{1-q}F(t,\omega),\\[2mm]
& \qquad\qquad        G=4x^\frac{6q+6}{1-q}u_{xx}^\frac{4q}{1-q}(3u_{xx}+2xu_{xxx})F(t,\omega)+x^\frac{2q+4}{1-q}u_{xx}^\frac{1+q}{1-q}G(t,\omega),\\[2mm]
& \qquad\qquad\qquad      \omega=x^\frac{2q+1}{1-q}u_{xx}^\frac{2q-1}{1-q}(3u_{xx}+xu_{xxx}),\\[2mm]
& A^{15}_{4.8}=\langle x\partial_u,\partial_u,x^2\partial_x+xu\partial_u,-x\partial_x+u\partial_u \rangle:\\[2mm]
& \qquad\qquad        F=\frac{F(t,x^3u_{xx})}{x^8(3u_{xx}+xu_{xxx})^4},\\[2mm]
& \qquad\qquad        G=\frac{3u_{xx}+2xu_{xxx}}{x^{10}(3u_{xx}+xu_{xxx})^4}F(t,x^3u_{xx})+\frac{G(t,x^3u_{xx})}{x^7(3u_{xx}+xu_{xxx})^2},\\[2mm]
& A^{16}_{4.8}=\langle x\partial_u,\partial_u,x^2\partial_x+xu\partial_u,x\partial_x+(tx+u)\partial_u \rangle:\\[2mm]
& \qquad\qquad        F=\frac{x^2}{u_{xx}^2}F(t,\omega),\\[2mm]
& \qquad\qquad        G=\frac{4(3u_{xx}+2xu_{xxx})}{u_{xx}^2}F(t,\omega)+x G(t,\omega)+\frac x2\ln{|x^3u_{xx}|},\\[2mm]
& \qquad\qquad\qquad      \omega=\frac{3u_{xx}+xu_{xxx}}{x^\frac12u_{xx}^\frac32},\\[2mm]
& A^{17}_{4.8}=\langle x\partial_u,\partial_u,x^2\partial_x+xu\partial_u,\partial_t-qx\partial_x+u\partial_u \rangle:\\[2mm]
& \qquad\qquad        F=x^8\mathrm{e}^{4qt}F(\omega_1,\omega_2),\\[2mm]
& \qquad\qquad        G=4x^6\mathrm{e}^{4qt}(3u_{xx}+2xu_{xxx})F(\omega_1,\omega_2)+x\mathrm{e}^{(1+q)t}G(\omega_1,\omega_2),\\[2mm]
& \qquad\qquad\qquad      \omega_1=x^3\mathrm{e}^{(q-1)t}u_{xx},\ \omega_2=x^4\mathrm{e}^{(2q-1)t}(3u_{xx}+xu_{xxx}),\\[2mm]
& A^{18}_{4.8}=\langle u\partial_x,\partial_x,\partial_t+xu\partial_x+u^2\partial_u,qt\partial_t+x\partial_x-qu\partial_u \rangle,\ q\neq0:\\[2mm]
& \qquad\qquad        F=\frac{u^5(1+t u)^3}{u_x^4}F(\omega_1,\omega_2),\\[2mm]
& \qquad\qquad        G=\frac{u^3(1+t u)^3}{u_x^6}[-5(2u_xu_{xxx}-3u_{xx}^2)u_{xx}u^2+8(u_xu_{xxx}-3u_{xx}^2)u_x^2u\\[2mm]
& \qquad\qquad\qquad     +8u_x^4u_{xx}]F(\omega_1,\omega_2)-u^{1-\frac1q}(1+t u)^\frac1q u_xG(\omega_1,\omega_2),\\[2mm]
& \qquad\qquad\qquad      \omega_1=\frac{u^{2+\frac1q}(1+t u)^{1-\frac1q}}{u_x^3}u_{xx},\\[2mm]
& \qquad\qquad\qquad      \omega_2=\frac{u^{6-\frac1q}(1+t u)^{\frac1q-2}}{u_x^5}[(u_xu_{xxx}-3u_{xx}^2)u+3u_x^2u_{xx}],\\[2mm]
& A^{19}_{4.8}=\langle u\partial_x,\partial_x,\partial_t+xu\partial_x+u^2\partial_u,x\partial_x \rangle:\\[2mm]
& \qquad\qquad        F=x^8F(\omega_1,\omega_2),\\[2mm]
& \qquad\qquad        G=4x^6(3u_{xx}+2xu_{xxx})F(\omega_1,\omega_2)+x^4u_{xx}G(\omega_1,\omega_2),\\[2mm]
& \qquad\qquad\qquad      \omega_1=t+\frac1x,\ \omega_2=\frac{x(3u_{xx}+xu_{xxx})}{u_{xx}}.\\[2mm]
\end{align*}

$A_{4.9}-$ invariant equations,
\begin{align*}
& A^1_{4.9}=\langle \partial_u,\partial_x,t\partial_x+x\partial_u,-(1+t^2)\partial_t+(q-t)x\partial_x+(-\frac{x^2}2+2qu)\partial_u \rangle:\\[2mm]
& \qquad\qquad        F=\mathrm{e}^{-4q\arctan t}(1+t^2)F(\omega_1,\omega_2),\quad G=\frac{\mathrm{e}^{-2q\arctan t}}{1+t^2}G(\omega_1,\omega_2)-\frac{u_x^2}2,\\[2mm]
& \qquad\qquad\qquad      \omega_1=(1+t^2)u_{xx}-t,\ \omega_2=\mathrm{e}^{-q\arctan t}(1+t^2)^\frac32u_{xxx}.\\[2mm]
\end{align*}

$A_{4.10}-$ invariant equations,
\begin{align*}
& A^1_{4.10}=\langle \partial_x,\partial_u,x\partial_x+u\partial_u,u\partial_x-x\partial_u \rangle:\\[2mm]
& \qquad\qquad        F=\frac{(1+u_x^2)^4}{u_{xx}^4}F(t,\omega),\\[2mm]
& \qquad\qquad        G=[15\frac{u_x^2}{u_{xx}}(1+u_x^2)^2-10\frac{u_xu_{xxx}}{u_{xx}^3}(1+u_x^2)^3]F(t,\omega)+\frac{(1+u_x^2)^2}{u_{xx}}G(t,\omega),\\[2mm]
& \qquad\qquad\qquad      \omega=\frac{u_{xxx}}{u_{xx}^2}(1+u_x^2)-3u_x,\\[2mm]
& A^2_{4.10}=\langle \partial_x,\partial_u,x\partial_x+u\partial_u,\partial_t+u\partial_x-x\partial_u \rangle:\\[2mm]
& \qquad\qquad        F=\frac{(1+u_x^2)^4}{u_{xx}^4}F(\omega_1,\omega_2),\\[2mm]
& \qquad\qquad        G=[15\frac{u_x(1+u_x^2)^2}{u_{xx}}(u_x-2\omega_1)-10\frac{u_{xxx}(1+u_x^2)^3}{u_{xx}^3}(u_x-\omega_1)]F(\omega_1,\omega_2)\\[2mm]
& \qquad\qquad\qquad      +\frac{(1+u_x^2)^2}{u_{xx}}G(\omega_1,\omega_2),\\[2mm]
& \qquad\qquad\qquad      \omega_1=t+\arctan{u_x},\ \omega_2=\frac{u_{xxx}}{u_{xx}^2}(1+u_x^2)+3(\omega_1-u_x),\\[2mm]
& A^3_{4.10}=\langle \partial_x,\partial_u,\partial_t+x\partial_x+u\partial_u,u\partial_x-x\partial_u \rangle:\\[2mm]
& \qquad\qquad        F=\frac{\mathrm{e}^{4t}}{(1+u_x^2)^2}F(\omega_1,\omega_2),\\[2mm]
& \qquad\qquad        G=-5\frac{u_xu_{xx}}{(1+u_x^2)^4}[2u_{xxx}(1+u_x^2)-3u_xu_{xx}^2]\mathrm{e}^{4t}F(\omega_1,\omega_2)+(1+u_x^2)^\frac12\mathrm{e}^tG(\omega_1,\omega_2),\\[2mm]
& \qquad\qquad\qquad      \omega_1=\frac{u_{xx}}{(1+u_x^2)^\frac32}\mathrm{e}^t,\ \omega_2=\frac{u_{xxx}(1+u_x^2)-3u_xu_{xx}^2}{(1+u_x^2)^3}\mathrm{e}^{2t},\\[2mm]
& A^4_{4.10}=\langle \partial_x,\partial_u,\partial_t+x\partial_x+u\partial_u,\partial_t+u\partial_x-x\partial_u \rangle:\\[2mm]
& \qquad\qquad        F=\frac{\mathrm{e}^{4(t+\arctan{u_x})}}{(1+u_x^2)^2}F(\omega_1,\omega_2),\\[2mm]
& \qquad\qquad        G=-5\frac{u_xu_{xx}}{(1+u_x^2)^4}[2u_{xxx}(1+u_x^2)-3u_xu_{xx}^2]\mathrm{e}^{3t+4\arctan{u_x}}F(\omega_1,\omega_2)\\[2mm]
& \qquad\qquad\qquad      +(1+u_x^2)^\frac12\mathrm{e}^{\arctan{u_x}}G(\omega_1,\omega_2),\\[2mm]
& \qquad\qquad\qquad      \omega_1=\frac{u_{xx}}{(1+u_x^2)^\frac32}\mathrm{e}^{t+\arctan{u_x}},\ \omega_2=\frac{u_{xxx}(1+u_x^2)-3u_xu_{xx}^2}{(1+u_x^2)^3}\mathrm{e}^{2(t+\arctan{u_x})},\\[2mm]
& A^5_{4.10}=\langle \partial_u,x\partial_u,u\partial_u,-(1+x^2)\partial_x-xu\partial_u \rangle:\\[2mm]
& \qquad\qquad        F=(1+x^2)^4F(t,\omega),\\[2mm]
& \qquad\qquad        G=4x(1+x^2)^2[2(1+x^2)u_{xxx}+3xu_{xx}]F(t,\omega)+(1+x^2)^2u_{xx}G(t,\omega),\\[2mm]
& \qquad\qquad\qquad      \omega=\frac{u_{xxx}}{u_{xx}}(1+x^2)+3x,\\[2mm]
& A^6_{4.10}=\langle \partial_u,x\partial_u,u\partial_u,\partial_t-(1+x^2)\partial_x-xu\partial_u \rangle:\\[2mm]
& \qquad\qquad        F=(1+x^2)^4F(\omega_1,\omega_2),\\[2mm]
& \qquad\qquad        G=-8(1+x^2)^2\int^t[\tan^2(\omega_1-s)+1][3(x-\tan(\omega_1-s))u_{xx}+(1+x^2)u_{xxx}]\mathrm{d}s\\[2mm]
& \qquad\qquad\qquad     F(\omega_1,\omega_2)+(1+x^2)^2u_{xx}G(\omega_1,\omega_2),\\[2mm]
& \qquad\qquad\qquad      \omega_1=t+\arctan{x},\ \omega_2=\frac{u_{xxx}}{u_{xx}}(1+x^2)+3(x-\omega_1),\\[2mm]
& A^7_{4.10}=\langle \partial_u,x\partial_u,\partial_t+u\partial_u,-(1+x^2)\partial_x-xu\partial_u \rangle:\\[2mm]
& \qquad\qquad        F=(1+x^2)^4F(\omega_1,\omega_2),\\[2mm]
& \qquad\qquad        G=4x(1+x^2)^2[2(1+x^2)u_{xxx}+3xu_{xx}]F(\omega_1,\omega_2)+(1+x^2)^\frac12\mathrm{e}^tG(\omega_1,\omega_2),\\[2mm]
& \qquad\qquad\qquad      \omega_1=(1+x^2)^\frac32\mathrm{e}^{-t}u_{xx},\ \omega_2=(1+x^2)^\frac32\mathrm{e}^{-t}[(1+x^2)u_{xxx}+3xu_{xx}],\\[2mm]
& A^8_{4.10}=\langle \partial_u,x\partial_u,\partial_t+u\partial_u,\partial_t-(1+x^2)\partial_x-xu\partial_u \rangle:\\[2mm]
& \qquad\qquad        F=(1+x^2)^4F(\omega_1,\omega_2),\\[2mm]
& \qquad\qquad        G=4x(1+x^2)^2[2(1+x^2)u_{xxx}+3xu_{xx}]F(\omega_1,\omega_2)+(1+x^2)^\frac12\mathrm{e}^{t+\arctan x}G(\omega_1,\omega_2),\\[2mm]
& \qquad\qquad\qquad      \omega_1=\frac{(1+x^2)^\frac32}{\mathrm{e}^{t+\arctan x} }u_{xx},\ \omega_2=\frac{(1+x^2)^\frac32}{\mathrm{e}^{t+\arctan x}}[(1+x^2)u_{xxx}+3xu_{xx}].\\[2mm]
\end{align*}

\end{document}